\documentstyle[12pt,equations,epsf]{article}
\input epsf.sty
\def\lsim{\mathrel{\raise.3ex\hbox{$<$\kern-.75em\lower1ex\hbox{$\sim$}}}}
\def\gsim{\mathrel{\raise.3ex\hbox{$>$\kern-.75em\lower1ex\hbox{$\sim$}}}}

    \def\fillboxx#1#2{\hbox to #1{\vbox to #2{\vfil}\hfil}    }

\def\ibid{{\it ibid.}}

\def\etal{{\it et al.}}

\def\dedx{dE/dx}

\def\mpi{m_{\pi}}

\def\cpmtwo{\wt \chi^{\pm}_2}

\def\dmchi{\Delta m_{\tilde\chi_1}}

\def\delgs{\delta_{GS}}

\def\etmiss{/ \hskip-7pt E_T}
\def\emiss{/ \hskip-7pt E}

\def\mslash{/ \hskip-7pt M}

\def\aeta{|\eta|}

\def\etc{{\it etc.}}

\def\eg{{\it e.g.}}

\def\gl{\wt g}
\def\mgl{m_{\gl}}

\def\tanb{\tan\beta}

\def\mz{m_Z}

\def\mgut{M_U}

\def\wp{W^+}

\def\wpm{W^{\pm}}

\def\cnone{\wt\chi^0_1}

\def\cntwo{\wt\chi^0_2}
\def\cnthree{\wt\chi^0_3}
\def\cnfour{\wt\chi^0_4}
\def\snu{\wt\nu}

\def\mcnone{m_{\cnone}}
\def\mcntwo{m_{\cntwo}}

\def\wt{\widetilde}

\def\cpone{\wt \chi^+_1}
\def\cmone{\wt \chi^-_1}
\def\cpmone{\wt \chi^{\pm}_1}
\def\cmpone{\wt \chi^{\mp}_1}
\def\mcpone{m_{\cpone}}
\def\mcpmone{m_{\cpmone}}

\def\cpmtwo{\wt \chi^{\pm}_2}

\def\stau{\wt \tau}
\def\mstau{m_{\stau}}

\def\wpm{W^{\pm}}

\def\wtil{\widetilde}

\def\gam{\gamma}

\def\anti{\overline}
\def\epem{e^+e^-}

\def\rts{\sqrt s}

\def\eg{{\it e.g.}}

\def\anti{\overline}
\def\wp{W^+}

\def\mz{m_Z}

\def\cm{~\mbox{cm}}
\def\fbi{~{\rm fb}^{-1}}
\def\fb{~{\rm fb}}
\def\pbi{~{\rm pb}^{-1}}

\def\mev{~{\rm MeV}}
\def\gev{~{\rm GeV}}
\def\tev{~{\rm TeV}}

\def\MPL #1 #2 #3 {{\sl Mod.~Phys.~Lett.}~{\bf#1} (#3) #2}
\def\NPB #1 #2 #3 {{\sl Nucl.~Phys.}~{\bf #1} (#3) #2}
\def\PLB #1 #2 #3 {{\sl Phys.~Lett.}~{\bf #1} (#3) #2}
\def\PR #1 #2 #3 {{\sl Phys.~Rep.}~{\bf#1} (#3) #2}
\def\PRD #1 #2 #3 {{\sl Phys.~Rev.}~{\bf #1} (#3) #2}
\def\PRL #1 #2 #3 {{\sl Phys.~Rev.~Lett.}~{\bf#1} (#3) #2}
\def\RMP #1 #2 #3 {{\sl Rev.~Mod.~Phys.}~{\bf#1} (#3) #2}
\def\ZPC #1 #2 #3 {{\sl Z.~Phys.}~{\bf #1} (#3) #2}
\def\IJMP #1 #2 #3 {{\sl Int.~J.~Mod.~Phys.}~{\bf#1} (#3) #2}
\def\NIM #1 #2 #3 {{\sl Nucl.~Inst.~and~Meth.}~{\bf#1} {#3} #2}
\def\JHEP #1 #2 #3 {{\sl JHEP}~{\bf#1} (#3) #2}

\newcommand{\nc}{\newcommand}
\nc{\beq}{\begin{equation}}   \nc{\eeq}{\end{equation}}
\nc{\bea}{\begin{eqnarray}}   \nc{\eea}{\end{eqnarray}}
\nc{\baa}{\begin{array}}      \nc{\eaa}{\end{array}}
\nc{\bit}{\begin{itemize}}    \nc{\eit}{\end{itemize}}
\nc{\ben}{\begin{enumerate}}  \nc{\een}{\end{enumerate}}
\nc{\bce}{\begin{center}}     \nc{\ece}{\end{center}}
\def\beqa{\begin{eqnarray}}
\def\eeqa{\end{eqnarray}}

\def\tightenlines{\def\baselinestretch{1.3}\small\normalsize}
\tightenlines
\begin{document}
\font\fortssbx=cmssbx10 scaled \magstep2
\hbox to \hsize{
$\vcenter{
\hbox{\fortssbx University of California - Davis}
}$
\hfill
$\vcenter{\normalsize
\hbox{\bf UCD-99-11} 
\hbox{\bf hep-ph/9906270}
\hbox{June, 1999}
}$
}
\begin{center}
{ \Large \bf A study of SUSY signatures at the Tevatron
in models with near mass
degeneracy of the lightest chargino and neutralino}
\rm
\vskip1pc
{\large\bf John F. Gunion and Stephen Mrenna}\\
\medskip
{\it Davis Institute for High Energy Physics}\\
{\it University of California, Davis, CA 95616}\\
\end{center}

\begin{abstract}
For some choices of soft SUSY--breaking parameters, the LSP is a stable
neutralino $\cnone$, the NLSP is a chargino $\cpmone$ almost degenerate in
mass with the LSP ($\dmchi\equiv\mcpmone-\mcnone\sim \mpi-$few GeV),
and all other sparticles are relatively heavy.  In this case, detection 
of sparticles in the usual, mSUGRA--motivated signals will be difficult, 
since the decay products in $\cpmone\to\cnone\ldots$ will be very soft,
and alternative signals must be considered. Here, we study the viability 
of signatures at the Tevatron based on highly--ionizing charged tracks, 
disappearing charged tracks, large impact parameters, missing transverse 
energy and a jet or a photon, and determine the
mass reach of such signatures assuming that only the $\cpmone$ and $\cnone$ 
are light. We also consider the jet+$\etmiss$ and $\gamma+\etmiss$ signatures 
assuming that the gluino is also light with $\mgl\sim \mcpmone$. We find 
that the mass reach is critically dependent upon $\dmchi$ and $\mgl-\mcpmone$. 
If $\dmchi$ is sufficiently big that $c\tau(\cpmone)\lsim$few cm and $\mgl$ 
is large, there is a significant possibility
that the limits on $\mcpmone$ based on LEP2 data cannot be extended at 
the Tevatron. If $c\tau(\cpmone)>$few cm, relatively background--free 
signals exist that will give a clear signal of $\cpmone$ production (for some 
range of $\mcpmone$) even if $\mgl$ is very large.
\end{abstract}

\section{Introduction}
In mSUGRA models, the soft SUSY--breaking parameters for the gauginos
satisfy a common boundary condition at the GUT scale, leading to a
relatively large mass splitting between the lightest chargino and 
the lightest neutralino (most often the LSP).
However, for different boundary conditions, this need not be the case,
and discovering SUSY may be more challenging or, at least, more difficult
to fully interpret.  Here, we focus on the possibility that
the $\cpmone$ and $\cnone$ are very degenerate in mass. This arises
naturally in two scenarios.
\ben
\item
{\boldmath$ M_2<M_1 \ll |\mu|$:} 
As reviewed in Ref.~\cite{bcgrunii2}, this hierarchy occurs when the 
gaugino masses are dominated by or entirely
generated by loop corrections. Models of this type include
the O--II superstring model \cite{ibanez,cdg1,cdg2,cdg3} 
and the closely related models in which SUSY--breaking
arises entirely from the conformal anomaly \cite{murayama,randall}.
The same hierarchy also occurs when SUSY is broken by
an $F$--term that is not an SU(5) singlet but
is rather is a member of the ${\bf 200}$ representation
contained in $({\bf 24}{\bf \times} 
{\bf 24})_{\rm symmetric}={\bf 1}\oplus {\bf 24} \oplus {\bf 75}
 \oplus {\bf 200}$ \cite{sm96nonuniv}.

It is demonstrative to give some 
specific results for the gaugino mass parameters
at the scale $\mgut$ ($\mz$). The O--II
model with $\delgs=-4$ yields $M_3:M_2:M_1=1:5:10.6$ ($6:10:10.6$),
the O--II model with $\delgs=0$ (equivalent to the 
simplest version of the conformal anomaly approach) yields
$M_3:M_2:M_1=3:1:33/5$ ($3:0.3:1$), 
while the ${\bf 200}$ model yields $1:2:10$ ($6:4:10$). As a result:
\bit
\item
In the $\bf 200$ model and the O--II $\delgs=0$ 
(or pure conformal anomaly) model,
$M_2$ is substantially below $M_1$ and $\dmchi\equiv \mcpmone- \mcnone$
can be very small. 
\item
In the O--II $\delgs=-4$ case, $M_2$ is only slightly
less than $M_1$ at low energies, but still $\dmchi<$ a few GeV is very typical
and $\dmchi<1\gev$ if $|\mu|\gsim 1\tev$ or if $\tanb$
is not large and RGE electroweak symmetry breaking is imposed \cite{cdg2}.
\item
In the $\bf 200$ model, and especially the O--II $\delgs=-4$ model, 
both $\mcntwo$ and $\mgl$ are typically quite close to the 
common $\cpmone,\cnone$ mass, and it is natural for the squark and slepton
masses to be much heavier than any of the gaugino masses. Typical values
of $|\mu|$ required by RGE electroweak symmetry breaking are large,
implying that the higgsino $\cpmtwo$, $\cnthree$ and $\cnfour$ states are very
heavy. 
\item
In the O--II $\delgs=0$ (or conformal anomaly) model, the gluino is typically
very heavy compared to the chargino.
\eit
\item
{\boldmath $ |\mu|\ll M_{1,2}$:}
In this case, the $\cpmone$, the $\cnone$ and the $\cntwo$ are all
closely degenerate and higgsino--like, while the gaugino states are
much heavier. Extreme degeneracy, $\dmchi<1\gev$, is only achieved for
$M_{1,2}\gsim 5\tev$. The squark and slepton masses might also be large.
Since currently there is less motivation for this scenario, we will
only make occasional remarks regarding it.

\een

The neutralino and chargino couplings to $W$ and $Z$ bosons
take the form $ig\gam^\mu[G_L P_L+G_RP_R]$,
where $P_L=(1-\gam_5)/2$ and $P_R=(1+\gam_5)/2$.
Ignoring CP violation, $G_L$ and $G_R$ are
\cite{ghinos}:
\bea
\wp\to\wtil\chi^+_i\wtil\chi^0_j:&& \quad 
G_L=-{1\over \sqrt 2}V_{i2}N_{j4}+V_{i1}N_{j2}\,,\quad
G_R=+{1\over \sqrt 2}U_{i2}N_{j3}+U_{i1}N_{j2}\,,\\
Z\to\wtil\chi^+_i\wtil\chi^-_j:&&\quad
G_L=V_{i1}V_{j1}+{1\over 2} V_{i2}V_{j2}\,,\quad
G_R=U_{i1}U_{j1}+{1\over 2} U_{i2}U_{j2}\,,\\\
Z\to\wtil\chi^0_i\wtil\chi^0_j:&&\quad
G_R=-G_L={1\over 2}\left(N_{i3}N_{j3}-N_{i4}N_{j4}\right)\,,
\eea
where the  $V,U$ matrices diagonalize the chargino mass matrix
(with $1,2$ referring to the $\wtil W^\pm,\wtil H^{\pm}$
basis) and $N$ diagonalizes the neutralino
mass matrix (with $1,2,3,4$ referring to the $\wtil B,\wtil W^0,
\wtil H_1^0,\wtil H_2^0$ basis). From these expressions we learn
the following.
\ben 
\item
When $M_2<M_1\ll|\mu|$, 
one finds $V_{11},U_{11}\sim 1$, $V_{12},U_{12}\sim 0$, 
$N_{11},N_{13},N_{14},N_{22},N_{23},N_{24}\sim 0$, and
$N_{12},N_{21}\sim 1$.  Focusing on the lighter
$\cpmone$, $\cnone$ and $\cntwo$ states,
the $Z\to \cnone\cnone$, $Z\to\cnone\cntwo$,
$Z\to\cntwo\cntwo$, and 
$\wpm\to \cpmone\cntwo$ cross sections are all small, while
$Z,\gam\to\cpone\cmone$ and $\wpm \to\cpmone\cnone$ can have large rates.
\item
When $|\mu|\ll M_{1,2}$, one has: 
$V_{11},U_{11}\sim 0$, $V_{12},U_{12}\sim {\rm
sgn}(\mu),1$;
$N_{11},N_{12},N_{21},N_{22}\sim 0$, $N_{13}=N_{14}=N_{23}=-N_{24}=1/\sqrt 2$.
In this case, the $Z,\gam\to \cpone\cmone$, $Z\to\cnone\cntwo$,
$\wpm\to\cpmone\cnone$, and $\wpm\to\cpmone\cntwo$ rates will be large
[but smaller than the unsuppressed channel rates in scenario (1)]
and $Z\to\cnone\cnone$, $Z\to \cntwo\cntwo$ are suppressed.
\een

\begin{figure}[ht!]
\leavevmode
\begin{center}
\epsfxsize=4.25in
\hspace{0in}\epsffile{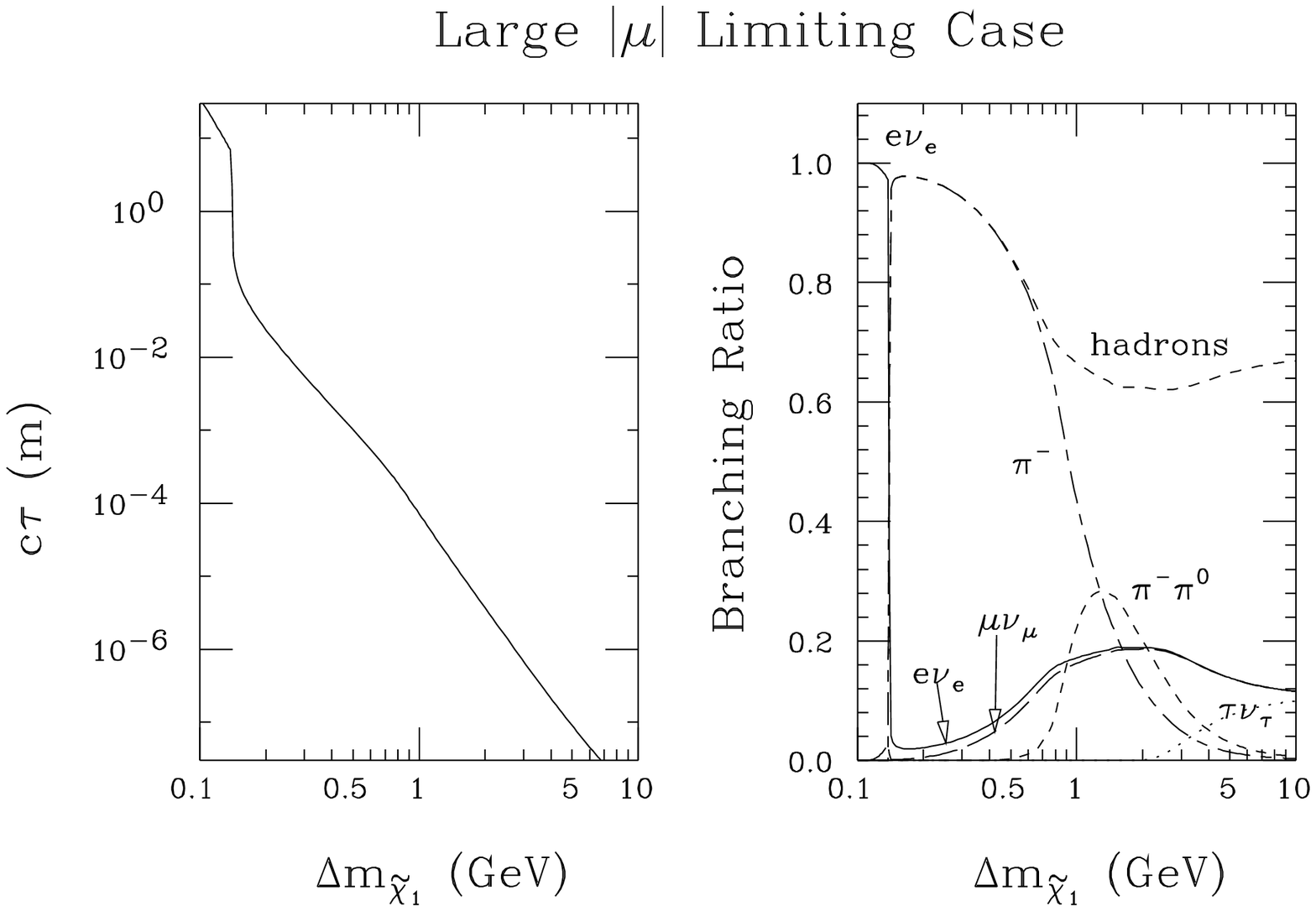}
\end{center}
\caption[]{The $c\tau$ and branching ratios
for $\cmone$ decay as a function of $\dmchi\equiv \mcpmone-\mcnone$
for the $M_2<M_1\ll|\mu|$ scenario. From Ref.~\cite{cdg3}.}
\label{lifebrsnew} 
\end{figure} 

The most critical ingredients in the phenomenology of such models
are the lifetime and the decay modes of the $\cpmone$, 
which in turn depend almost entirely
on $\dmchi$ when the latter is small. The $c\tau$ and branching ratios
of the $\cpmone$ as a function of $\dmchi$ have been computed
in Ref.~\cite{cdg3} (see also the closely related computation
for a nearly degenerate heavy lepton pair $L^\pm,L^0$ in Ref.~\cite{wells}) and
are illustrated in Fig.~\ref{lifebrsnew} for scenario (1).
For $\dmchi<\mpi$, only $\cpmone\to e^{\pm}\nu_e\cnone$ is important and
$c\tau>10$~m. Once $\dmchi>\mpi$, the $\cpmone\to \pi^\pm\cnone$ mode
turns on and is dominant for $\dmchi\lsim 800\mev$, at which
point the multi--pion modes start to become important:
correspondingly, one finds $c\tau\lsim 10-20$~cm
for $\dmchi$ just above $\mpi$ decreasing to $c\tau\sim 100~\mu$m
by $\dmchi\sim 1\gev$. In general, the exact value of $\dmchi$ is 
model dependent.  However, it is generally true that $\dmchi<\mpi$
is difficult to achieve. Even in scenario (1),
where the tree--level mass splitting can be extremely small, the electroweak
radiative corrections \cite{radcor} increase the mass splitting significantly;
one finds (see e.g. the $\dmchi$ graphs in \cite{cdg2}) that $\dmchi<\mpi$
is only possible for very special parameter choices. 
Most typically $\dmchi$ is predicted to lie 
in the range from slightly above $\mpi$ to several GeV.
As noted earlier, the tree--level value of $\dmchi$
in scenario (2) is normally substantially larger than $\mpi$,
and the one--loop corrections do not have much influence on the phenomenology.
For later reference, we present in Table~\ref{ctaus} the specific $c\tau$
values as a function of $\dmchi$ that we have employed in our Monte Carlo
studies.

\begin{table}[ht]
  \begin{center}
    \begin{tabular}[c]{|c|c|c|c|c|c|c|c|} \hline
$\dmchi(\mev)$ & 125 & 130 & 135 & 138 & 140 & 142.5 & 150 \\
$c\tau(\mbox{cm})$ & 1155 & 918.4 & 754.1 & 671.5 & 317.2 & 23.97 & 10.89 \\
\hline
$\dmchi(\mev)$ & 160 & 180 & 185 & 200 & 250 & 300 & 500 \\
$c\tau(\mbox{cm})$ & 6.865 & 3.719 & 3.291 & 2.381 & 1.042 & 0.5561 & 0.1023 \\
\hline
    \end{tabular}
    \caption{Summary of $c\tau$ values as a function of $\dmchi$
as employed in Monte Carlo simulations.}
    \label{ctaus}
  \end{center}
\end{table}

In order to set the stage for later, detailed studies, 
let us temporarily assume that only the
$\cpmone$ and $\cnone$ are light and briefly
preview the types of signals that will be important
for different ranges of $\dmchi$. For this discussion, we ask
the reader to imagine a canonical detector (e.g. CDF or D\O~at RunII) with
the following components ordered according to increasing radial distance
from the beam.
\bit
\item  An inner silicon vertex (SVX) detector extending radially
from the beam axis. The CDF RunII vertex detector has layers at 
$r\sim 1.6$, 3, 4.5, 7, 8.5 and $11\cm$
(the first and second layers are denoted L00 and L0, respectively) 
extending out to $z=\pm 45$ cm
\cite{cdfextra}. The D\O~SVX has 4 layers (but 2 are double--sided),
with the first at 2.5 cm and the last at 11 cm.
\item
A central tracker (CT) extending from $15\cm$ to $73\cm$ (D\O)
or from roughly $20\cm$ to $130\cm$ (CDF).
\item
A thin pre--shower layer (PS).
\item
An electromagnetic calorimeter (EC) and hadronic calorimeter (HC).
\item
The inner--most muon chambers (MC), starting just beyond the HC.  
The D\O~inner central muon chambers form (very roughly) a box, 
the ends of which (through
which the beam passes) are a $5.4~\mbox{m}\times 5.4~\mbox{m}$ square
and the sides of which are $8~\mbox{m}$ in length.
The sides (parallel to the beams)
cover $\aeta<1$, while the ends are instrumented out to $\aeta<2$.
The CDF inner muon chambers form roughly a barrel at a radial
distance of 3.5 m with length of about $5~\mbox{m}$.
There is no muon detection capability on the ends of the barrel, so
only $\aeta<0.6$ is covered by the inner chambers.
\item
Both CDF and D\O~will have a precise time--of--flight measurement
(TOF) for any charged particle that makes it to the muon chambers.
\eit
It is important to note that the SVX, CT and PS can all give (independent)
measurements of the $dE/dx$ from ionization of a track passing through them.
This will be important to distinguish a heavily--ionizing chargino
(which would be $\geq$ twice minimal ionizing [2MIP] for $\beta\gamma\leq
0.85$) from an isolated minimally ionizing particle [1MIP]. For example,
at D\O~the rejection against isolated 1MIP tracks will be
${\rm few}\times 10^{-3}$, ${\rm few}\times 10^{-3}$, and $\sim 10^{-1}$
for tracks that pass through the SVX, CT and PS, respectively, with an
efficiency of 90\% for tracks with $\beta\gamma<0.85$
\cite{glandsberg}.\footnote{It is a combination
of Landau fluctuations, electronic noise 
and, most importantly in hadron collisions, overlapping soft tracks 
that is responsible for these discrimination factors
being worse than one might naively expect.}
At CDF, the discrimination factors have not been studied in detail
but should be roughly similar \cite{dstuart}.
Because of correlations, one cannot simply multiply these numbers
together to get the combined discrimination power of the SVX, CT
and PS for an isolated track that passes through all three; for such
a track with $\beta\gamma<0.85$, 
the net discrimination factor would probably be
of order ${\rm few}\times 10^{-5}$.
A summary of our shorthand notations for detector components appears
in Table~\ref{detector}. 
At LEP/LEP2, the detector structure is somewhat different and important
features will be noted where relevant.
We now list the possible signals.

\begin{table}[ht]
  \begin{center}
    \begin{tabular}[c]{|c|p{5.5in}|} \hline
Component & Description \\
\hline
SVX & {Silicon vertex detector from close to beam pipe to $\sim$11 cm.}\\
CT  & {Central tracker starting just past SVX.}\\
PS  & {Pre--shower just outside the tracker.}\\
EC  & {Electromagnetic calorimeter.}\\
HC  & {Hadronic calorimeter.}\\
TOF & {Time--of--flight measurement after HC and 
just before MC.}\\
MC  & {Muon chamber with first layer after the HC and
just beyond the TOF.}\\
\hline
    \end{tabular}
    \caption{Summary of detector components referred to in the text.} 
    \label{detector}
  \end{center}
\end{table}

\clearpage
\noindent {\bf (a) LHIT and TOF signals:} 
\medskip

For $\dmchi<\mpi$, 
a heavy chargino produced in a collision travels a distance of order
a meter or more and will often penetrate to the muon chambers. 
If it does, the chargino may be distinguished from a muon by
heavy ionization in the SVX, CT and PS
(e.g., if $\beta\gamma<0.85$ the track will be at least 2MIP).
There should be no hadronic energy deposits associated
with this track, implying that the energy deposited in the hadronic
calorimeter should be consistent with ionization energy losses
for the measured $\beta$.
With appropriate cuts, such a signal should be background--free. This type of
long, heavily--ionizing track signal will be denoted as an LHIT signal.

If the chargino penetrates to the muon chambers, its large mass
will also be evident from the time delay of its TOF signal.
This delay can substitute for the heavy ionization requirement.
The passage of the chargino through the muon
chamber provides an adequate trigger
for the events. In addition, the chargino will be clearly visible
as an isolated track in the CT, and this track could
also be used to trigger the event. In later analysis (off--line even),
substantial momentum can be required for the track without
loss of efficiency. (The typical transverse
momentum of a chargino when pair--produced 
in hadronic collisions is of order 1/2 the mass.)

After a reasonable cut on $p_T$, the LHIT and TOF signals will
be background free. In addition,
we will find that if the chargino mass is near the upper limits
that can be probed by the LHIT and TOF signals, the LHIT and TOF
requirements can be simultaneously imposed
without much loss of efficiency.

\medskip
\noindent {\bf (b) DIT signals:} 
\medskip

For $\dmchi$ above but near $\mpi$, 
the chargino will often appear as an
isolated track in the central tracker but it will decay before
the muon chamber. (The appropriate mass range for which this
has significant probability is
$\mpi<\dmchi< 145\mev$, for which $c\tau\gsim 17\cm$.)
As such a chargino passes part way through the calorimeters
beyond the CT, it will deposit little energy. In particular,
any energy deposit in the hadronic calorimeter should be
no larger than that consistent with ionization energy deposits 
for the $\beta$ of the track as measured using ionization
information from the SVX+CT+PS. (In general,
the chargino will only deposit ionization energy up to the point of its decay.
Afterwards, the weakly--interacting neutralino will carry away most
of the remaining energy, leaving only a very soft pion or lepton remnant.)
Thus, we require that the track effectively disappear once it exits the CT.
(The point at which the ionization energy deposits end would typically
be observable in a calorimeter with sufficient radial segmentation,
but we do not include this in our analysis.)
Such a disappearing, isolated track will be called a DIT. The DIT
will have substantial $p_T$, which can be used to trigger the event.
A track with large $p_T$ from a background process
will either be a hadron, an electron or a muon. The first two will
leave large deposits in the calorimeters (EC and/or HC) and the latter
will penetrate to the muon chamber. Thus, the signal described
is very possibly a background--free signal. If not, a requirement
of heavy ionization in the SVX, CT and PS will certainly eliminate 
backgrounds, but with some sacrifice of signal events.

We will also consider the possibility of requiring that the DIT
track be heavily ionizing. In the most extreme case, we require
that the average ionization measured in the SVX, CT and PS correspond
to $\beta<0.6$ ($\beta\gamma<0.75$).  For a DIT signal, this is a very
strong cut once $\dmchi$ is large enough that the average $c\tau$
is smaller than the radius of the CT.
This is because rather few events will have both large enough $\beta\gamma
c\tau$ to pass all the way through the CT and small enough $\beta$
to satisfy the heavy ionization requirement.

\medskip
\noindent {\bf (c) STUB and KINK signals:} 
\medskip

For $145\mev<\dmchi<160\mev$, $17\cm>c\tau>7\cm$.
For such $c\tau$, the probability for the chargino to pass all the
way through the central tracker will be small, but the chargino will
be at least fairly likely to pass all the way through the SVX.
An off--line analysis will find a track in the SVX that
does not make it through the CT, and is certainly
not associated with any calorimeter energy deposits.
Such a short track we term a STUB. One may be able to detect the soft
pion that is emitted by the decay of the STUB chargino. At a hadron collider,
the primary difficulty associated with a STUB signal is that
it will not provide its own Level--1 trigger. (At both CDF and D\O,
information from the SVX can only be analyzed at Level--3 or later.)
One must trigger the event
by, for example, requiring missing transverse energy ($\etmiss$) above
some appropriate value and/or extra jets in the event.

Once an interesting event is triggered,
off--line analysis will provide a measurement of the ionization deposited
by the STUB in the SVX. However, $\beta\gamma<0.85$ (the 2MIP requirement)
can conflict with the requirement that the chargino pass all
the way through the SVX. For a given $c\tau$, 
the minimum $\beta$ required for a $\cpmone$ with small $|\eta|$
to reach the final SVX layer at $r=11\cm$ is given by
$\beta\gamma c\tau=11\cm$. In comparison, a chargino
with $\dmchi = 160\mev$ has
$c\tau=7\cm$ (on average) and $\beta\geq 0.85$ is typically required
to reach 11 cm; the corresponding $\beta\gamma$ value would be substantially
above 0.85 and the ionization level would be $\leq 2$MIP.
For smaller $\dmchi$, the $c\tau$ of the chargino is larger and smaller 
$\beta$ values will typically allow both $\beta\gamma<0.85$
and $\beta\gamma c\tau>11\cm$. Of course, the actual decay time
is characterized by an exponential distribution, so that for a given
$\dmchi$ some charginos will reach 11 cm with substantially lower
velocity than that needed on average. Similarly, charginos that decay
`late' can have a track that extends into the central tracker
even for $\beta\gamma$ substantially below $0.85$. 
Still, it would be better to find requirements that do
not employ heavy ionization but still leave us with a background--free signal.

In particular, it might be possible to see (in the CT) 
the charged pion that emerges
from the ``disappearing'' track seen as a STUB. 
In the rest frame of the chargino, the pion energy is given
by $E_\pi^*\simeq \dmchi$ and $p_\pi^*\simeq\sqrt{\dmchi^2-\mpi^2}$.
For $\mpi\leq\dmchi\leq 160\mev$, one has $0\leq p_\pi^*\leq 77\mev$,
but there will be some boost of the pion in going to the laboratory
frame. In this frame, the (transverse) radius of curvature of the pion is 
$R({\rm cm})={p_\pi^T(\mev)\over 3 B({\rm T})}$. For the CDF detector,
$B\sim 1.4$~T and $p_\pi^T\sim 77\mev$ yields $R\sim 18$~cm. Thus,
most of the soft pion tracks from charginos that decay
after passing through the vertex detector will be seen in the tracker.
Typically, the soft pion track that intersects the STUB track will
do so at a large angle, a signature we call a KINK.

\medskip
\noindent {\bf (d) HIP + KINK signals:}
\medskip

For $160\mev<\dmchi<190\mev$, $7\cm>c\tau>3\cm$.
Some of the produced charginos will decay late compared to $c\tau$
and yield a STUB signature of the type discussed just above. More typically,
however, the $\cpmone$ will pass through two to three layers of the SVX.
The $\cpmone$ track will then end and turn into a single charged pion 
with substantially different momentum. Both the sudden disappearance of
and the lack of any calorimeter energy deposits associated
with the $\cpmone$ track will help to distinguish it from 
other light--particle tracks that would normally
register in all layers of the SVX and in the calorimeters.

For $160\mev<\dmchi<190\mev$, $p_\pi^*\sim 77-130\mev$.
The corresponding transverse impact parameter resolution of the SVX,
$\sigma_b$, is approximately  $300-170~\mu$m 
(taking $p_\pi^T\sim p_\pi^*$ and applying the $1\sigma$ values from 
Fig.~2.2 of \cite{cdfextra} when L00 is included),
and is much smaller than the typical impact
parameter (which is a sizeable fraction of $\sim c\tau>3\cm$). 
Thus, we will consider a signal
consisting of an appropriate trigger (we will use large missing transverse
energy) and an isolated pion with high impact parameter (HIP) that
forms a KINK with a short track in the SVX. 

In the present study, we do not explicitly look for KINK's. This
would require going beyond the transverse impact parameter
and performing a three--dimensional reconstruction of the 
point at which the chargino decays and tracking the soft pion
(and all other charged tracks) through the magnetic field. 
Nor do we attempt to include discrimination against backgrounds
coming from ionization deposit measurements on the few SVX layers
that the chargino does pass through. In other words, we only
make use of the HIP signature in our estimates. Thus, sensitivity
based on our HIP studies in this mass range will be overly
conservative. Presumably, the actual experiments will do better.

\medskip
\noindent {\bf (e) HIP signals:}
\medskip

For $\dmchi>230\mev$, $c\tau<1.6\cm$
and the typical $\cpmone$ will not even pass through the innermost SVX layer
unless $\beta$ is very large . However, $p_\pi^*>180\mev$ and
the impact parameter resolution for the single emitted pion
moves into the $<150~\mu$m range.
For example, if $\dmchi=240,300,500,1000\mev$, $c\tau\sim 1.2,
0.37,0.09,0.007\cm$ while $p_\pi^T\sim 195,265,480,990\mev$ yields
$1\sigma$ impact parameter resolutions of $\sigma_b\sim 120,90,50,25~\mu$m. 
We will explore a signal based on events defined by an appropriate trigger
and the presence of one or more large--$b$ charged pions. For the trigger,
we will employ a requirement of substantial missing energy.
So, we are once again dealing with the HIP signature.

Once $\dmchi>1\gev$
a HIP signature will not be useful and we must consider the chargino
decay to be prompt. This is because the largest possible impact parameter
is only a few times the $1\sigma$ value for the resolution and will be
dominated by fakes.
This leads us to one of two completely different types of analysis.

\medskip
\noindent {\bf (f) \boldmath $\gam+\etmiss$ and jets+$\etmiss$ signals:}
\medskip

For some interval of $\dmchi$ (e.g. $200\mev\lsim \dmchi\lsim
300\mev$ at the DELPHI LEP/LEP2 detector --- see later --- or, perhaps,
$1\gev\lsim\dmchi\lsim 10-20\gev$ at the Tevatron)
the decay products (hadron(s) or $\ell\nu$) produced along with the $\cnone$
will be too soft to be distinctively visible in the main part
of the detector and at the same time high--impact--parameter tracks associated
with chargino decay will
not be apparent. One will then have to detect chargino
production as an excess of events with an isolated photon or missing energy
above a large $\gam+\etmiss$ or jet(s)+$\etmiss$ background.
In the jet(s)+$\etmiss$ case, the most reliable signal will
result from requiring exactly one jet, i.e. monojet+$\etmiss$.
For some values of the chargino mass and $\dmchi$, an
excess in these channels could confirm the SVX signals discussed earlier.

\medskip
\noindent {\bf (g) standard mSUGRA signals:}
\medskip

For large enough $\dmchi$, the extra lepton or hadron 
tracks from $\cpmone$ decay will be sufficiently energetic to be detected
and will allow identification of chargino production events
when associated with a photon or missing energy trigger.
A detailed simulation
is required to determine exactly how large $\dmchi$ needs to be
for this signal to be visible above backgrounds. At LEP/LEP2,
backgrounds are sufficiently small that the extra
tracks are visible for $\dmchi\gsim 300\mev$ in association
with a photon trigger while standard mSUGRA searches
based on missing energy and jets/leptons require $\dmchi\gsim 3\gev$.
At a hadron
collider we estimate that $\dmchi\gsim 10-20$ GeV will be necessary to 
produce leptons or jets sufficiently energetic to produce a distinctive event
assuming a missing energy trigger.

In order to assist the reader with our shorthand notations,
Table~\ref{signals} gives shortened definitions for our signals.

\begin{table}[p]
  \begin{center}
    \begin{tabular}[c]{|c|p{5.5in}|} \hline
Signal & Definition \\
\hline
LHIT & {Long, heavily--ionizing ($\geq$ 2MIP's as measured by SVX+CT+PS),
large--$p_T$ track that reaches the MC.
The energy deposit in the HC in the track direction must be consistent
with expected ionization energy deposit
for the $\beta$ measured (using TOF and/or SVX+CT+PS), i.e.
no hadronic energy deposit.}\\
\hline
TOF  & {A large--$p_T$ track seen in the SVX and CT along with a 
signal in the TOF delayed by 500 ps or more (vs. a particle with
$\beta=1$). HC energy deposit (in the direction of the track)
is required to be consistent with the ionization expected
for the measured $\beta$ (i.e. no hadronic deposit).}\\
\hline
DIT  & {An isolated, large--$p_T$ 
track in the SVX and CT that fails to reach the MC
and deposits energy in the HC no larger than that consistent with
ionization energy deposits for the measured (using SVX+CT+PS) $\beta$.
Heavy ionization in the SVX+CT+PS, corresponding
to $\beta<0.8$ or $\beta<0.6$ (DIT8 or DIT6), may be required.}\\
\hline
KINK & {A track that terminates in the CT, turning into a soft,
but visible,
charged--pion daughter--track at a substantial angle to parent.}\\
\hline
STUB & {An isolated, large--$p_T$ (as measured using SVX) track 
that registers in all SVX layers, but does not pass
all the way through the CT. 
Energy deposits in the EC and HC in the direction of the track
should be minimal.}\\
\hline
SNT  & {One or more STUB tracks with no additional trigger. 
Heavy ionization of the STUB in the SVX,
corresponding to $\beta<0.6$ (SNT6), may be required.}\\
\hline
SMET & {One or more STUB tracks with an $\etmiss>35\gev$ trigger.
Heavy ionization of the STUB in the SVX,
corresponding to $\beta<0.6$ (SMET6), may be required.}\\
\hline
HIP  & {A high--impact--parameter ($b\geq 5\sigma_b$)
track in the SVX, with large $\etmiss$
triggering, perhaps in association with a visible KINK in the SVX.}\\
\hline
$\gamma+\etmiss$ & {Isolated, large--$p_T$ photon and large $\etmiss$.}\\
\hline
monojet+$\etmiss$ & {Large--$p_T$ jet and large $\etmiss$.}\\
\hline
mSUGRA--like & {jet(s)+$\etmiss$, tri--leptons, like--sign di--leptons, 
$\etc$,
except that the cross section for the $\cpmone\cntwo$ tri--lepton signal can be
suppressed.}\\
\hline
    \end{tabular}

    \caption{Summary of signals. MIP refers to a minimally--ionizing--particle
such as a $\beta=1$ muon. For detector component notation, see
Table~\ref{detector}.}
    \label{signals}
  \end{center}
\end{table}

The backgrounds to and efficiencies for all 
these various signals will be different
at a hadron collider vs. an $\epem$ collider, and will be
detector dependent. Further, at
a hadron collider, gluino production can greatly enhance the various
signals outlined above. In particular, gluino production
leads to a missing energy signal when the gluinos decay invisibly,
and to LHIT, TOF, DIT, CT, STUB and/or HIP signals 
when $\gl\to \cpmone q^\prime\anti q$ and the $\cpmone$ is long--lived.

\section{Collider Phenomenology of degenerate models}

Although our main focus will be on Tevatron RunII, it is useful
to summarize which of the above signals have been 
employed at LEP2 and the resulting constraints on the degenerate
scenarios we are considering.

\subsection{Lepton Colliders}

As discussed above and in Refs.~\cite{cdg1,cdg2,cdg3},
collider phenomenology depends crucially on $\dmchi$.
Most importantly, SUSY detection depends on which aspects (if any)
of the $\cpone\cmone$ final state are visible. 

\bit
\item If the $\cpmone$ decay products are soft and the $\cpone\cmone$
production is otherwise untagged, the event may be 
indistinguishable from the large
$\epem\to\epem\gam\gam\to\epem+{\rm soft}$ background.
In this case, one will need to tag $\cpone\cmone$ 
production. 
The proposal of Ref.~\cite{cdg1} is to employ a photon tag.  Such
a photon can arise from 
initial or final state radiation of a photon, denoted as ISR.
Even with an ISR tag, 
it is possible that the $\cpone$ and $\cmone$ will both be effectively
invisible because of the softness of their decay products and
the lack of a vertex detector signal. 
In this case, $\gam\cpone\cmone$ production is observable
as a $\gamma\mslash$ signature, which is distinguishable 
from the $\gam\nu\anti\nu$ process by the threshold in the missing
mass variable~$\mslash=\sqrt{(p_{e^-}+p_{e^+}-p_{\gam})^2}$
at~$\mslash=2\mcpmone$. 
The exact mass reach in $\mcpmone$                         
depends upon available luminosity and machine energy. Estimates
were presented in Ref.~\cite{cdg1}, 
which we summarize for the $M_2<M_1\ll|\mu|$ 
scenario (1). At LEP2, for $L=125\pbi$ per experiment,
no improvement was found over the $\mcpmone<45\gev$
limit coming from LEP1 $Z$--pole data on $Z\to$invisible decay channels.
At the NLC, the prospects are better: with $L=50\fbi$,
the $\gam\mslash$ channel will be sensitive
up to $\mcpmone\sim 200\gev$.
In scenario (2), both $\gam\cpone\cmone$
and $\gam\cnone\cntwo$ will have significant rates and a common
threshold in~$\mslash$, and the discovery reach is similar to that in 
scenario (1).
\item
The experimental situation is greatly improved if the LHIT
signal can be employed or if the soft pions from the $\cpmone$
decays in $\gam\cpone\cmone$ events can be detected.
This is most clearly illustrated
by summarizing the analysis from DELPHI at LEP2 \cite{delphideg}.
This analysis  employs (in order of increasing radius from the
beampipe)
their central ID and TPC tracking devices and the ring--imaging Cherenkov
device RICH. (For details regarding these devices, please refer
to Ref.~\cite{delphideg}.)
\bit
\item
For scenario (1) or (2), when $\dmchi\lsim 200\mev$, the 
charginos are sufficiently long--lived to produce 
one of two signals for $\cpone\cmone$ production.
\begin{description}
\item{(a)} For $\dmchi\lsim \mpi$, the charginos
produce heavily--ionizing tracks (LHIT's) recognizable by high
specific ionization in the TPC or by the absence of Cherenkov
light in the RICH. 
\item{(b)} For $\mpi\lsim\dmchi\lsim 200\mev$,
both the charginos and their soft pion daughters
yield visible tracks in the central tracking devices (the ID and the TPC,
located in the region $10~{\rm cm}<r<1~{\rm m}$). A clean signal
is provided by demanding two primary 
particle tracks emitted in opposite hemispheres, each decaying 
to a soft, charged daughter moving at a substantial angle to
the primary track. This type of signal is called a KINK.
\end{description}
Note that no additional trigger is required for either signal. As
a result, by combining (a) and (b), DELPHI is able to exclude
$\mcpmone$ out to nearly the kinematic limit (currently 90 GeV).
\item
When $\dmchi\gsim$ 3 GeV, the decay products of the $\cpmone$
become easily visible, and the standard mSUGRA search results apply; the
$\cpmone$ is excluded out to the kinematic limit (90 GeV for the data
sets analyzed), except for the case of a relatively
light sneutrino, for which the $\cpone\cmone$ cross
section is smaller and the limit does not extend past 75 GeV.
\item
For $200\mev\lsim\dmchi\lsim3\gev$, the chargino tracks are not long
enough to use the ID/TPC kink signature, and the chargino decay
products are too soft to provide a clear signature on their own.
In this case, one must overcome the very large $\gamma\gamma$ collision
background rate for events containing only soft tracks by tagging
the chargino pair production event. As proposed in Ref.~\cite{cdg1},
DELPHI employs an ISR photon tag. The photon
is required to have energy above $4\gev$ and the recoil mass $\mslash$
is required to be above $96\gev$ (which
eliminates all but the virtual $Z$ tail of
$\gam Z^*\to\gam \nu\anti\nu$ events and the non--resonant contributions).
Visible energy (excluding the photon) is required to be less than
a few percent of $\sqrt s$
(the exact value depends upon the $\dmchi$ value being probed). 
Finally, in order to essentially eliminate the $\gam\nu\anti\nu$
background, the event is required to contain soft charged tracks
consistent with the isolated pions expected from the chargino decays.
DELPHI observed no events after all cuts. 
For scenario (1) and a heavy (light) sneutrino, 
this excludes $\mcpmone\lsim 62\gev$ ($49\gev$)
for $0.3\lsim\dmchi\lsim 3\gev$ ($0.5\lsim\dmchi\lsim 3\gev$).
The gap from $0.2-0.3\gev$ ($0.2-0.5\gev$) arises because of the low
efficiency for detecting very soft pions.\footnote{With the ISR tag,
the $\gam\gam$ background is completely negligible.}
Finally, for scenario (2), the ISR signature excludes $\mcpmone\lsim 48\gev$
for $0.3\lsim\dmchi\lsim 3$ GeV and $\mcpmone\lsim 50\gev$
for $1\lsim\dmchi\lsim 3$ GeV.

Thus, there is a gap from just above $\dmchi\sim 200\mev$ to at least 
$300\mev$ for which the chargino is effectively invisible. DELPHI finds
that the $\gam\mslash$ signature, discussed earlier, is
indeed insufficient to improve over the $\mcpmone\gsim 45\gev$
limit from $Z$ decays. We are uncertain whether DELPHI explored
the use of high--impact--parameter tracks in their vertex detector
(in association with the ISR trigger) to improve their sensitivity
(by sharply reducing the $\gam\nu\anti\nu$ background) in these gap regions.

\eit
\eit

\subsection{Hadron Colliders}

At hadron colliders, typical signatures of mSUGRA are tri--lepton events
from neutralino--chargino production, like--sign di--leptons from
gluino pair production, and multijets$+\etmiss$ from squark and
gluino production. The tri--lepton signal from $\cpmone\cntwo$
production and the like--sign di--lepton signal from $\gl\gl$ production
are both suppressed when $\dmchi$ is small by the softness of the leptons
coming from the $\cpmone$ decay(s). In $M_2<M_1\ll|\mu|$ scenarios,
the tri--lepton signal is further diminished
by the suppression of the $\cpmone\cntwo$ cross section.
In $|\mu|\ll M_2,M_1$ scenarios, $\mcntwo\simeq\mcnone$ and
even though the $\cpmone\cntwo$ cross section is not suppressed
the $\cntwo$ decay products, like those of the $\cpmone$, are very soft,
yielding further suppression of the tri--lepton signal.
Provided that $\mgl$ is light enough, the most obvious signal
for SUSY in degenerate models is jet(s) plus missing energy. 
Even if the gluino is rather degenerate with the $\cpmone$ and $\cnone$,
it has been shown \cite{cdg2} (see also \cite{bcgrunii2})
that the Tevatron and LHC will probe
a significant (albeit reduced compared to mSUGRA boundary conditions)
range of $\mgl$.  This is true since initial state gluon radiation
can be used to `tag' the missing energy.  This search can also be augmented
by the $\gamma\gl\gl$ process, where the $\gamma$ is the tag.
As $\mgl-\mcpmone$ increases, the jets from $\gl\to q\anti q\cpmone+q\anti
q\cnone$ decays become visible and the jet(s)+$\etmiss$ signature
initially becomes stronger \cite{cdg2} despite the decrease in the $\gl\gl$
production cross section. However, it is entirely possible that the gluino is
much heavier than the light $\cpmone,\cnone$ states
and that the $\gl\gl$ production rate (at the Tevatron at least)
will be quite suppressed.  In this case, the ability to detect events
in which the only directly produced SUSY particles
are light neutralino and chargino states could prove critical.  
In the remainder of this paper, we consider the
sfermion, and heavier chargino and neutralino states to be 
extremely heavy, and investigate methods to probe degenerate models
at the Tevatron.  Expectations for scenarios where the gluino
has a mass comparable to $\cpmone$ will be given less discussion.
First, we study whether photon
tagging (which we noted above is useful at a lepton collider)
or jet tagging (as employed in many studies)
might provide a viable signal when the
$\cpmone$ decay is effectively prompt 
and its decay products are too soft to be visible in the detector.
Later, we consider the modifications to this picture when the $\cpmone$
decay is not prompt.

In the following, we perform particle level studies using either
the processes contained in the PYTHIA 6.125 \cite{pythia}
event generator or by adding
external processes (several of the $\gamma+X$ processes considered here)
into PYTHIA.  
A calorimeter is defined out to $\eta=4.4$ with a Gaussian $E_T$ resolution
of $\sigma_{E_T}=80\%/\sqrt{E_T}$.  Jets with $E_T>5$ GeV and $R=0.5$ are
reconstructed to define $\etmiss$.
Non--Gaussian contributions will be estimated as described later.
Charged track momenta and impact parameters $b$ are unsmeared, but the
effects of detector resolution on $b$ are included.

\subsubsection{Pure $\gamma$ or jet and $\etmiss$  signatures}

One of the most challenging possibilities in degenerate models is 
when the $\cpmone$ decay is prompt and its decay products are 
too soft to be visible. At leading order in perturbation theory, 
$\cpmone(\to\cnone+{\rm soft})\cnone$ and 
$\cpone(\to\cnone{+\rm soft})\cmone(\to\cnone+{\rm soft})$ production
provide no good signature since the missing transverse momenta
of the LSP's essentially cancels and the soft decay products are
obscured by detector resolution and the combined effect of
the underlying event and fragmentation/hadronization.
However, it may still be possible to observe the 
transverse momentum of the LSP's
if a high--$p_T$ jet or photon is also produced in an event.

In the absence of mismeasurements,
the major physics background to $\gamma+\etmiss$ at the Tevatron is
$\gam Z(\to \nu\anti\nu)$ and $\gam\tau^\pm\nu_\tau$ production, where
$\tau(\to\etmiss+{\rm soft})$.
In reality, mismeasurements of jets can produce
a false $\etmiss$, and the loss of a track can cause
an electron to fake a photon.  A realistic search strategy
should account for these potential backgrounds and/or rely on
cuts that reduce them to a manageable level.  If we ignore
such backgrounds, we need to understand under what conditions this
is reasonable.
We can gain some insight into the relative importance of mismeasurement
backgrounds from RunI analyses.  
The D\O~RunI measurement of the $\gamma Z(\to\nu\bar\nu)$ signal
(which is a {\it background} to our signature) has a background
from $W(\to e^\pm\nu_e)$ when the $e$ fakes a $\gamma$.  For RunI,
the fake probability was roughly a
constant with magnitude $R_{e\to\gamma}=5\times 10^{-3}$.
The background estimate in the D\O~analysis is well reproduced
by generating $W(\to e^\pm\nu_e)$ events with PYTHIA, replacing
the $e$ with a $\gamma$, weighting the event by an additional
factor $R_{e\to\gamma}$, and performing all other cuts.
The value of $S/B$ is about $0.3$, but the $W$ contribution becomes
negligible once $p_T^{\gamma}\gsim 50-60$ GeV, which is beyond
the Jacobian peak for the electron $p_T$ spectrum.
Another significant background arises from $\gamma+$jet, where
the jet fakes $\etmiss$.  For $\etmiss>40$ GeV, this probability
can be conservatively estimated\footnote{Note
that $R_{j\to\,\etmiss}$ represents a non--Gaussian component to the $\etmiss$
resolution; the Gaussian component is already included in our calorimeter
simulation.} at $R_{j\to\,\etmiss}=10^{-4}$. 
This background can also be reproduced by generating
$\gamma+q,\gamma+g$ events with PYTHIA, demanding only one
reconstructed jet with $E_T>15$ GeV, discarding this jet,
weighting the event by an additional factor $R_{j\to,\etmiss}$,
calculating $\etmiss$ from the sum of all remaining jets,
and performing all remaining cuts.  Once $\etmiss>50$ GeV,
this background is roughly 5\% of the $\gam Z(\to\nu\anti\nu)$
signal, and decreases quickly with increasing $\etmiss$.  Since we
can reproduce the mismeasurement backgrounds in a simple manner,
we feel confident that we can reasonably estimate the full
background.  Additionally, we will set our cuts so that
the mismeasurement backgrounds are smaller than the physics ones,
which, for the chargino signal, will be dominated by $\gam Z(\to\nu\anti\nu)$.
To reflect the detector improvements in Run II,
we use the factor $R_{e\to\,\gamma}=10^{-3}$.

\begin{figure}[ht]
\leavevmode
\begin{center}
\epsfxsize=3.25in
\epsfysize=3.25in
{\hspace{0in}\epsffile{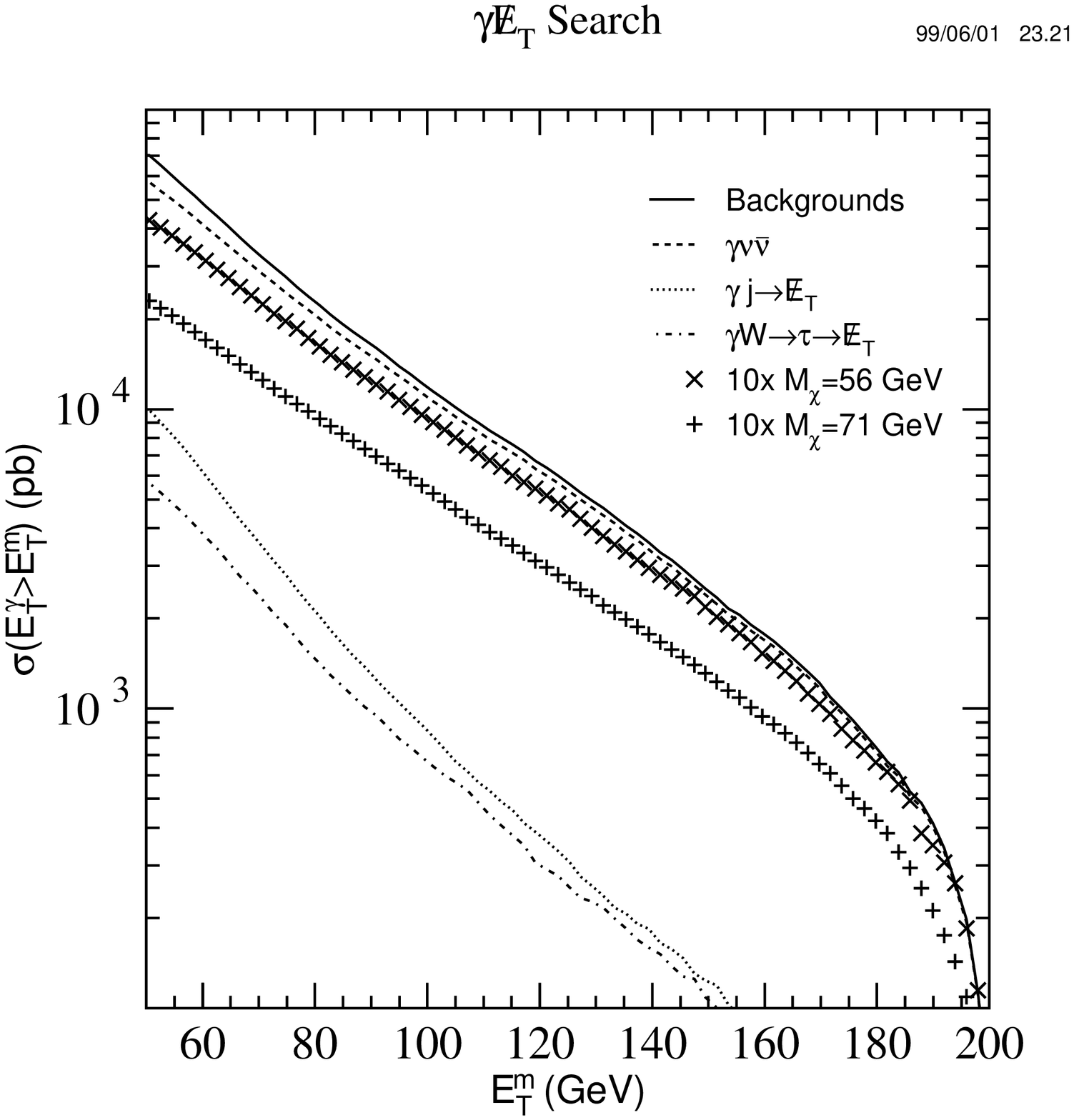}
\epsfxsize=3.25in
\epsfysize=3.25in
\hspace{0in}\epsffile{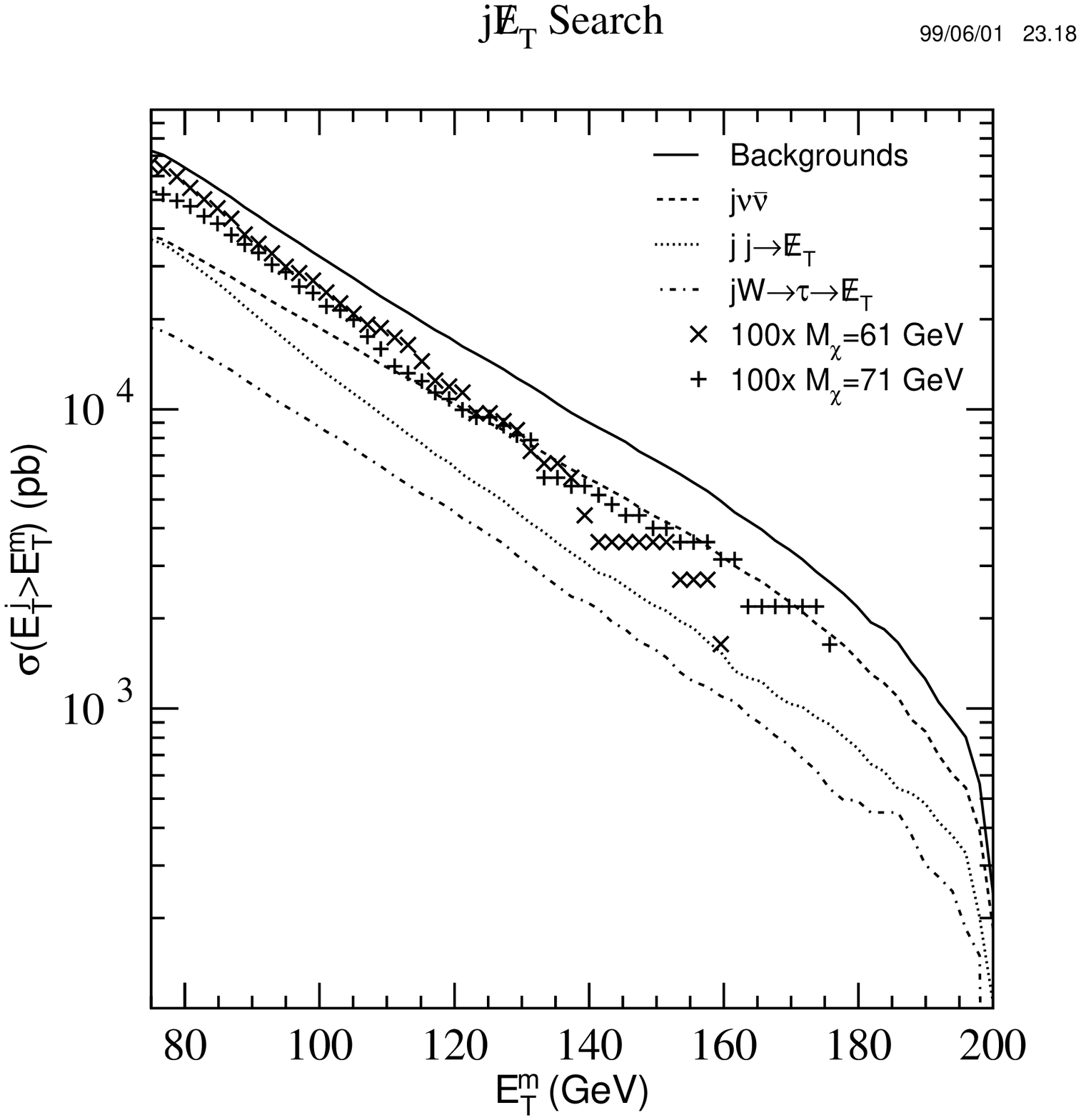}}
\end{center}
\caption[]{Comparison of the signal and backgrounds
for $\gam+\etmiss$ and $j+\etmiss$ searches
at the Tevatron for $\rts=2\tev$. In (a), we
plot the sum of the $\gam\cpone\cmone$ and
$\gamma\cpmone\cnone$ cross sections integrated over 
$E_T^\gamma,\etmiss>E_T^{\rm min}$; the additional cuts imposed
are given in of Eq.~(\ref{gamcuts}). 
Results are given for $\mcpmone\simeq\mcnone\sim 56\gev$
and $\sim 71\gev$; the signal has been multiplied by a factor of 10.
In (b), we plot the $j\cpone\cmone$ and $j\cpmone\cnone$ `monojet'
cross sections integrated over $E_T^j,\etmiss>E_T^{\rm min}$, after imposing
the additional cuts given in Eq.~(\ref{jcuts}). Results are presented for
$\mcpmone\simeq\mcnone\sim 61\gev$ and $\sim 71\gev$; the signal
has been multiplied by a factor of 100.}
\label{gamorjtagxsec}
\end{figure} 

We have studied $\gam\cpone\cmone$ and $\gam\cpone\cnone$ production
(computed for various values of $M_2$ taking $M_1=1.5M_2$, $\tanb=5$ 
and $\mu=1\tev$ --- typically, $\mcpmone\simeq\mcnone$ is close to $M_2$)
and the expected backgrounds identified above.
Because only $\etmiss$ (and not $\emiss$) can be measured,
we cannot perform a cut that eliminates the real $Z$
background from $\gam Z$ production. 
At best, the distributions for signal and background in $E_T^\gam$ may be
sufficiently different that a cut requiring high $E_T^\gam$
will allow a reasonable signal--to--background ratio while retaining
adequate cross section for the signal.
To demonstrate this, we plot in Fig.~\ref{gamorjtagxsec}(a) the 
$\gam\cpone\cmone$ and $\gam\cpmone\cnone$ integrated
signal and $\gam+\etmiss$ background (and some components thereof) 
as a function of a minimum accepted value for $E_T^\gam$.
(Note that the signal is multiplied by a factor of 10 in the figure.) 
Our nominal cuts are:
\bea
&E_T^\gam>E_T^{\rm min}\,,\quad \etmiss>E_T^{\rm min}\,,\quad
|\eta^\gamma|<2.0\,,& \nonumber\\
&\mbox{no jets with}~ E_T>15\gev,\,|\eta|<3.5\,,&\nonumber\\
&\mbox{no}~ e's~\mbox{or}~ \mu's ~\mbox{with}~ p_T>5\gev,\,
|\eta|<2.0\,.& 
\label{gamcuts}
\eea
While the signal is somewhat flatter, $S/B>0.1$ is only achieved
if a very high $E_T^{\rm min}$ cut is imposed.
However, the signal cross sections are rapidly decreasing
as the cut is increased, so we cannot take too high a $E_T^{\rm min}$ cut.
To quantify the difficulty, consider $\mcpmone\sim 60\gev$. For 
$E_T^{\rm min}=50\gev$ ($100\gev$), one finds
$\sigma(\gam\cpone\cmone+\gam\cpmone\cnone)\sim 10\fb$ ($1.5\fb$)
compared to $\sigma_B(\gam \etmiss)\sim 211\fb$ ($27\fb$),
so that these cuts yield $S/B\sim 0.05$ (0.06) and $S/\sqrt B=3.8$ ($1.6$).
Thus, we will be able to see a signal in the $\gam+\etmiss$ channel
for $\mcpmone\sim 60\gev$ only if systematics are understood
at the $S/B\sim 0.05$ level. If $S/B\gsim 0.1$ is required, 
the $\gam+\etmiss$ signal can probe only $\mcpmone\lsim
50\gev$. Either value is only a marginal improvement 
over the $45\gev$ lower bound deriving from LEP $Z$--pole 
data when the chargino decay products cannot be detected
(i.e. when $200\mev\lsim\dmchi\lsim 300-500\mev$). Even more importantly,
both values are below the limits set by DELPHI once $\dmchi\geq 300-500\mev$.
In scenario (2), the signal cross section sum will be somewhat
smaller than in scenario (1), and
$S/B$ will typically be too small to extract a signal from the data.

Given the importance of achieving a very small systematic error level
in order to extend the LEP/LEP2 limits on an invisible chargino,
it is worth noting that systematic errors
{\it do} decrease with integrated luminosity, and many RunI analyses
have systematic errors that are smaller than or the same size as the
statistical error.  The RunII situation should be much better
than in RunI.  Furthermore,
we have not exploited any difference in shape between the signal and background,
which may increase the significance of the signal.  If any other distinguishing
characteristics of the signal can be observed, or if there are other sources
of chargino production, then the upper limit of $\mcpmone$ 
for which the $\gamma+\etmiss$ signature is viable could be significantly
larger than estimated here.

Given the somewhat pessimistic results for the $\gam+\etmiss$ signal,
it is worth exploring the standard SUSY jets+$\etmiss$ signal, 
which will have a larger event rate for comparable cuts.
As compared to the normal mSUGRA scenario, the softness
of the $\cpmone$ decay products implies that the jets+$\cpone\cmone$
events will have much lower jet multiplicity.
After including the effects of initial
state gluon radiation, many events have a monojet nature.  The published 
RunI analyses have taken advantage of the jet multiplicity to control 
the QCD backgrounds, and are of little help in understanding potentially
large monojet backgrounds.  Thus, we will consider
applying the standard mSUGRA 3--jet+$\etmiss$ cuts to a Monte Carlo
prediction of the signal, using the published
background estimates to set limits.  Since the parton showering
machinery can generate several jets per event,
some signal events will pass the cuts.
However, as discussed below, we find that there are
substantial uncertainties in the Monte Carlo predictions for the 
multijet signal rate when the multiple jets are generated
from parton showering and not by sparticle decays. 
Thus, we also consider a monojet signature
for which we believe that the monojet+$\etmiss$ Monte Carlo signal
rate computed using parton showering will be more reliable. This 
necessitates a study of the mismeasurement
background. As discussed below, we believe that this and other
monojet+$\etmiss$ backgrounds can be understood and convincingly controlled.

We consider the monojet+$\etmiss$ signal first.
To illustrate the size of the signal from
jet+$\cpone\cmone$ and jet+$\cpone\cnone$ 
compared to background,
we proceed much as in the case of the $\gam$ tag.
Our specific cuts are:
\bea
&E_T^j>E_T^{\rm min}\,,\quad \etmiss>E_T^{\rm min}\,,\quad
|\eta^j|<3.5\,,& \nonumber\\
&\mbox{no other jets with}~ E_T>15\gev,\,|\eta|<3.5\,,&\nonumber\\
&\mbox{no}~ e's~\mbox{or}~ \mu's ~\mbox{with}~ p_T>5\gev,\,
|\eta|<2.0\,.& 
\label{jcuts}
\eea
We claim that these cuts are such that the physics backgrounds
from $j\nu\anti\nu\to j\etmiss$ and $jW(\to\tau\to\etmiss)$ 
are larger than the mismeasurement background.
The major mismeasurement background to a monojet+$\etmiss$ search is
jet$+$jet production, where one jet fakes $\etmiss$.  We will only
consider values of $E_T^{\rm min}$ above $75\gev$, which means that
$\etmiss$ will always be required to be above the threshold employed for the
RunI multijet analyses. To estimate the mismeasurement background,
we have generated all QCD two parton processes with
PYTHIA, and retained only those events containing only
one or two jets with $E_T>15$ GeV.  
If there are two jets, we then randomly discard one of
the two and weight the event by a factor $2\times R_{j\to\,\etmiss}$.
We then impose the cuts of Eq.~(\ref{jcuts}).
For $E_T^j>E_T^{\rm min}=75\gev$,
we arrive at a cross section estimate of 4 pb.
The dominant physics backgrounds, $Z(\to\nu\bar\nu)+$jet and
$W(\to\tau\nu_\tau)+$jet, contribute $4$ and $1.6$ pb for the same
cuts (note, we are far beyond the Jacobian peak, so that
$W(\to\ell\nu_\ell)+$no jet, where $\ell=e,\mu,\tau$, can be
ignored).  Thus, even if our estimate of the QCD background is off
by a factor of $2$, this will not substantially bias 
an exclusion limit obtained using Gaussian statistics.  After including 
$t\bar t$, single top, gauge boson pairs, and $W(\to e\nu_e,\mu\nu_\mu)+$jet
backgrounds, the full background for the cuts of Eq.~(\ref{jcuts})
and $E_T^{\rm min}=75\gev$ is about 10 pb.

Figure~\ref{gamorjtagxsec}(b) shows the integrated cross section for
the background and signal (signals are multiplied by 100 in the figure)
as a function of $E_T^{\rm min}$. It is clear that the background is so severe
that the monojet+$\etmiss$ channel will be 
much less useful than the $\gamma+\etmiss$ channel.
The relative behavior of the two channels is easy to 
understand.  At $E_T^{\rm min}=75$ GeV, we observe that the $Z$ backgrounds
differ by a factor of about 63.  Naively, we estimate that they should
differ by $2\alpha_s/(\alpha Q_u^2)\simeq 58$, where $\alpha=1/128$ and
the factor of 2 accounts for the 2 different topologies, $qg\to Zq$
and $q\bar q\to Zg$.  On the other hand, the signal is not so
suppressed in switching from the $j+\etmiss$ to $\gamma+\etmiss$ channel,
since final state, photon radiation off the charginos is important.
Also, since we generated the $j+\etmiss$ signal using parton showering,
we underestimate the signal cross section at
high $\etmiss$.  Finally, we note that the mismeasurement backgrounds
are much more relevant for the $j+\etmiss$ channel.  To see
if there is any hope for this discovery channel, we have varied
the $E_T^{\rm min}$ cut in search of a value such that 
$S/B>0.1$ and such that there are at least 5 events for $L=30\fbi$.
We never satisfy
these constraints for $\mcpmone>45$ GeV, so no limit beyond that
from LEP $Z$--pole data can be set using this channel.
Nonetheless, as
explained below, a significant signal may appear in the $j+\etmiss$
channel if there are other sources of chargino production.

We will now turn momentarily to the multijet+$\etmiss$ signal.  At the same
time, we will also consider the more optimistic possibility
that the gluino mass is small enough that $\gl\gl$ pair production
has a reasonable rate at the Tevatron. In particular, we
consider the limit, previously analyzed in Ref.~\cite{cdg2}
and motivated in the O--II model,
that the gluino is almost degenerate in mass with $\cpmone$.
The results of Ref.~\cite{cdg2} were that
$\mgl=150$ GeV could be excluded with $L=2\fbi$
of data, and that this reach could not be extended
using higher $L$ if one demanded
$S/B>0.2$.  The exclusion was based on background estimates
from D\O~and CDF for their RunI 3 jet+$\etmiss$ searches.  
We have repeated the analysis of Ref.~\cite{cdg2} using PYTHIA
instead of ISAJET \cite{isajet}. We find that we cannot reproduce all of the Ref.~\cite{cdg2}
results, and the reasons (to be discussed below) suggest that one
may not wish to trust results obtained via Monte Carlo for a
multijet+$\etmiss$ signal of the type considered here, in which
the jets are generated entirely by parton showering.
For example, consider $\mgl=75$ GeV. Using PYTHIA,
we find roughly half the signal cross section (compared to \cite{cdg2})
after the D\O~cuts.  This discrepancy arises because of the details
of parton showering used in ISAJET (Ref.~\cite{cdg2}) as compared to
PYTHIA (our study).
For the first, soft--gluon emission in a shower, PYTHIA restricts
the polar angle of the branching to be smaller than
the angle of the color flow, while ISAJET does not.
As a result, the soft gluons in ISAJET are more widely distributed, and
the resultant jet multiplicity is higher.
Indeed, when we turn off the angular ordering 
effect in PYTHIA, we reproduce the
ISAJET results.  For larger $\mgl$, however, the discrepancy remains,
and is not entirely resolved.  If we use the PYTHIA results, the
Ref.~\cite{cdg2} Tevatron limit on $\mgl$ is reduced to $\mgl\sim 95-100$ GeV.
However, the comparison of the two Monte Carlo programs 
suggests that one cannot trust
a parton showering result in degenerate scenarios
for 3 hard, well--separated jets.

\begin{figure}[ht]
\leavevmode
\begin{center}
\epsfxsize=3.25in
\epsfysize=3.25in
{\hspace{0in}\epsffile{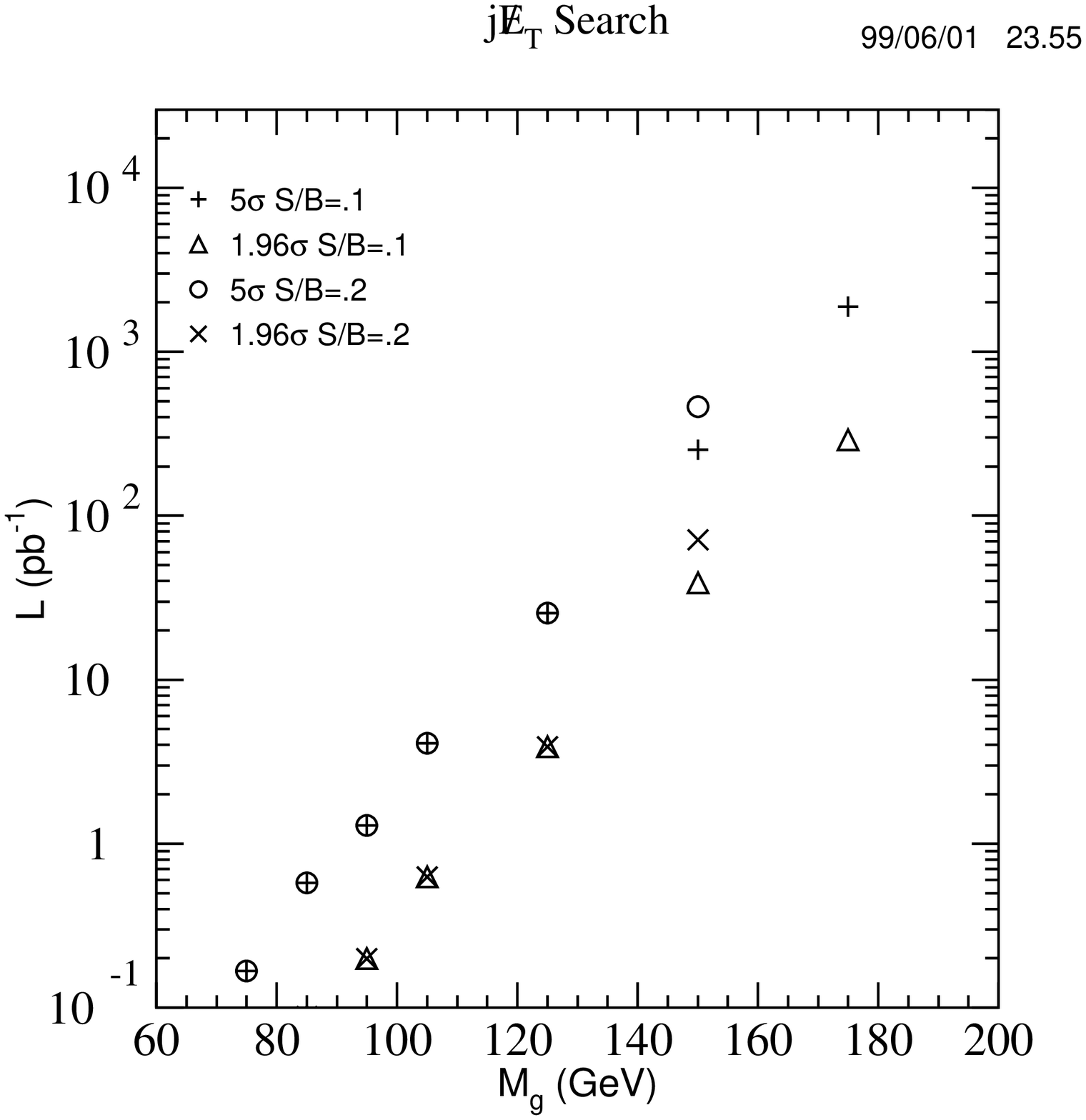}
\epsfxsize=3.25in
\epsfysize=3.25in
\hspace{0in}\epsffile{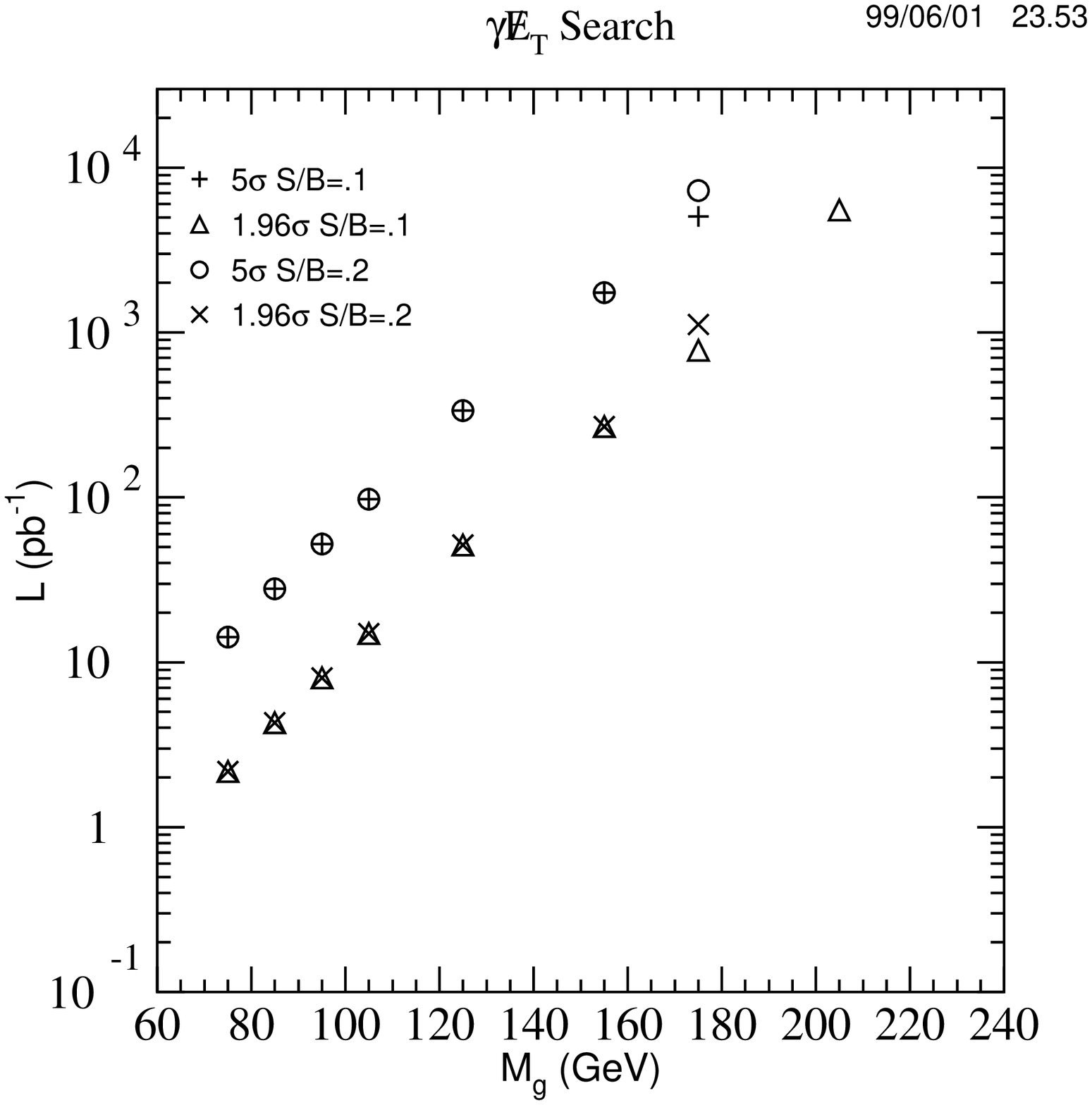}}
\end{center}
\caption[]{Integrated luminosity (in pb)
required to observe ($S/\sqrt B=5$) or exclude
at 95\% CL ($S/\sqrt B=1.96$) a monojet+$\etmiss$ or $\gamma+\etmiss$ signal
from $\gl\gl$ production, when $\mgl\simeq\mcpone$.  
Predictions are shown for signal to background ratios of $0.1$ and $0.2$.
For $\mgl$ values above those for which points are plotted, $S/B$ is
below the required value. For the monojet+$\etmiss$ [$\gamma+\etmiss$]
signal we impose the cuts of Eq.~(\ref{jcuts}) [Eq.~(\ref{gamcuts})],
optimizing $S/\sqrt B$ by scanning over $E_T^{\rm min}>75\gev$ [$>50\gev$].
}
\label{gljorgamtagxsec}
\end{figure}

Thus, we return to our proposed monojet signature
to estimate the potential of the Tevatron
for probing the $\mgl\sim \mcpmone\simeq\mcnone$ scenario. 
As already noted, the $\etmiss$
signature is enhanced by $q\bar q,gg\to \gl\gl,$ where 
$\gl\to q^\prime\anti q\cpmone,q\anti q\cnone$.
For small $\mgl-\mcpmone$ the $q,q^\prime,\anti q$
are typically too soft to be counted as jets and, of course, in the study
of this section 
we are assuming that the $\cpmone$ decay products are not visible. 
The monojet still comes from parton showering.
In Fig.~\ref{gljorgamtagxsec}(a) we plot the luminosity required for
$S/\sqrt B=1.96$ or $5$ and $S/B>0.1$ or $0.2$ as a function of $\mgl$.
We have employed the cuts of Eq.~(\ref{jcuts}), searching for the
$E_T^{\rm min}>75\gev$ that maximizes $S/\sqrt B$ while
satisfying the given $S/B$ criteria. (For lower
$\mgl$, $E_T^{\rm min}=75\gev$ is always best; for the highest $\mgl$ values
the best $E_T^{\rm min}$ increases.)
With the enhanced production cross sections, we observe that 
it is much easier to achieve
$S/B>0.2$ and a gluino with $\mgl=150\gev$ should be discovered or excluded
early in RunII. However, discovering or excluding $\mgl=175\gev$ will
require reducing systematics to the extent that an $S/B=0.1$ signal
can be trusted. Specifically, for $S/B>0.1$, $\mgl=175\gev$ can be excluded at
95\% CL with $L=0.3\fbi$ or discovered at the $5\sigma$ level with $L=2\fbi$. 

The process $\gamma\gl\gl$, where $\gl\to$ soft, yielding
a $\gam+\etmiss$ signal, is complimentary to the monojet+$\etmiss$
signal.  We follow the same procedure as discussed for
the monojet+$\etmiss$ signal, except that we employ the
cuts of Eq.~(\ref{gamcuts}) and require $E_T^{\rm min}>50\gev$.   
The luminosity required to discover
or exclude a given $\mgl$ using this signal is plotted in
Fig.~\ref{gljorgamtagxsec}(b).
Even though the $\gam+\etmiss$ signal requires more integrated luminosity to
establish a signal for low $\mgl$, $S/B$ is larger, allowing exclusion
out to a larger value of the common chargino/gluino mass; 
$\mcpmone\sim \mgl\leq 175\gev$ can be excluded at 95\% CL
with $L=1\fbi$ of integrated luminosity even if $S/B>0.2$ is required.

For purposes of comparison, we note that in an mSUGRA scenario the tri--lepton
signature from $\cpmone\cntwo$ production allows one to probe
chargino masses up to about $160\gev$ for $L=30\fbi$ when the 
scalar soft--SUSY--breaking mass is large \cite{trilepref}.

We note that the monojet+$\etmiss$ and $\gam+\etmiss$ signatures should
persist (and perhaps even improve somewhat) for $\mgl-\mcpmone\sim
\mbox{few}-10\gev$ and/or $\dmchi\sim \mbox{few}-10\gev$.

Before concluding this subsection, we should comment that 
there are potential contributions from $\gl\cpmone$ and
$\gl\cnone$ production that have not been included here.
These will depend on the exact values of the squark masses,
which are assumed to be heavy.  For the remainder of the paper,
we assume that the gluino is also very heavy.

\subsubsection{LHIT and TOF signatures}

In the previous section, we considered the case where the mass splitting
was large enough for the chargino decay to be very prompt, but yet too
small for the chargino decay products to be visible. In this section,
we consider the opposite extreme, namely $\dmchi$ sufficiently small
that the chargino is so long--lived 
that it passes through the TOF and enters the muon chambers.
For instance, if $\dmchi<\mpi$, then the average $c\tau$ is of order
a meter or more.  Of course, as noted earlier, the radiatively
generated mass splitting makes $\dmchi<\mpi$ somewhat unlikely
in the context of the existing models.
But, even for $\dmchi>\mpi$ there is 
a tail of events with large enough $\beta\gamma c\tau$ values
for the chargino to reach the muon chambers. 
Thus, from the experimental point of view it is important
to consider signals based on a muon--chamber or TOF signal
as a function of $\dmchi$.

To distinguish a chargino that reaches the muon chambers
from an actual muon without using the TOF, we employ the procedures used
by CDF in RunI for identifying a penetrating particle that is
sufficiently heavily--ionizing that it cannot be a muon. 
However, because the D\O~inner muon chambers are closer to the interaction
point and cover more range in $\aeta$, 
it is advantageous to employ the D\O~muon chamber configuration (see earlier
description).
In analogy to the CDF RunI procedure, we first demand a trigger
for the event using one track (Track I) that penetrates
to the muon chambers. We then examine the triggered
events for a track (Track II) that is heavily--ionizing and
penetrates to the muon chambers. The specific cuts/requirements
we impose in our study are:\footnote{Note that since the CDF procedure
was originally designed for looking for massive quarks, they did
not impose a requirement of small hadronic energy deposit
in the track direction(s). 
We have not imposed this requirement either. However, for the chargino
signal of interest here it could be imposed with little loss of
signal event rate were this useful for reducing backgrounds.}
\bea
&\mbox{Track I and II:}~ & (\aeta<1,\beta_{\perp}\gamma
c\tau>2.7~\mbox{m})~\mbox{or}~(1<\aeta<2,|\beta_{z}|\gamma
c\tau>4~\mbox{m}),~ \beta>\beta_{\rm min} \nonumber\\
&\mbox{Track I:}~ & p_T>15\gev \nonumber\\
&\mbox{Track II:}~ &  |\vec{p}|>35~{\rm GeV}, \beta\gamma<0.85\,,
\label{lhitcuts}
\eea
where $\beta_{\rm min}$ is the minimum velocity ($\simeq 0.4-0.5$)
required for the $\cpmone$ to penetrate to the muon chambers.
In Eq.~(\ref{lhitcuts}), $\beta_{\perp}$ is the velocity perpendicular
to the side of the box formed by the inner muon chambers (see earlier
description) that the chargino eventually passes through,
and $\beta\gamma<0.85$ is the 2MIP requirement.
Tracks I and II may be the same track.
An event satisfying these cuts will be called an LHIT event
and is expected to be background--free.

\begin{figure}[ht!]
\leavevmode
\begin{center}
\epsfxsize=3.25in
\hspace{0in}\epsffile{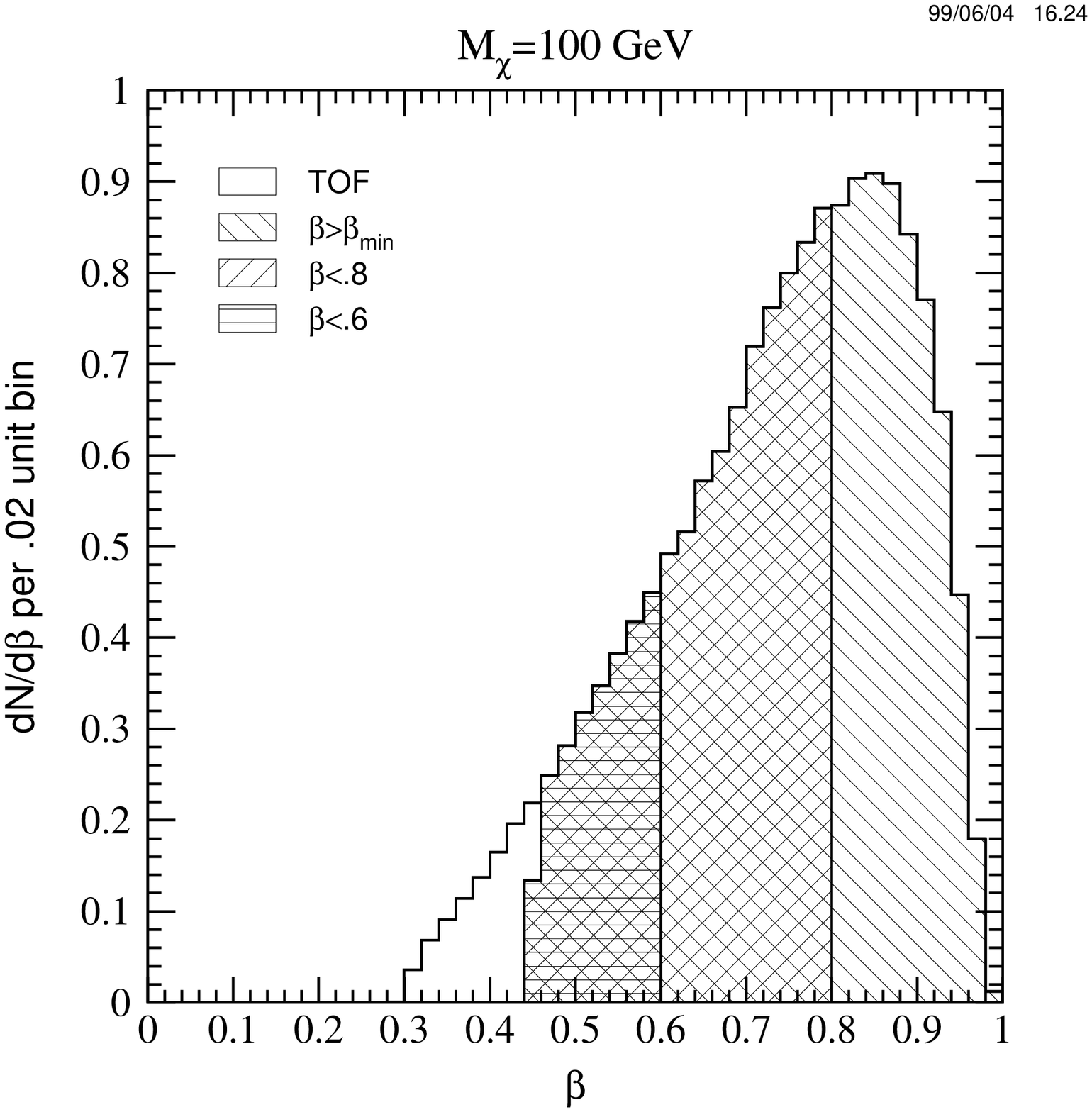}
\epsfxsize=3.25in
\hspace{0in}\epsffile{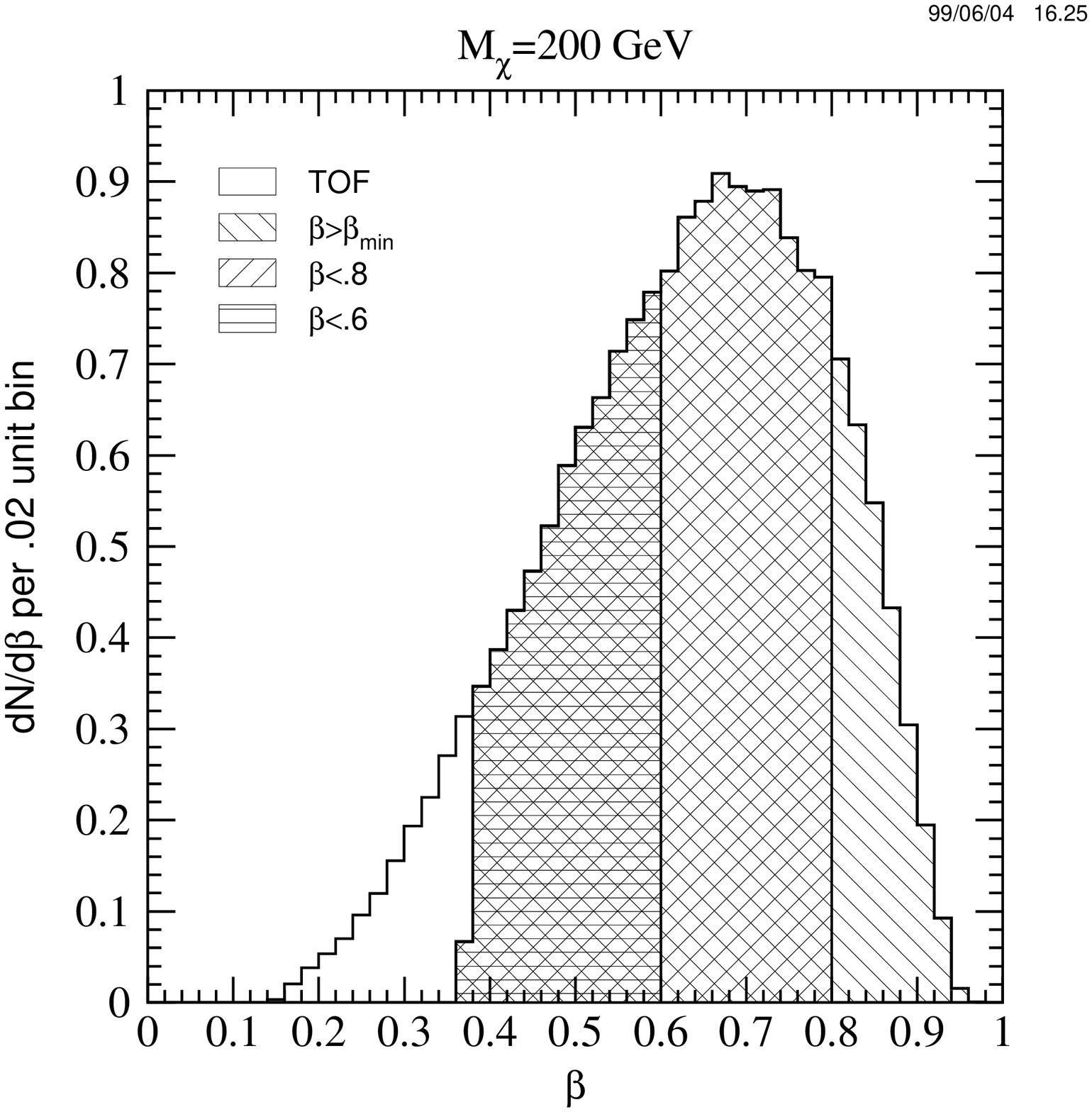}
\epsfxsize=3.25in
\hspace{0in}\epsffile{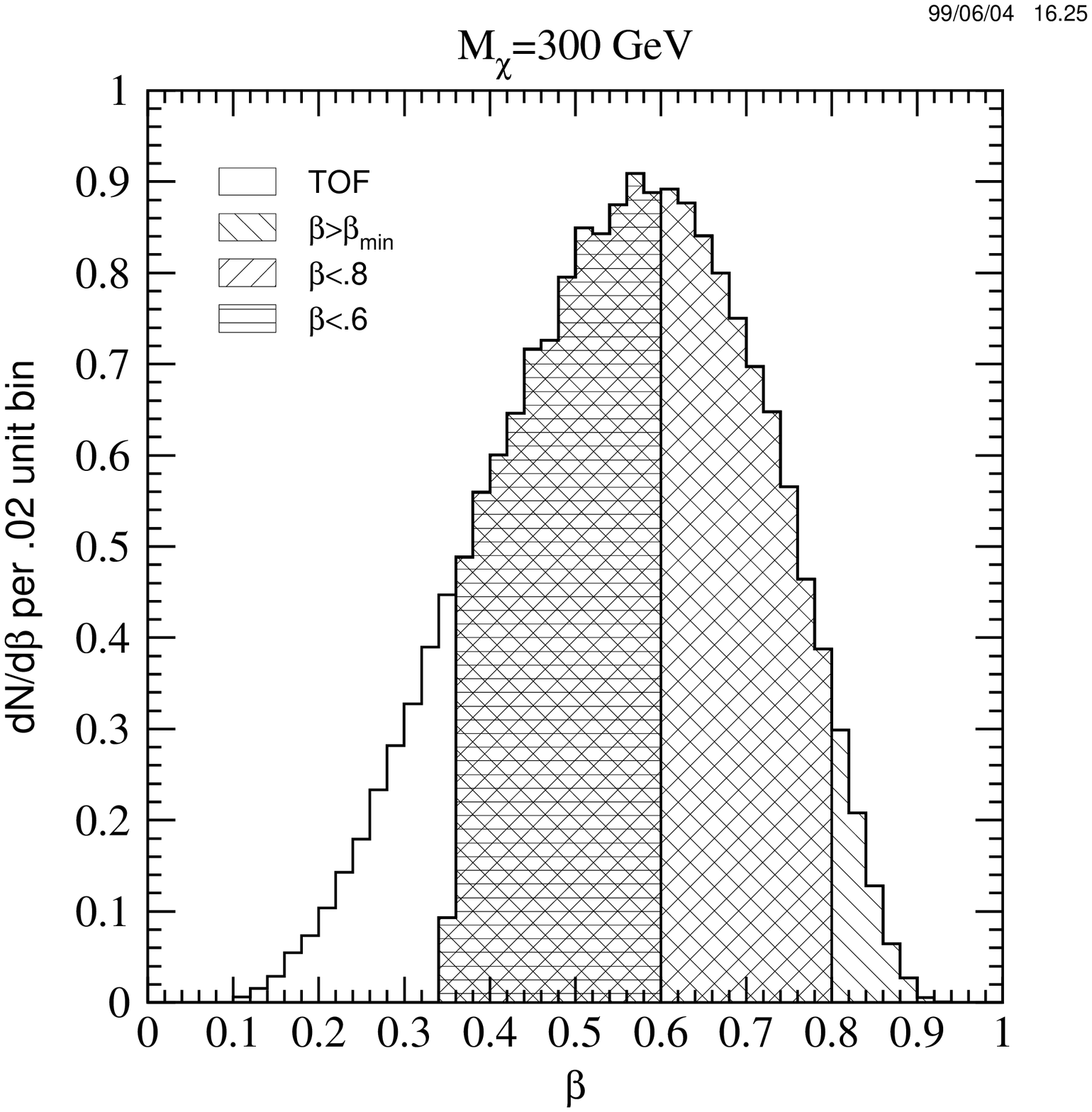}
\epsfxsize=3.25in
\hspace{0in}\epsffile{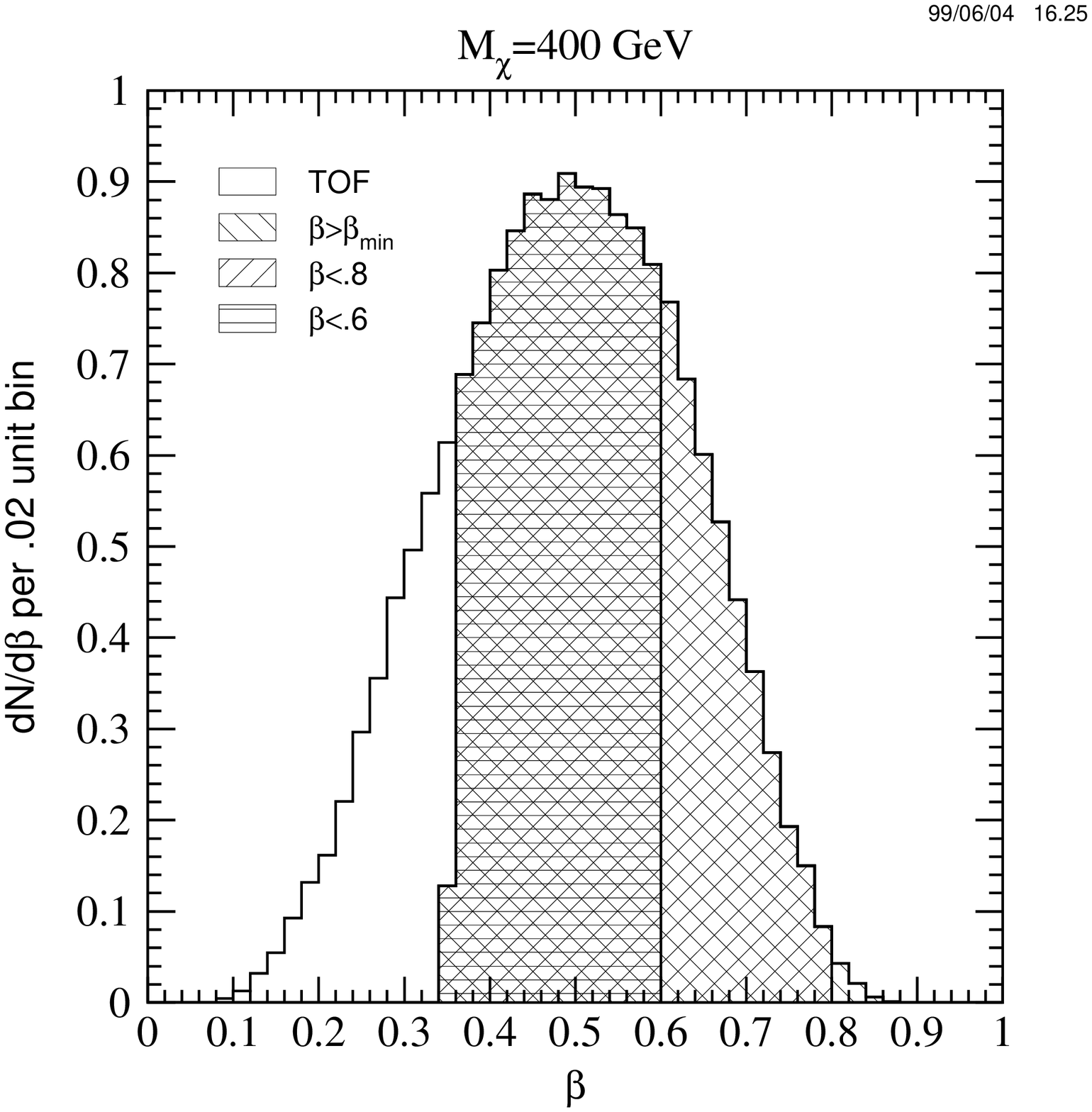}
\end{center}
\caption[]{Unnormalized $\beta$ distributions for
events accepted by the non-$\beta$ cuts of Eq.~(\ref{lhitcuts})
after requiring that the TOF time delay be $>500$~ps.
The segments left after imposing various more restrictive $\beta$
cuts are also shown. Distributions are given for $\mcpmone=100$, 200,
300 and $400\gev$.}
\label{betadist} 
\end{figure}

For small $\mcpmone$, the $\cpmone\cnone$ and $\cpone\cmone$
production cross sections are large before
cuts, but the $\beta_{min}$ and $\beta\gamma$ requirements accept
only a small portion of the full rate.  For larger $\mcpone$, the
cross section decreases, but $\beta\gamma$ is typically smaller.  
These trends are illustrated in Fig.~\ref{betadist}.
This figure shows the full $\beta$ distributions for 
$\mcpmone=100$, 200, 300 and $400\gev$ that remain after requiring that
the chargino pass through the TOF device at least 500~ps later
than a particle with $\beta=1$ (as required in the TOF signal discussed below).
The impact of the $\beta_{\rm min}$ cut and of various
requirements on the maximum value of $\beta$ is also shown. We see that for
$\mcpmone=100\gev$ a relatively small slice of the $\beta$ distribution
is retained after requiring both $\beta>\beta_{\rm min}$ and $\beta<0.6$ 
($\beta\gamma<0.75$) or $\beta<0.65$ ($\beta\gamma<0.85$). 
The slice accepted by such cuts is much larger for $\mcpmone=400\gev$.

\begin{figure}[ht!]
\leavevmode
\begin{center}
{\epsfxsize=3.25in
\epsfysize=3.25in
\hspace{0in}\epsffile{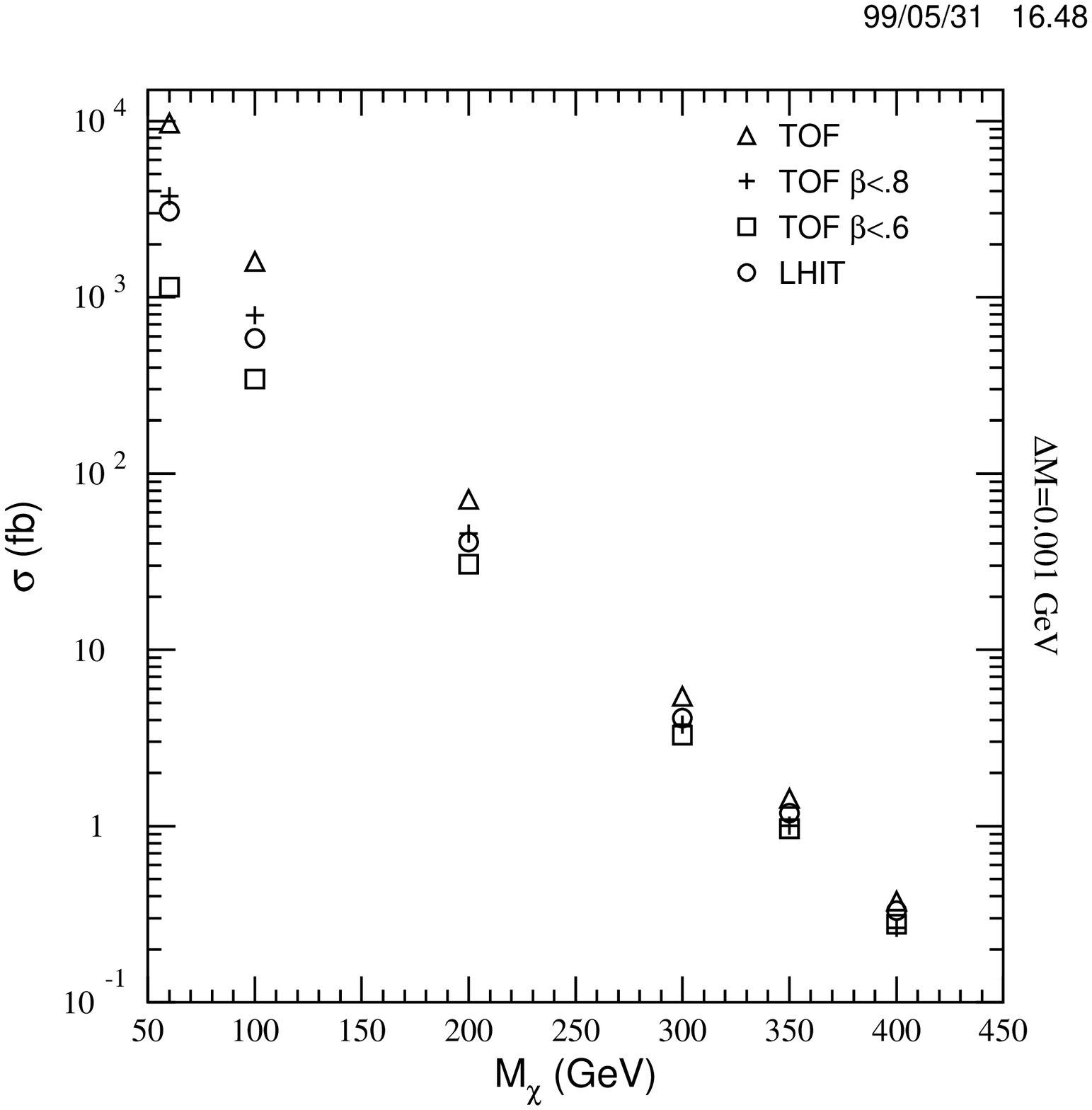}
\epsfxsize=3.25in
\epsfysize=3.25in
\hspace{0in}\epsffile{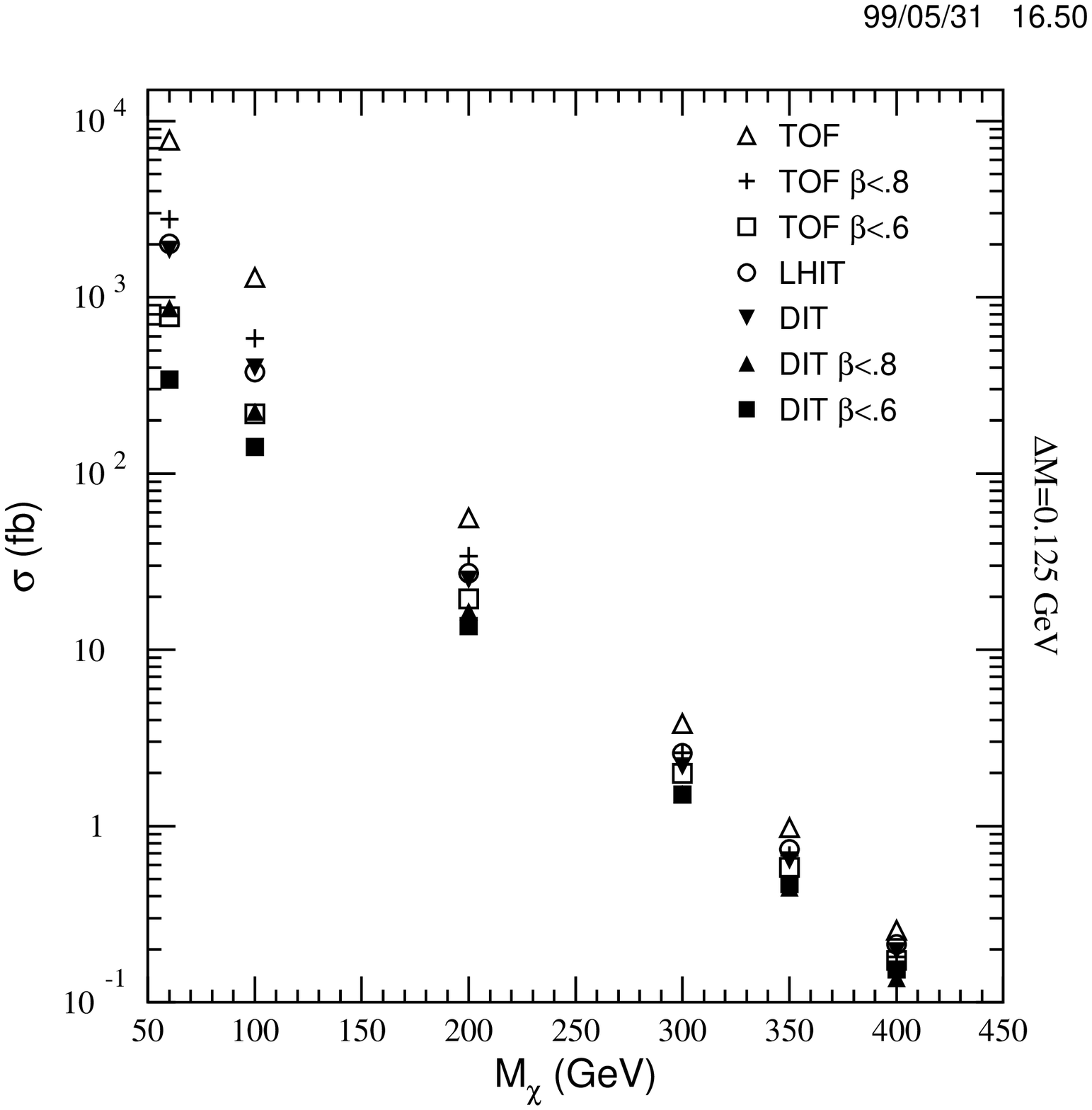}}
{\epsfxsize=3.25in
\epsfysize=3.25in
\vspace{0.in}\hspace{0in}\epsffile{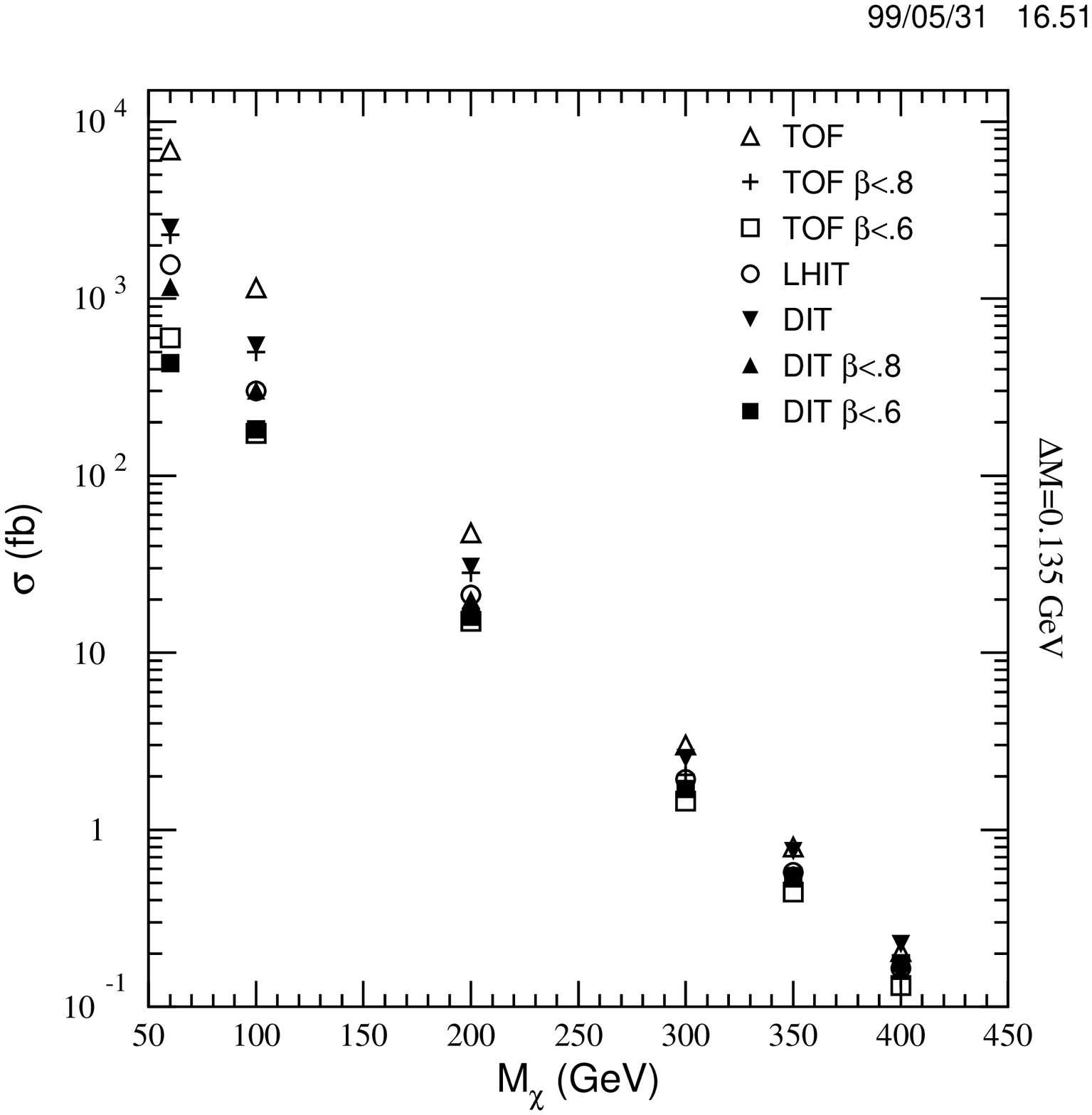}
\epsfxsize=3.25in
\epsfysize=3.25in
\hspace{0in}\epsffile{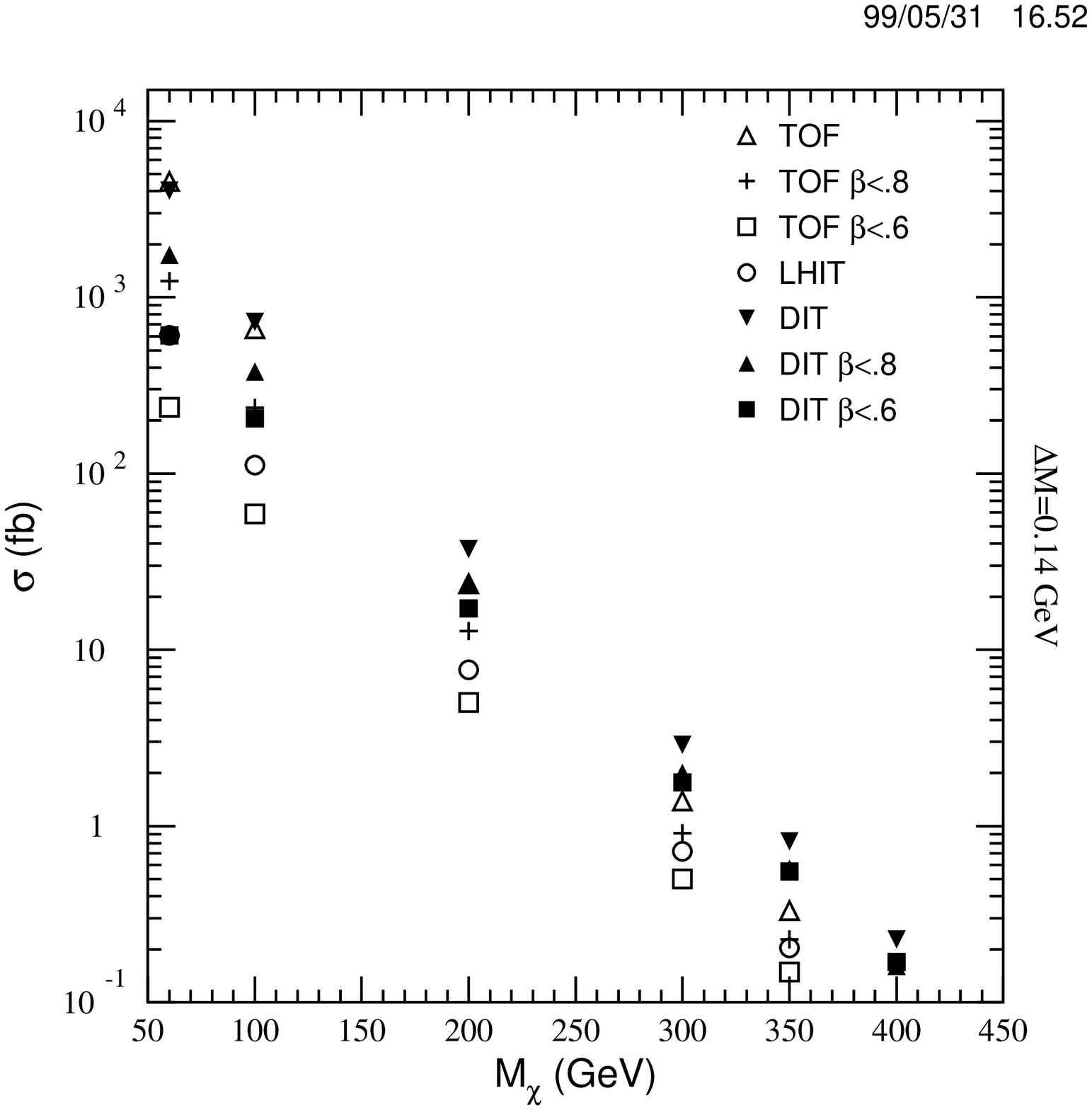}}
\end{center}
\caption[]{Cross sections for LHIT, TOF and DIT ``background--free''
signatures at RunII: $\dmchi=0,125,135,140\mev$.}
\label{sigsa} 
\end{figure} 

The cross section obtained after imposing all 
the cuts of Eq.~(\ref{lhitcuts}) is plotted as the open circles in
Fig.~\ref{sigsa} for a selection
of $\dmchi$ values. For $c\tau=\infty$, $\mcpone=350\gev$ ($450\gev$)
can be excluded at the 95\% CL (3 events predicted,
none observed) with 2 (30) $\fbi$ of data.  
A 5 event discovery would require $\mcpmone\leq 325\gev$ ($\leq 430\gev$)
for $L=2\fbi$ ($30\fbi$). The 3 event limits for various $\dmchi$
values are summarized in Fig.~\ref{limits:summary}. As expected,
the LHIT signal mass reach declines as $\dmchi$ increases, and the
LHIT signal has vanished by $\dmchi=142.5\mev$.

Let us now turn to the time--of--flight (TOF) signal for identifying
a slow--moving, long--lived chargino. Our procedure for triggering
will be exactly as for Track I in the LHIT procedure.
We then look for a Track II that satisfies
the same cuts as in the LHIT case except that the $\beta\gamma<0.85$
requirement is replaced by the requirement that Track II arrive
at the TOF device at least 500~ps later than would a relativistic track.
For the expected 100~ps time resolution of the TOF signal, this
corresponds to a $5\sigma$ delay in arrival
compared to a particle with $\beta\sim 1$. 
Thus, we replace the $\beta\gamma<0.85$ requirement by
\beq
d_{\rm TOF}/(\beta c)-d_{\rm TOF}/c>500~\mbox{ps}\,,
\label{tofcuts}
\eeq
where $d_{\rm TOF}$ is the distance to the muon chamber along the direction
of flight. 

As mentioned previously, for smaller $\mcpmone$ values the TOF signal 
accepts a significantly larger range of $\beta\gamma$ 
than does the heavy--ionization $\beta\gamma<0.85$ requirement of the
LHIT signal. This is clearly apparent from the $\mcpmone=100$
and $200\gev$ windows of Fig.~\ref{betadist}
by comparing the total $\beta>\beta_{\rm min}$ region to
the $\beta_{\rm min}<\beta<0.65$ region. However, for large
$\mcpmone$ near the upper limit that can be probed by the LHIT signal
(see the $\mcpmone=400\gev$ window of Fig.~\ref{betadist})
the $\beta<0.65$ ($\beta\gamma<0.85$) LHIT cut is not that much less
efficient than the TOF cut. Thus, we can anticipate that the TOF
signal will be viable for lower luminosity than the LHIT signal
if $\mcpmone$ is not large, but that the TOF signal will not be
viable for $\mcpmone$ values much beyond those reachable by the LHIT signal.

The TOF cross sections as a function of $\mcpmone$ are given
in Fig.~\ref{sigsa} for the same $\dmchi$ values for which the LHIT
cross sections were plotted. As expected,
Fig.~\ref{sigsa} shows that the TOF signal is much more efficient
than the LHIT signal at lower masses,
but the upper mass limit attained using the
TOF and LHIT signals is the same, e.g. $\mcpmone\sim 430\gev$ for $L=30\fbi$
and $\dmchi=125\mev$.

As already noted, the fact that the TOF and LHIT upper mass limits
are the same is due to the fact that the largest $\mcpmone$ masses 
that can be probed are such that the $\beta\gamma<0.85$ requirement
is not very restrictive. That such large masses can be probed is due
to the $s$--wave nature of the $\cpone\cmone$
and $\cpmone\cnone$ production subprocesses, which, in turn, implies
large cross sections in $p\anti p$ collisions out to high $\mcpmone$.
This can be contrasted with the result for long--lived staus.
The $p$-wave nature of the $\stau^+\stau^-$ production subprocesses
implies much smaller $p\anti p$ cross sections, for which
the LHIT signal mass reach is limited to $\mstau\lsim 145\gev$. 
For such masses, the LHIT requirement of $\beta\gamma<0.85$ is very
restrictive. Due to its acceptance of $\beta\gamma$ values substantially 
beyond 0.85, the TOF signal improves the mass reach to $175\gev$ \cite{dstuart}.

We have not performed a study as to where additional TOF devices should
be placed in order to further optimize the TOF signal. However, we believe
that if $c\tau\lsim 1$m there would be significant 
gain if there were a TOF device in between
the EC and the HC (in addition to the TOF device next to the inner
muon chambers). With an appropriate electronics design,
events with a chargino that reached the inner TOF device
but not the muon chamber could be
triggered by the inner TOF signal and the presence of a stiff chargino
track in the tracker. The time delay
of the TOF signal would indicate the mass of the particle, and such
events would be background free. However, the mass reach
would only improve over the DIT signal discussed below if
a heavy--ionization requirement has to be imposed in order that the latter
be background free.

\subsubsection{DIT signatures}

As $\dmchi$ increases above $\mpi$ and $c\tau$
becomes too small to produce a LHIT or TOF signature with large efficiency.
The next signature of interest is an isolated
track that passes all the way through the CT
but disappears before reaching the TOF and MC. 
The disappearing, isolated track signature is denoted by DIT. 
For our study, we employ the D\O~detector CT radius
of $73\cm$ (which gives greater coverage for this signal than does
the CDF detector with CT radius of $130\cm$). The D\O~trigger
logic is well adapted to this type of signal
in that the CT track itself can be used to trigger the event provided
it is sufficiently isolated. The isolation required by D\O~for
a track trigger is that no other track be in the same azimuthal wedge
as the trigger track.  Each azimuthal wedge is of size $\Delta\phi\sim 0.1$.
The specific triggering requirements we impose are: 
\beq
\beta_{T}\gam c\tau>73\cm\,,\quad
p_T^{\rm trigger}>11\gev\,,\quad |\eta^{\rm trigger}|<2\,,\quad
\sum_{|\Delta\phi|<0.1} p_T^{\rm tracks}-p_T^{\rm trigger}<2\gev\,, 
\label{dittrigcuts}
\eeq
where $\Delta\phi=\phi^{\rm track}-\phi^{\rm trigger}$.
Once the event is triggered, we require (off--line) that it have
high $p_T$ and decay before reaching the TOF device and
the muon chamber. Without this latter
requirement, the track would be confused with a muon, unless we impose
a further requirement that it be heavily--ionizing.  We hope to avoid such
a requirement as it significantly reduces the signal event rate.
Our specific cuts are then:
\beq
p_T^{\rm track}>30\gev\,,\quad 73\cm<\beta\gam c\tau< d_{\rm TOF}\,,\quad
E_{\rm cal}(\Delta R<0.4)-E_{\rm cal}^{\cpmone}(\beta)\leq 2\gev \,,
\label{ditsigcuts}
\eeq
where $d_{\rm TOF}$ is the distance to the TOF device, e.g. to the
box of the inner D\O~muon chamber, 
$E_{\rm cal}(\Delta R<0.4)$ is the total energy deposited in the EC and HC 
calorimeters in the indicated cone surrounding the track,
and $E_{\rm cal}^{\cpmone}(\beta)$ is the average ionization energy
that the chargino would be expected to deposit in the EC and HC 
calorimeters for its (measured) $\beta$ in the given event, {\it assuming
it does not decay before exiting the HC}.
Given that in some events the chargino will decay soon after entering the EC,
this latter cut is quite conservative. 
A more optimal approach when the calorimeters are sufficiently
segmented in the radial direction might be to look for events with
chargino ionization energy deposits in a few inner segments but no corresponding
energy deposits along the track direction in the outer segments.
If the termination of the track could be seen despite the small size
of the ionization energy deposits ($2-3$ MIP's, typically) and if
``hot--spot''/$K^0$/$\ldots$ backgrounds are not large,
such events would be clearly distinct from background events,
especially given
that the trigger track must have $p_T>30\gev$. We have not attempted
to implement this approach in our studies.
In the absence of using the radial segmentation,
the $E_{\rm cal}$ cut may be very important for eliminating backgrounds. 
Fortunately, it is highly efficient for the signal.
Although, $\cpmone\cnone$ and $\cpone\cmone$ production 
will have the usual hadronic
(initial state radiation, mini--jet, $\ldots$) activity associated
with a hard scattering event, the probability of having 
more than $5\gev$ of $E_T$ of such activity in $\aeta<1$ is only
about 30\%, implying that small $E_{\rm cal}$ near the chargino,
in addition to the ionization energy deposits of the chargino itself,
will be automatic for most signal events. The signature defined
by Eqs.~(\ref{dittrigcuts}) and (\ref{ditsigcuts}) is called the 
disappearing, isolated track (DIT) signal.

\begin{figure}[ht!]
\leavevmode
\begin{center}
{\epsfxsize=3.25in
\epsfysize=3.25in
\hspace{0in}\epsffile{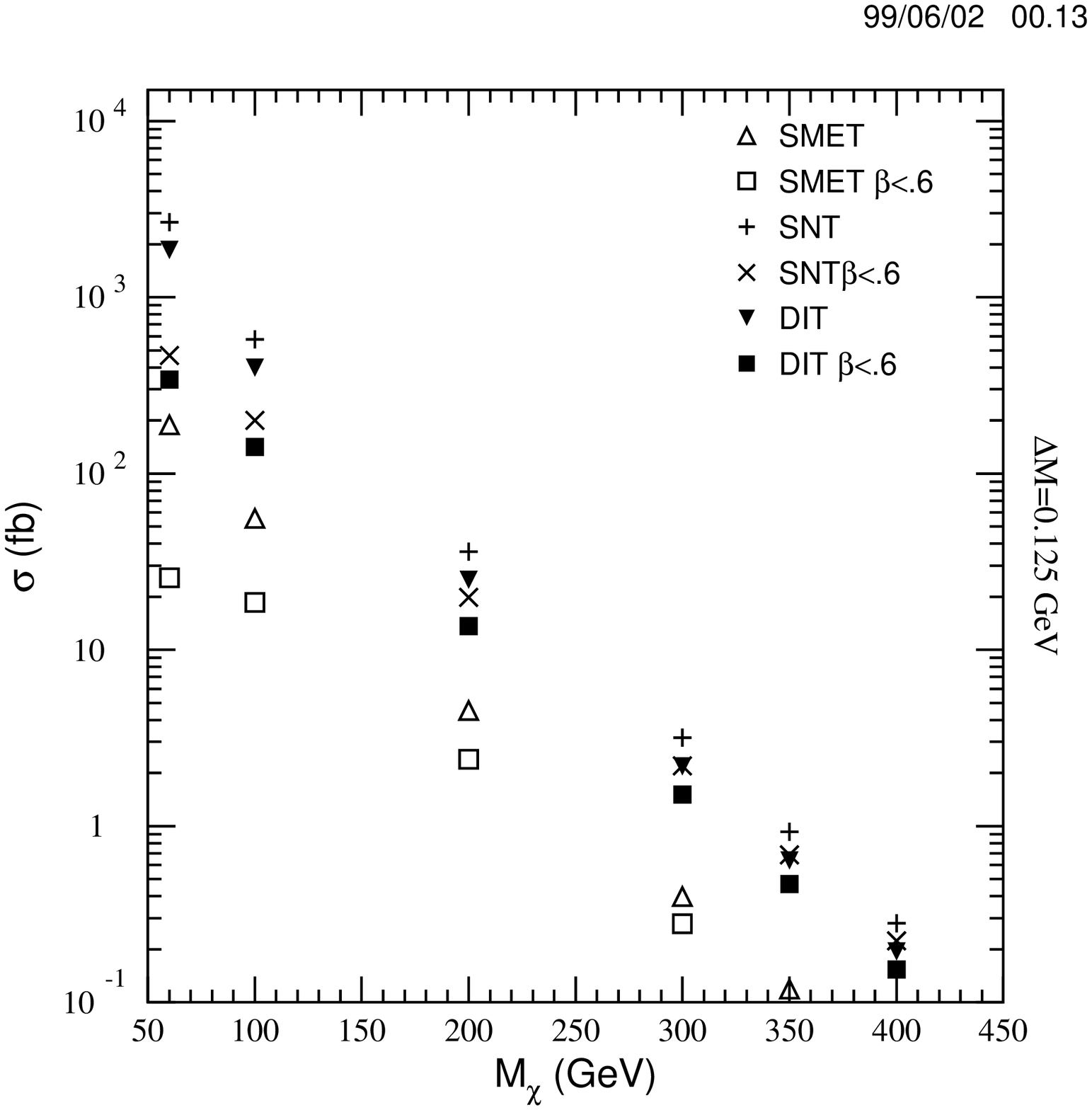}
\epsfxsize=3.25in
\epsfysize=3.25in
\hspace{0in}\epsffile{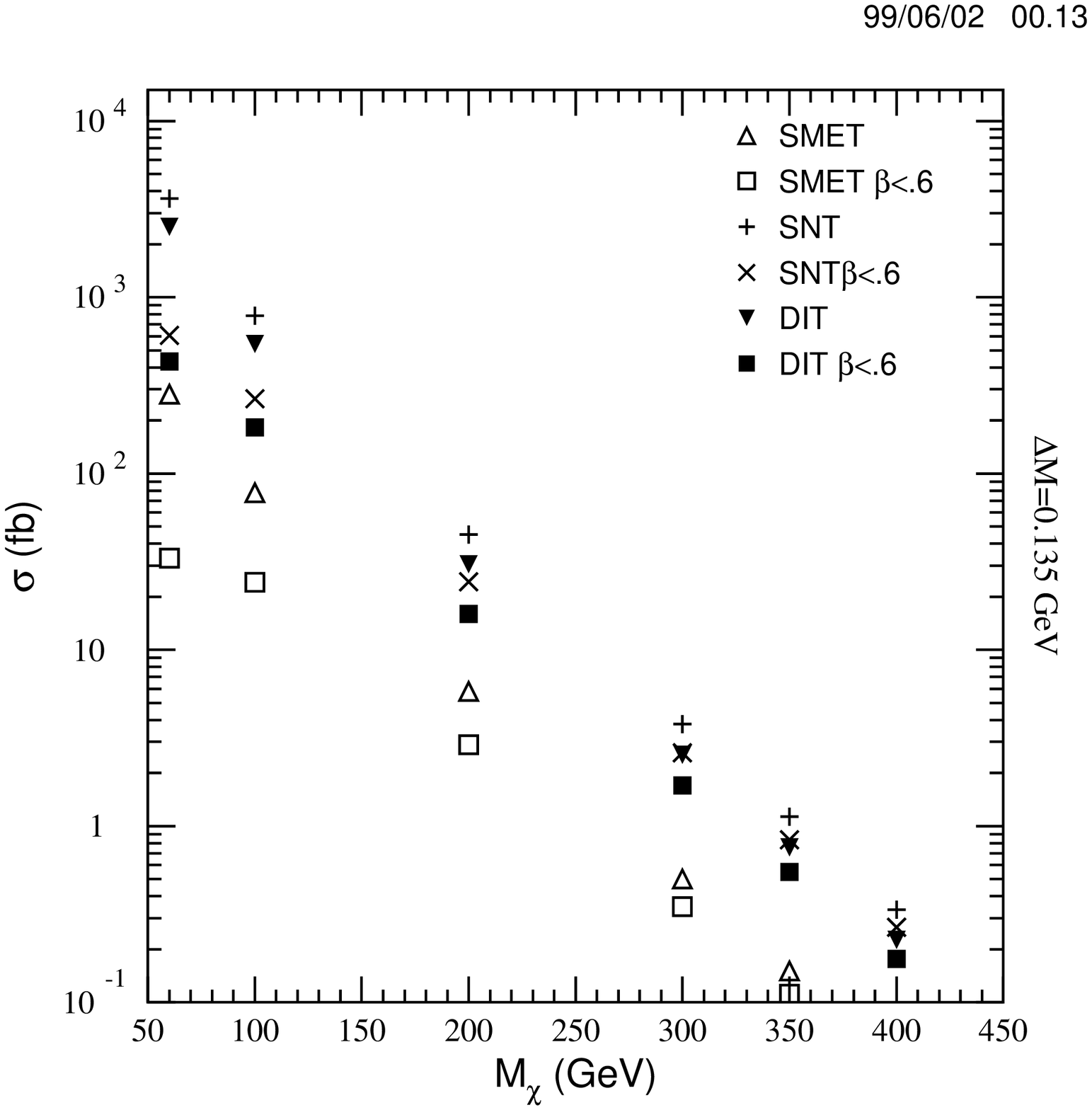}}
{\epsfxsize=3.25in
\epsfysize=3.25in
\vspace{0.in}\hspace{0in}\epsffile{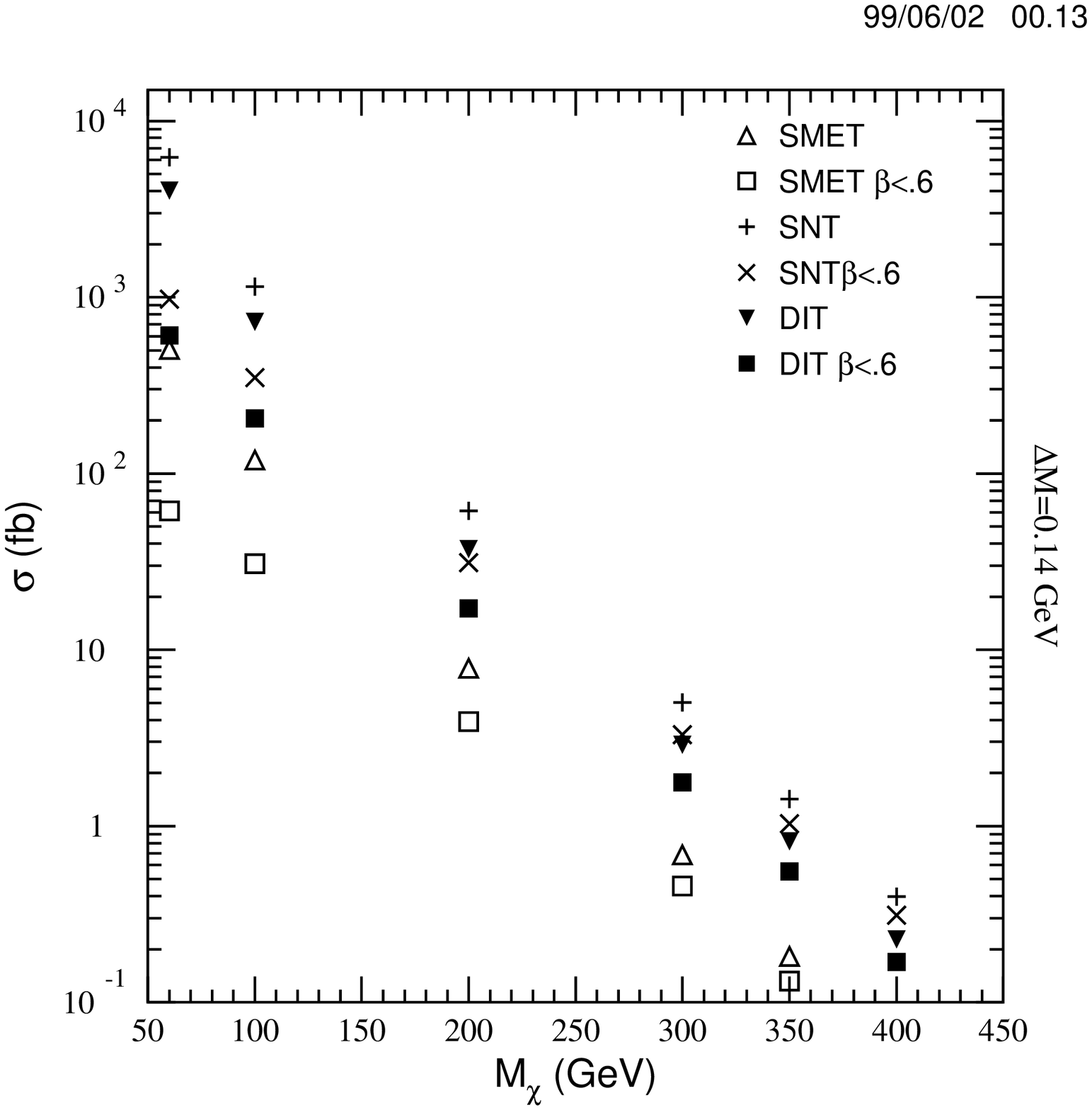}
\epsfxsize=3.25in
\epsfysize=3.25in
\hspace{0in}\epsffile{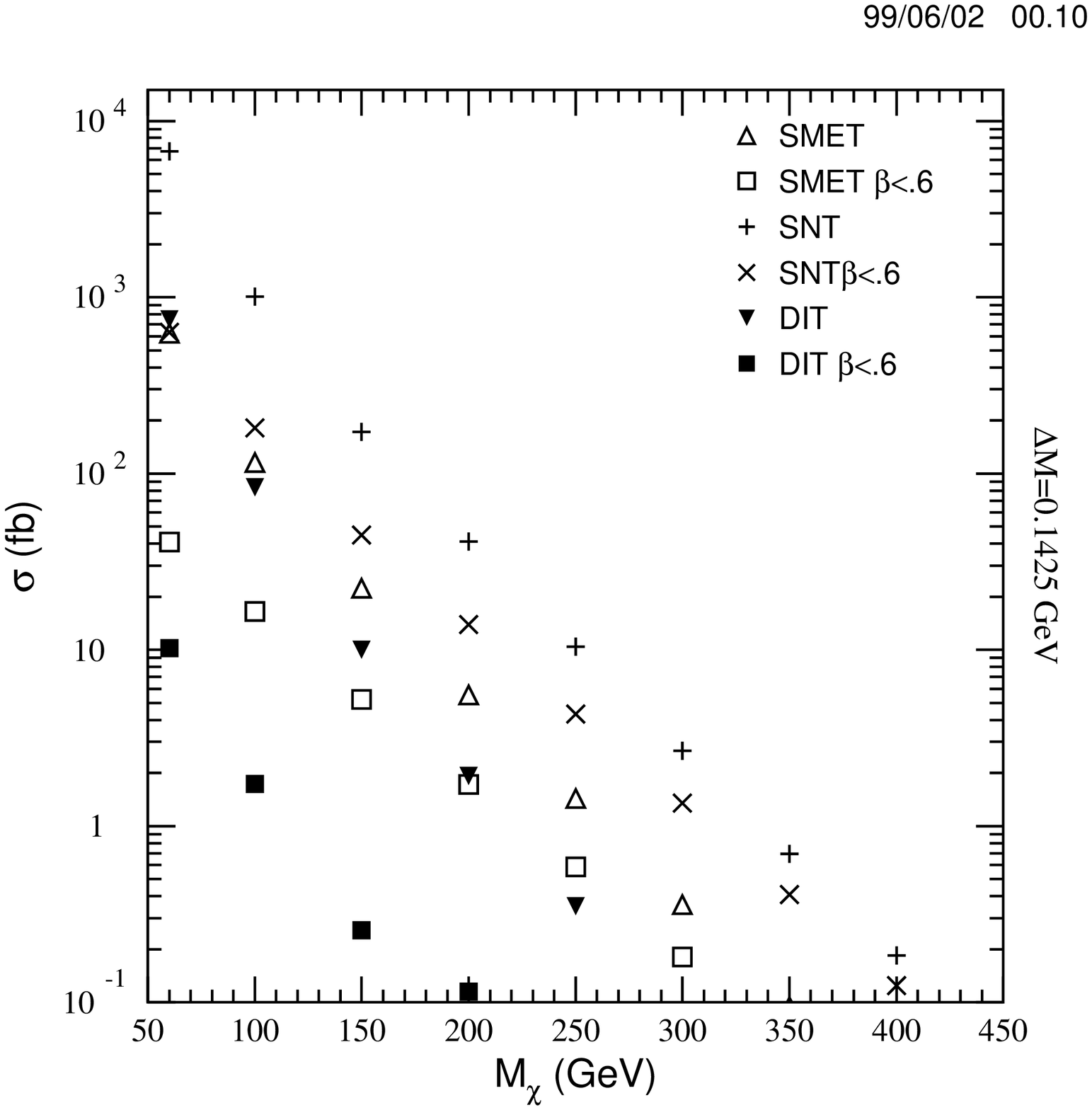}}
\end{center}
\caption[]{Cross sections for DIT and STUB ``background--free''
signatures at RunII: $\dmchi=125,135,140,142.5\mev$.}
\label{sigsb} 
\end{figure} 

\begin{figure}[ht!]
\leavevmode
\begin{center}
{\epsfxsize=3.25in
\epsfysize=3.25in
\hspace{0in}\epsffile{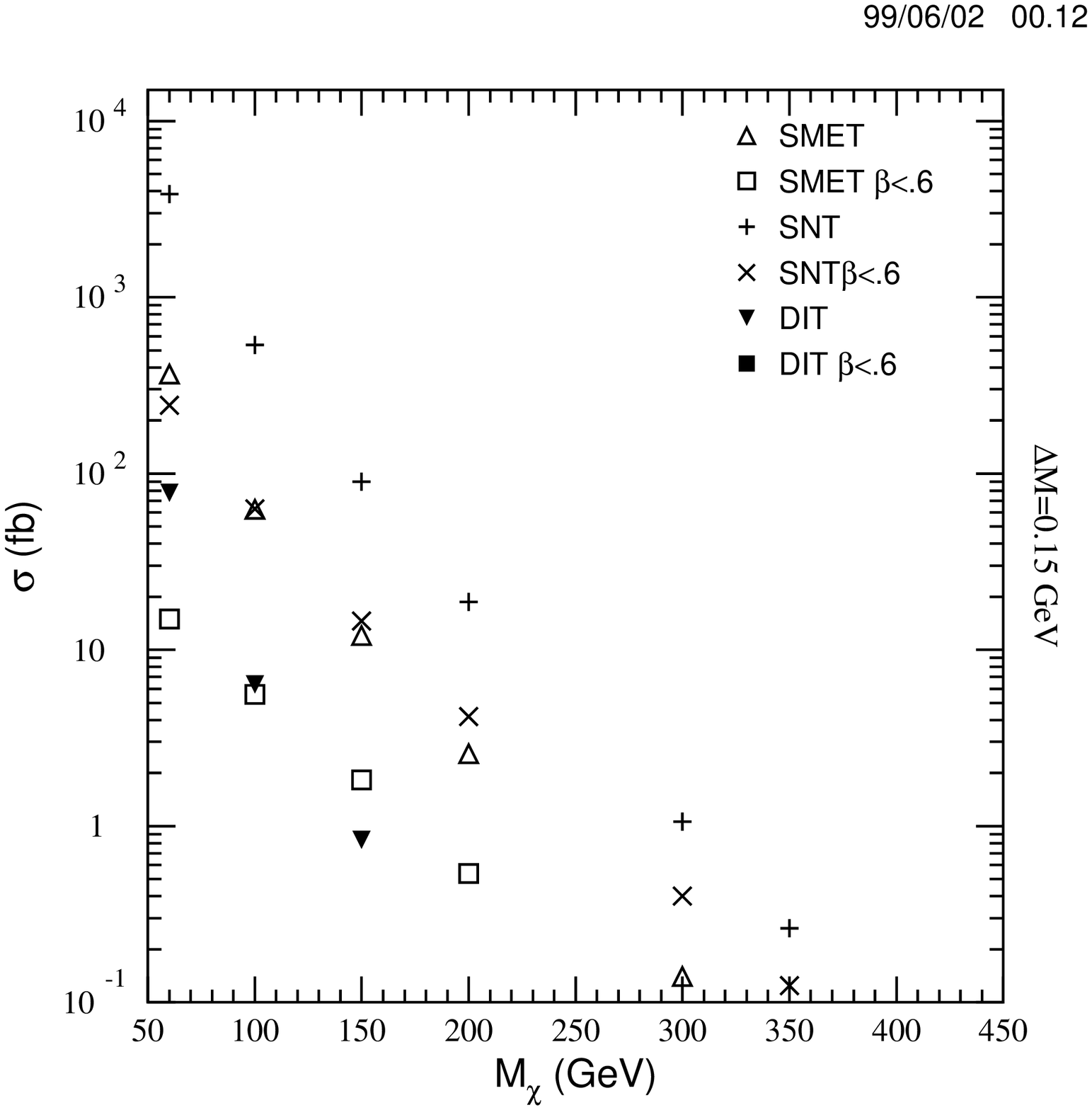}
\epsfxsize=3.25in
\epsfysize=3.25in
\hspace{0in}\epsffile{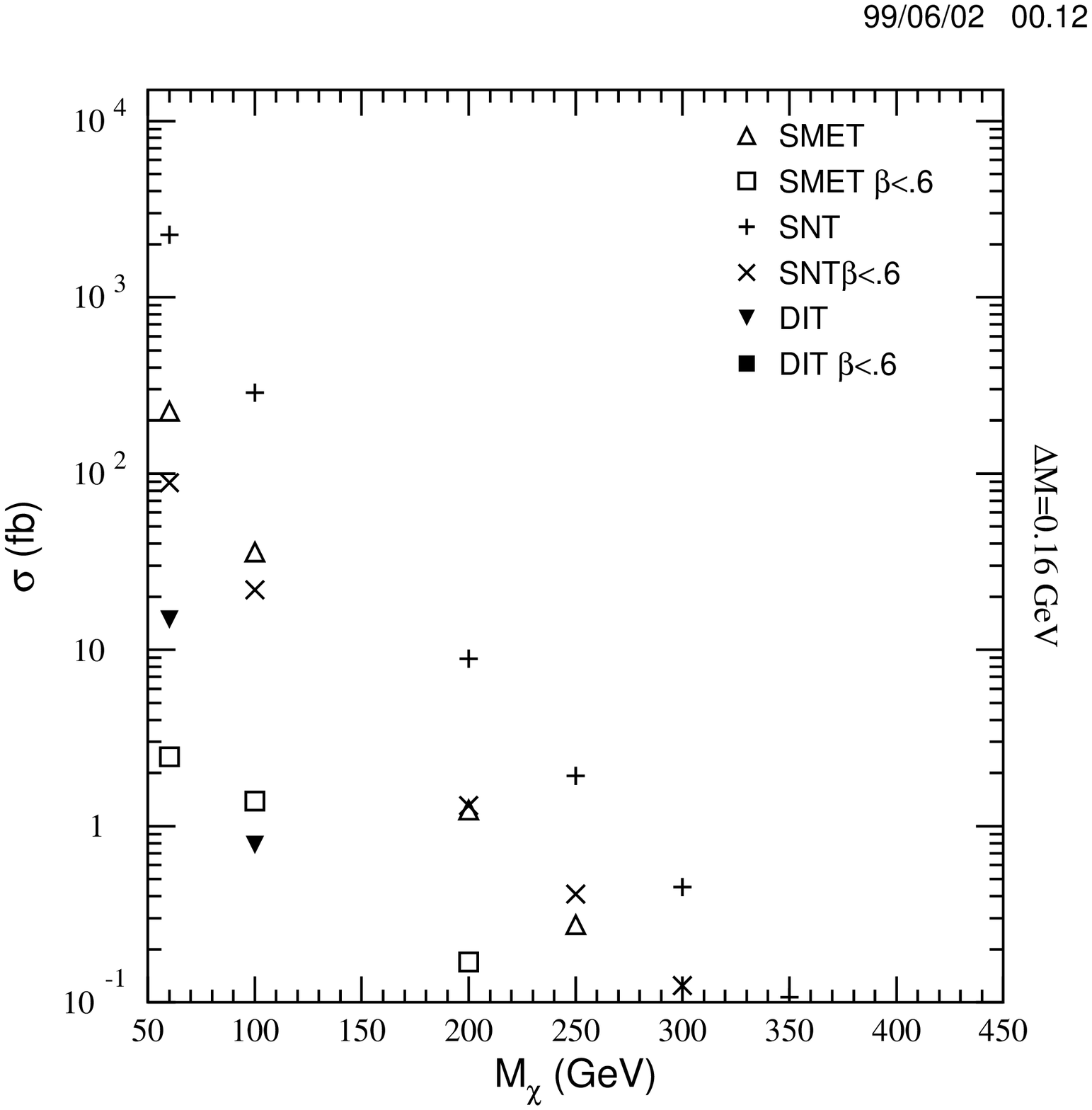}}
{\epsfxsize=3.25in
\epsfysize=3.25in
\vspace{0.in}\hspace{0in}\epsffile{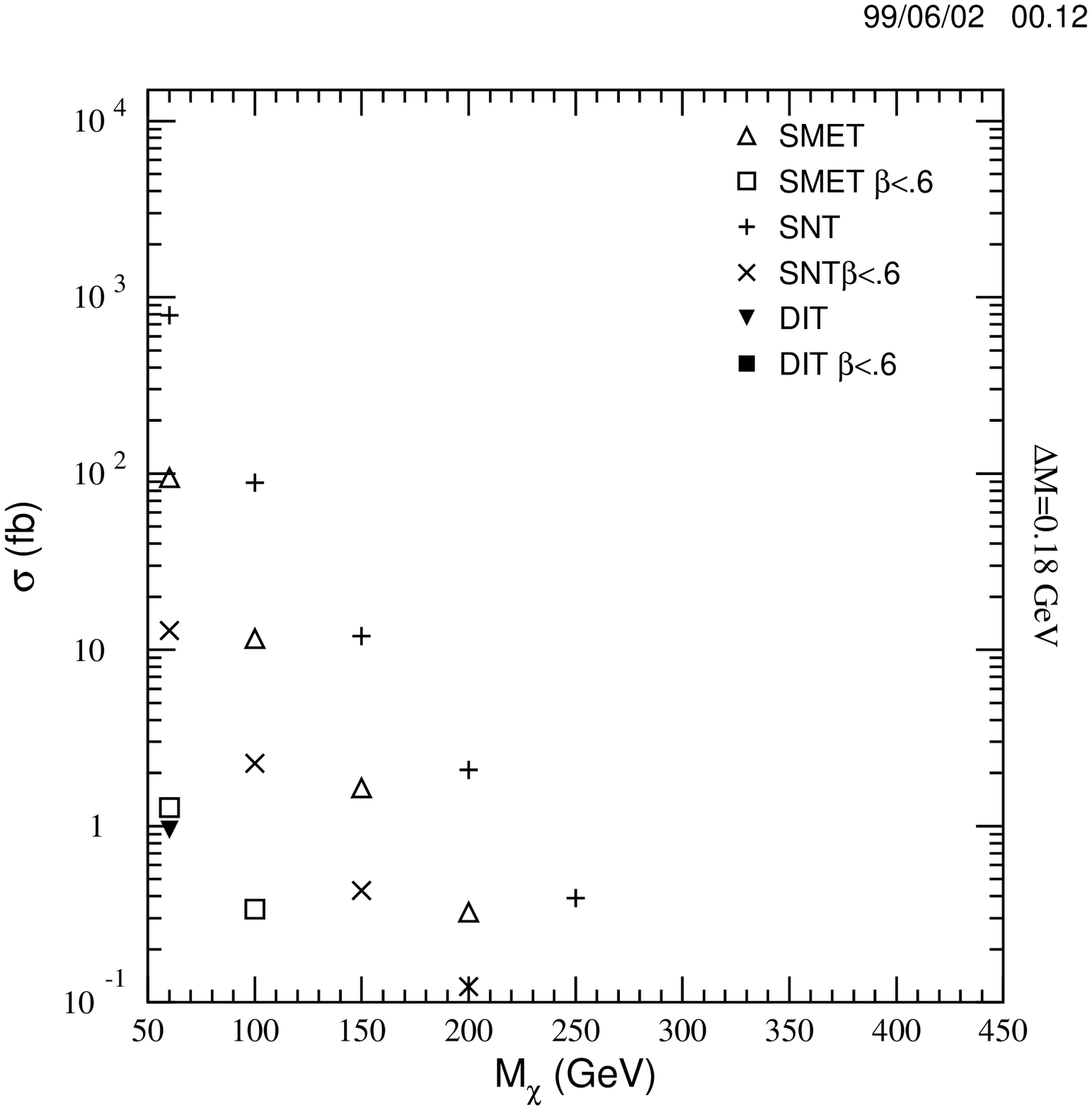}
\epsfxsize=3.25in
\epsfysize=3.25in
\hspace{0in}\epsffile{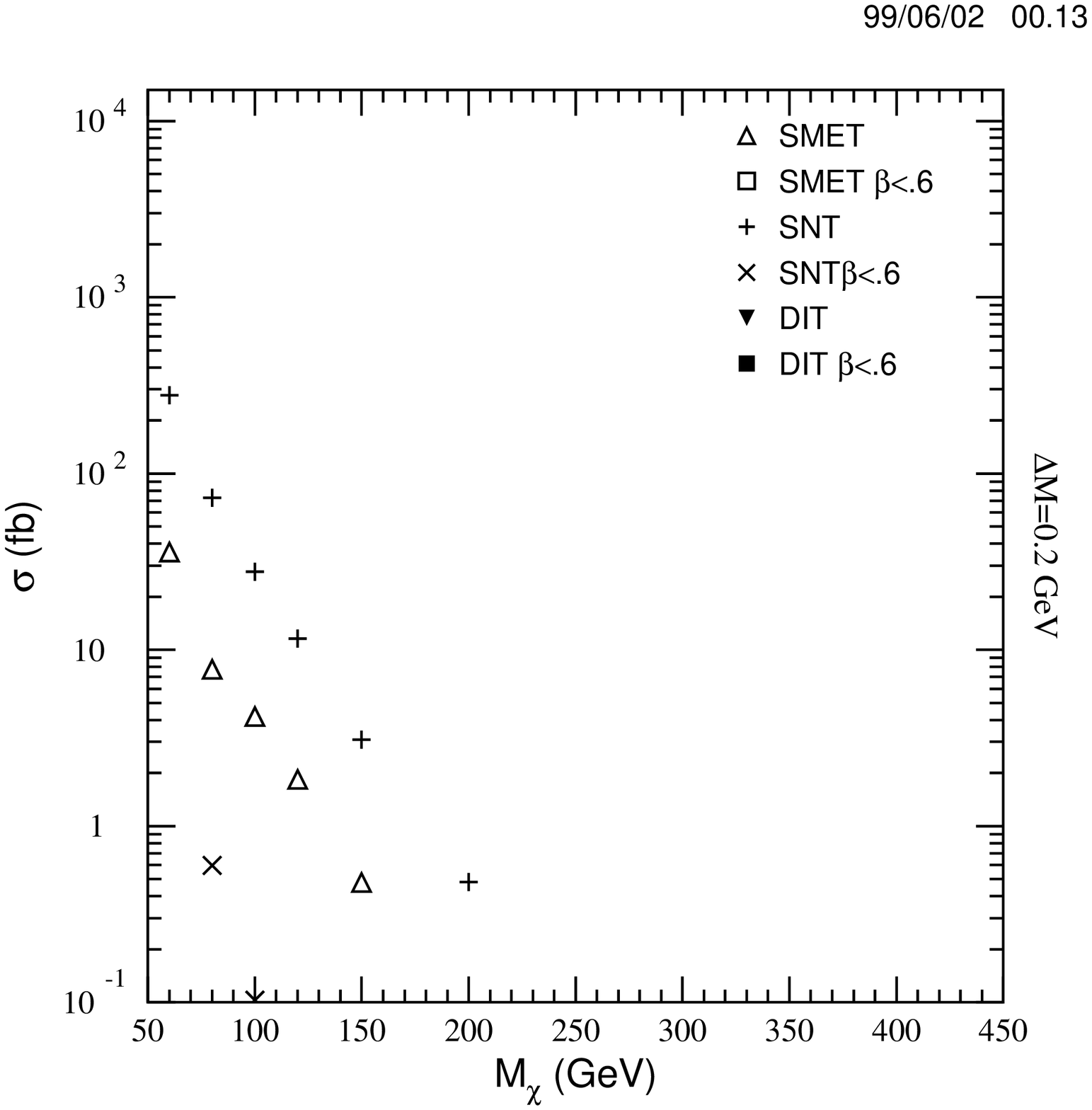}}
\end{center}
\caption[]{Cross sections for DIT and STUB ``background--free''
signatures at RunII: $\dmchi=150,160,180,200\mev$.}
\label{sigsc} 
\end{figure}

As discussed earlier, the requirements of Eq.~(\ref{ditsigcuts}) may 
on their own be sufficient to yield a background--free signal. 
In this regard, the $E_{\rm cal}(\Delta R<0.4)$ 
cut is probably critical for eliminating backgrounds. For example,
an event in which a $\Sigma^+$ or $\Sigma^-$ is 
produced and makes an isolated track in the tracker
would be removed by this cut.
Even if the $\Sigma^\pm$ decays just outside the central
tracker, its decay products are strongly interacting
and will produce substantial deposits in the calorimeters, especially
the hadronic calorimeter.
Still, even after the $E_{\rm cal}(\Delta R<0.4)$ cut, 
one should consider the possibility that the DIT signal will
not be entirely free of background. If not, one can impose a heavy--ionization
requirement. The ionization of the DIT
will be measured in the SVX, CT and PS.
We will consider cuts requiring $\beta<0.6$ ($\beta\gamma< 0.75$)
or $\beta<0.8$ ($\beta\gamma<1.33$).  The former is roughly
equivalent to requiring 3 MIP's of ionization. As illustrated
in Fig.~\ref{betadist}, the latter is a much
weaker cut that would accept many more signal events 
(at least for lower chargino masses), but we estimate
that it would still reduce
the number of background events containing 1MIP tracks 
by at least a factor of 10.
The DIT signals with the above $\beta$ cuts are denoted by DIT6 and DIT8.

Cross sections for the DIT, DIT6 and DIT8 signals are plotted
as solid triangles, upside--down triangles and squares, respectively,
in Figs.~\ref{sigsa}, \ref{sigsb} and \ref{sigsc}.  
From Fig.~\ref{sigsa}, one finds that
for $\dmchi>125\mev$ the DIT signal is as good or better than the LHIT signal.
From the $\dmchi=140\mev$ window of Fig.~\ref{sigsa}, we see 
that even the DIT6 signal becomes superior
to the LHIT and TOF signals as soon as $\dmchi$ exceeds $\mpi$.
Fig.~\ref{sigsb} repeats some of the small $\dmchi$
DIT results, but now in comparison to the STUB and STUB+KINK signals
discussed in the next section. Also shown in Fig.~\ref{sigsb}
is a $\dmchi=142.5\mev$ window. One sees that the DIT signals survive
crossing the $\cpmone\to\pi^\pm\cnone$ decay threshold. In contrast,
the LHIT and TOF cross sections are already very small at this $\dmchi$ value.
Fig.~\ref{sigsc} gives results for the DIT signals for still larger $\dmchi$
values. Assuming no background, the 95\% CL reaches of the DIT and DIT6
signals are given in Fig.~\ref{limits:summary} for a range of $\dmchi$ values
using the 3 event (no background) criterion.

\subsubsection{STUB and KINK signatures}

As the chargino lifetime becomes still shorter, 
the probability for triggering an interesting event 
using LHIT, TOF or DIT signals becomes small.  In this case, one good 
strategy appears to be to use the $\etmiss$ generated by initial--state--radiation of jets
to trigger the event.
Such jets are inevitably present in association with pair--production
of massive particles at a hadron collider. In this section,
we will then identify a chargino event by looking for a
track that passes all the way through the SVX but disappears prior
to reaching the outer radius of the central tracker, i.e. a STUB track. 
The $c\tau$ range of interest is thus roughly 
${\rm 50~cm}\gsim c\tau(\cpmone)\gsim \mbox{few~cm}$. 
From Fig.~\ref{lifebrsnew} and Table~\ref{ctaus}, 
we observe that such $c\tau$ values
are predicted as $\dmchi$ ranges from just slightly
above $\mpi$ up to about $190\mev$. In this $\dmchi$ range,
the chargino decays primarily to a single soft charged pion plus
the $\cnone$. The soft pion might be visible in the 
tracker (where it would be emitted at substantial angle relative to the STUB
track, resulting in a KINK type of signature). The neutralino
takes most of the energy of the decay and is invisible. There are
no calorimeter deposits associated with the STUB. Thus, interesting
events can potentially be identified by demanding that
the STUB be heavily--ionizing, be connected to a KINK, and/or
have no associated calorimeter deposits.

For this study, we assume the detector capabilities and
structure of the CDF detector, including the upgraded SVX described
in Ref.~\cite{cdfextra}. We define a STUB track
by the requirement that the chargino pass through all
layers of the vertex detector (we assume that L00 is present)
and that it have large $p_T$ (as determined off--line
using the SVX track). We also demand
that there be very little calorimeter activity in a cone surrounding
the STUB and that the track not make it to the end of the CT or, equivalently,
to the PS. (In particular, it does not enter the calorimeters.) 
Our specific requirements are
\beq
  \label{stubcuts}
p_T>30\gev\,,\quad E_{\rm cal}(\Delta R<0.4)<2\gev\,,\quad 
\beta_{T}\gamma c\tau>11\cm\,,\quad
|\beta_z|\gamma c\tau<45\cm \,,\quad \beta\gamma c\tau<d_{\rm PS}\,,
\eeq
where $c\tau$ is generated for each chargino following the exponential
distribution determined by the proper lifetime for the given $\dmchi$ value.
There is some chance that a signal requiring one
or more STUB's might be background free, but 
such events cannot be triggered on in the present CDF
and D\O~designs by virtue of the fact that the SVX information
is not analyzed until Level--3. Still, should some sign of this scenario
become apparent in RunII data, perhaps via a very weak DIT signal,
an upgrade of the trigger to include this possibility might be feasible. 

Additional handles are available for ensuring that a STUB signal
is background free.  First, one can search for the KINK created
when the chargino responsible for the STUB decays to a charged
pion inside the tracker.  For $c\tau$ values near 11 cm, this
will be very probable. We will not explicitly explore
the efficiency for searching for KINK's here. However, we have 
computed STUB cross sections after requiring that the chargino
decay a significant distance prior to 
reaching the outer radius of the CT.  Specifically,
we will give STUB cross sections for decay
prior to a radial distance of 50 cm or 1.1 m. (The former
is appropriate for the D\O~tracker that ends at 73 cm --- the
latter is appropriate for the CDF tracker that
extends to 1.3 m.) This type of signature will be denoted by SK (for STUB+KINK).

Finally, we have considered the additional requirement of 
heavy ionization deposit in the SVX. Thus, we also present
results requiring $\geq 1$ STUB with $\beta<0.6$.
The $\geq 3$ MIP ionization of a $\beta<0.6$ track, accompanied
with the high $p_T$ requirement and the lack of associated calorimeter
activity would certainly make this a background--free signal.
Note that the STUB requirement that the chargino pass through 
all six layers of the SVX is critical to obtaining enough $dE/dx$    
samples for a reliable determination of whether or not the track
is heavily--ionizing. Samplings from just a couple of layers would
not be adequate.

Results for the cross sections obtained by requiring $\geq 1$ STUB,
possibly with $\beta<0.6$ imposed, and
no additional trigger (NT), are denoted by SNT and SNT6, respectively.
The cross sections for these signals as a function of $\mcpmone$ are
plotted in Fig.~\ref{sigsb} and \ref{sigsc} for a series of $\dmchi\leq
200\mev$ values. 
Of course, they are always larger than the DIT, DIT8 and DIT6 cross sections,
and certainly remain substantial out to much larger $\dmchi$ values.
The 95\% CL limits based on 3 events (no background) are given for
the SNT and SNT6 signals for a selection of $\dmchi$ values in
Fig.~\ref{limits:summary}. Very significant mass reach results for
the smallest of the $\dmchi$ values, but the mass reach decreases significantly
as $\dmchi$ increases. In particular, we note that
for $\dmchi\geq 250\gev$, only the SNT signal (and
the SMET signal discussed below) have cross sections above 0.1 fb
for $\mcpmone\geq 50\gev$. The corresponding 95\% CL upper limits are shown in
Fig.~\ref{limits:summary}, but we do not give the corresponding
cross section plots. 

\begin{figure}[ht]
\leavevmode
\begin{center}
\epsfxsize=3.25in
\epsfysize=3.25in
\hspace{0in}\epsffile{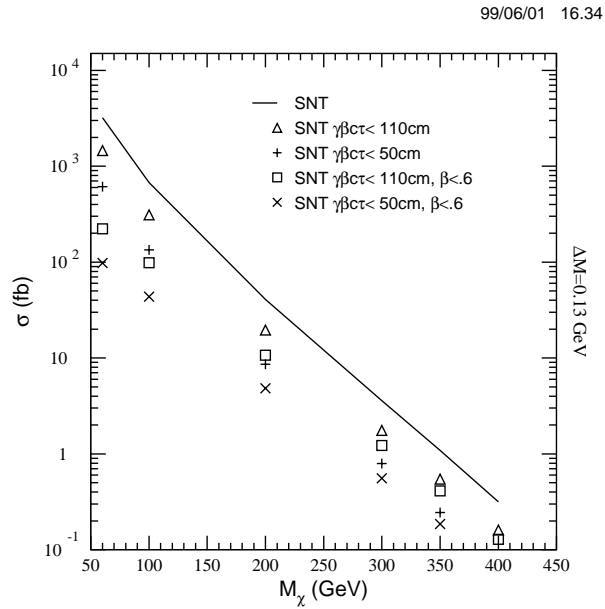}
\end{center}
\caption[]{Cross sections for SKNT and SKNT6 signals
where the KINK distance is either 50 cm, as appropriate for D\O,
or 110 cm, as appropriate for CDF. The solid curve is the SNT
(no $\beta$ cut) cross section.}
\label{kinksa}
\end{figure} 

\begin{figure}[ht]
\leavevmode
\begin{center}
{\epsfxsize=3.25in
\epsfysize=3.25in
\hspace{0in}\epsffile{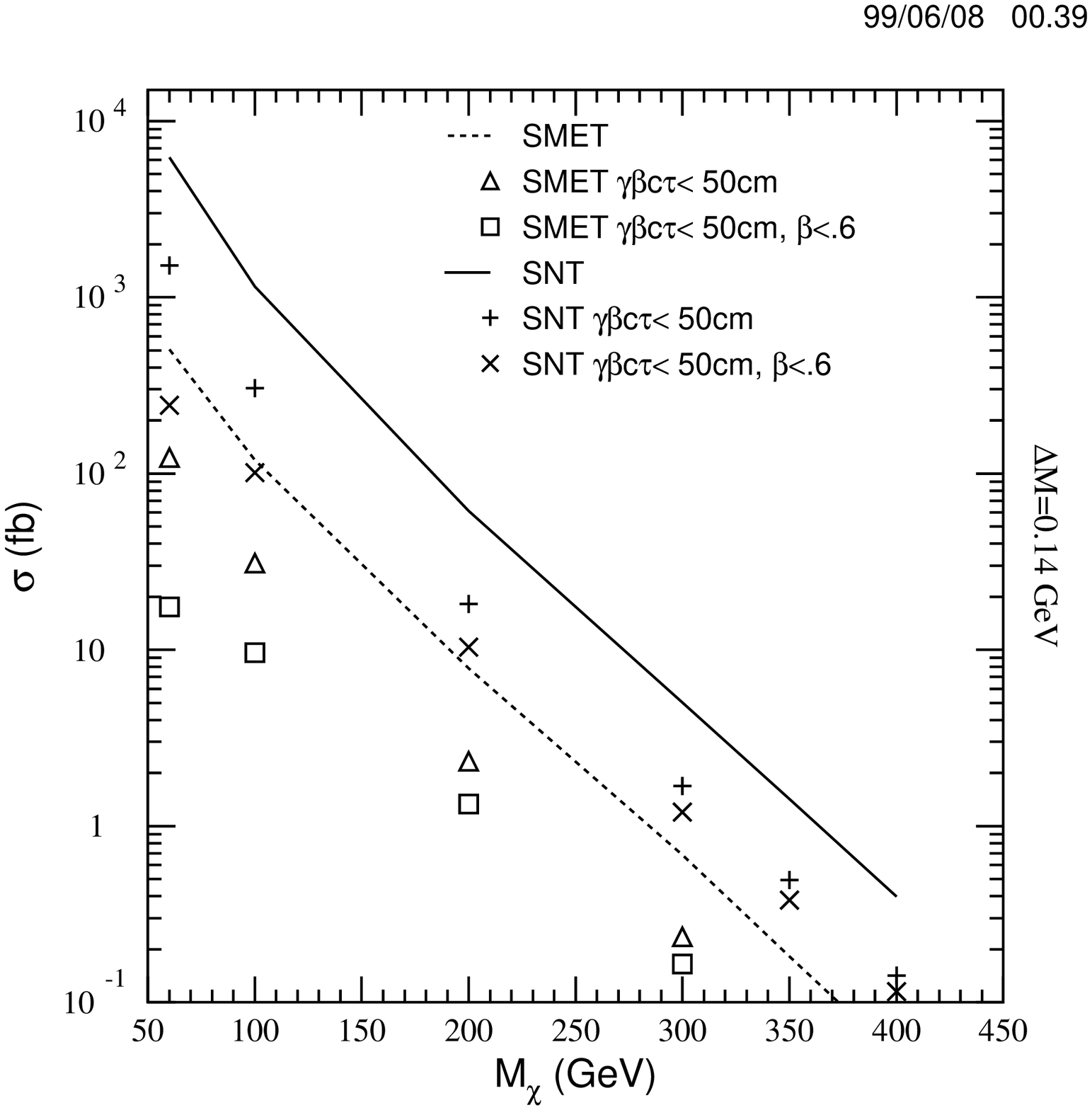}
\epsfxsize=3.25in
\epsfysize=3.25in
\hspace{0in}\epsffile{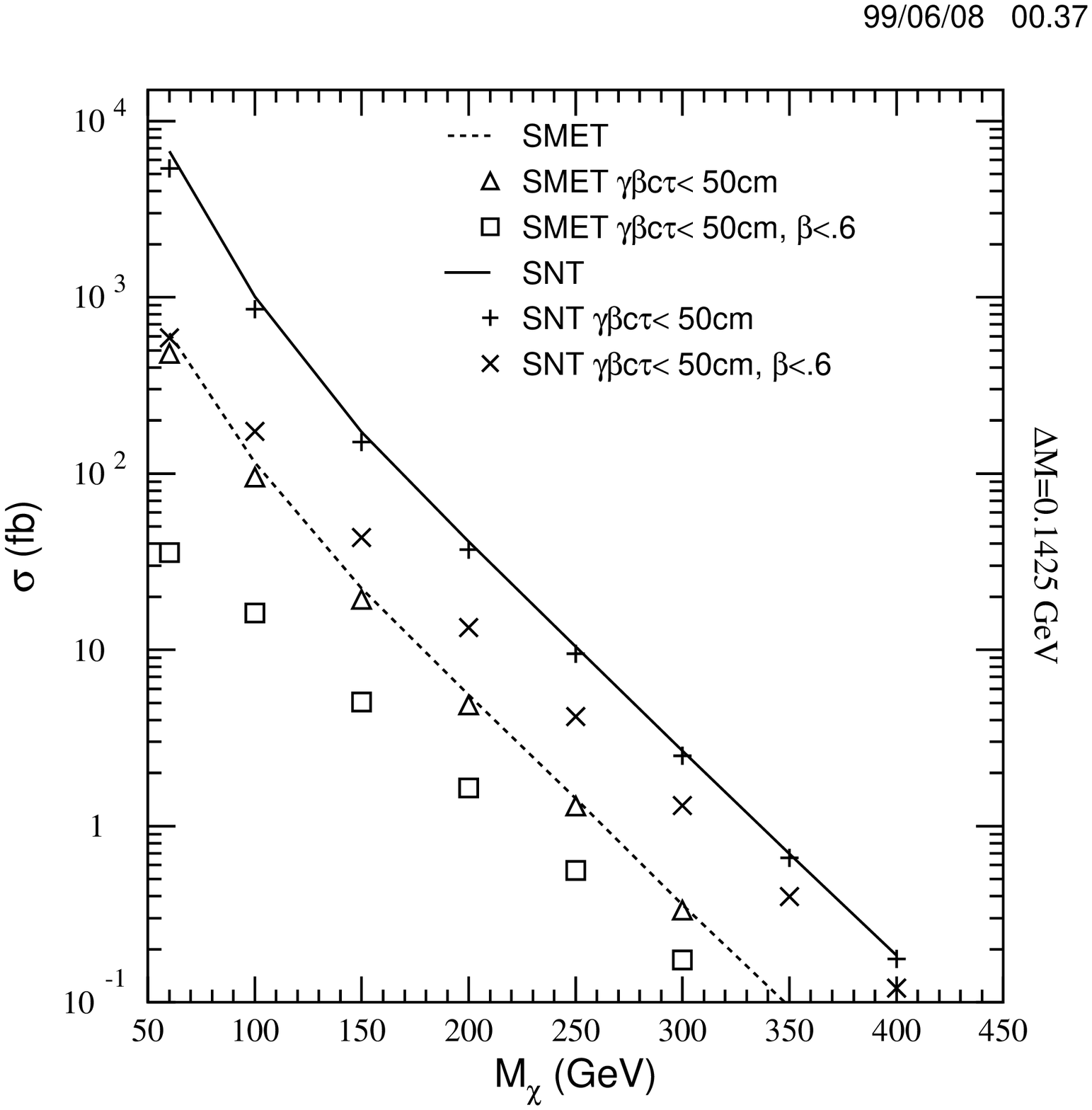}}
\end{center}
\caption[]{Cross sections for SKNT, SKNT6, SKMET and SKMET6 signals
at D\O~in which the K refers to the requirement that the chargino
decay prior to a radial distance of 50 cm. Also shown for
comparison, as the solid and
dotted lines, respectively, are the SNT and SMET (no $\beta$ cut)
cross sections at D\O.}
\label{kinksb}
\end{figure} 

In order to assess the efficiency for seeing KINK's in association with
STUB's, we present Figs.~\ref{kinksa} and \ref{kinksb}.
In Fig.~\ref{kinksa}, we give $\dmchi=130\mev$
cross sections for the 50 cm and 110 cm
maximum radii (as appropriate for D\O~and CDF, respectively),
both before and after a $\beta<0.6$ cut, in comparison to the full
SNT cross section (no $\beta$ cut). For any $\dmchi<\mpi$,
the relative efficiencies
for these different cross sections are essentially the same.
But, observation of a KINK decay for $\dmchi<\mpi$ will be
very difficult since the electron in the dominant $\cpmone\to e^\pm \nu_e
\cnone$ decay is very soft.
However, once $\dmchi>\mpi$ we will be looking for a somewhat harder
(but still soft) charged pion daughter track.
In Fig.~\ref{kinksb}, we present
the SKNT and SKNT6 cross sections for the D\O~KINK
distance of 50 cm for $\dmchi=140\mev$ and $\dmchi=142.5\mev$. 
We observe that the $\dmchi=142.5\mev$ SNT and SKNT results are
essentially the same. This is because, for $\dmchi>142.5\mev$,
the $c\tau$ of the $\cpmone$ is sufficiently short
that the decay always occurs before reaching 50 cm. 
Similarly, results for the CDF KINK distance of 1.1~m differ
very little from the SNT results for any $\dmchi\geq 140\mev$.

Unfortunately, as we have already emphasized, 
the above signals are not available for the current
CDF and D\O~trigger designs.  Thus, we now consider STUB
and STUB+KINK type signatures using an $\etmiss>35\gev$ trigger
for the event ($\etmiss$ is computed assuming $\aeta<4$ calorimeter
coverage and standard smearing and without including any SVX or tracker
information). These signals will be generically denoted by SMET and SKMET.
The $\etmiss$ trigger selects events with initial state gluon radiation.
We compute rates using the PYTHIA parton showering approach. 
For reference, the $\etmiss>35$ GeV trigger requirement 
retains $8-13$\% of all $\cpone\cmone$ and $\cpmone\cnone$
events. While this is not a large efficiency, it has the advantage of
further reducing backgrounds from the very beginning.
A photon tag trigger was also considered, but was not found to be
competitive with the $\etmiss$ trigger.

The main physics background after the trigger, but before any
STUB requirements, is $Z(\to\nu\bar\nu)+$jet,
which has an effective cross section after our triggering requirements
of $\sigma_{\rm eff}\sim 10^3\fb$.
Before STUB requirements, pure QCD backgrounds
are two orders of magnitude larger than the $Z(\to\nu\bar\nu)+$jet background
after requiring $\etmiss>35\gev$, i.e. $\sigma_{\rm eff}\sim 10^5\fb$.
The requirement of an isolated, charged track reduces this
background by at least a factor $10^{-3}$ \cite{rfield}.
A further requirement of $\geq 2$ MIP energy
deposit on all 6 SVX layers contributes
another factor of $\sim 10^{-3}$. 
Therefore, we estimate a background
cross section below about $0.1\fb$.
A cut of $p_T>30$ GeV on the track may be sufficient without
the 2MIP requirement.
For a first estimate
of sensitivity, we assume that the backgrounds are negligible 
after requiring one or more STUB tracks.

The cross sections for the SMET and SMET6 (i.e. $\beta<0.6$ being
required for the latter) signals are plotted as a function of $\mcpmone$
in Figs.~\ref{sigsb} and \ref{sigsc} in comparison to the
DIT and DIT6 signals. The corresponding 3 event
mass limits are given in Fig.~\ref{limits:summary}, including
results for $\dmchi=250\mev$ and $300\mev$. These latter points
show that only the SNT and SMET signals will give a background--free
cross section for $\dmchi$ as large as $300\mev$.

We have also compared the SMET cross section to the cross section
obtained by requiring 2 STUB's without any cut on missing energy (not plotted).
One finds that the SMET efficiency is higher than that for 2 STUB's. Thus,
assuming that the SMET signal is background--free,
it is only if the 1 STUB, i.e. SNT, signal is also background--free
that one would gain by modifying the triggering systems at CDF and D\O~so
that a STUB could be directly triggered on using the SVX alone.

Finally, Fig.~\ref{kinksb} shows the SKMET and SKMET6 cross sections at 
D\O~obtained by adding to the SMET and SMET6 cuts the requirement that
the chargino decay prior to 50 cm, so that one could see in the tracker
the KINK produced by the chargino decay to a soft pion.
For both $\dmchi=140\mev$ and $142.5\mev$, we see very little difference
between these two cross sections.  Thus, one could look for KINK's
with little sacrifice in efficiency.

\subsubsection{HIP signatures}

As $\dmchi$ increases above $250\mev$,
the chargino, on average, passes through fewer and fewer layers of the SVX.
Consequently, it becomes increasingly difficult to  reconstruct the SVX
track and determine its $p_T$. In addition, the number of
$dE/dx$ samplings decreases and
it becomes progressively more difficult
to determine whether or not it is heavily--ionizing.
The STUB signatures become very inefficient. 
The precise point at which the SMET and SMET6 signals (that can
be implemented using current D\O~and CDF trigger designs) become untenable
must be determined by the experimental groups. One could be hopeful
that the reach in $\mcpmone$ of these signals might
be increased for $\dmchi<300\mev$ or so by looking for tracks
that penetrate some, but not all, of the SVX layers. 
Such signals might be relatively
clean if one could also see the KINK of the $\cpmone\to \pi^\pm \cnone$ decay
in the SVX. But, it seems very unlikely
that one could go much beyond $\dmchi=500\mev$ ($c\tau=0.1$~cm).
Above some point in the $\dmchi=300-500\mev$ range, the only visible
sign of the chargino will be the high--impact--parameter of the soft
charged pion emitted in $\cpmone$ decay.
Aside from needing a means for triggering on HIP tracks, we will see that
substantial additional
requirements must be imposed to control the backgrounds.
As a baseline for this analysis, we use the impact parameter resolution
$\sigma_b$ of the upgraded SVX of the CDF detector, described earlier
and detailed in Ref.~\cite{cdfextra}.
As before, we assume that the SVX will have the proposed, extra layer L00
at a radius of roughly 1.6 cm.  If the chargino decays before
this radius, we use the L00 parameterization of $\sigma_b$.
Otherwise, if a decay occurs between 1.6 and 3.0 cm, we use the
L0 resolution.
For a pion track of $p_T=75$ MeV, this corresponds to $\sigma_b$
of .28 (.37) mm using L00 (L0); the corresponding
large--$p_T$ limits are roughly
.012 (.014) mm.  We require $b/\sigma_b > 5$ to eliminate fakes,
which means the detector is not sensitive to $b< 0.06$ (0.07) mm. 
Such charged tracks, with $b$ larger than 5 times
the resolution, will be denoted as HIP's.

Unlike the STUB signature, the HIP signature has irreducible backgrounds.
The best results are obtained using events that pass
our $\gamma+\etmiss$ requirements. The HIP backgrounds for the
monojet+$\etmiss$ event sample are much larger.
In any hard scattering process, fragmentation and
hadronization of hard jets and beam remnants can produce particles
in the central rapidity region
with $\gamma c\tau$ on the order of 0.1 to 10 cm that decay
to charged tracks: $K^0_S, D, B, \Lambda, \Sigma, \Xi, \Omega$.
To reduce this background without
substantially reducing the signal, we impose a number of additional cuts:
\beq
75\mev<p_T^{\rm HIP}<1\gev\,,\quad E_T(\Delta R<0.4)<2\gev\,,\quad
N_{\rm tracks}=1\,,
\label{hipcuts}
\eeq

\bit
\item The $p_T<1$ GeV cut is not optimized. It is
100\% efficient for the soft charged pions emitted in chargino
decays in the models considered here, but strongly
suppresses the many backgrounds that tend to yield HIP's with
large $p_T$.
\item
The $p_T>75$ MeV cut is imposed because $\sigma_b$ is increasing quickly
below this value.
\item
The $E_T(\Delta R<0.4)<2\gev$ cut is designed to
remove HIP's directly associated with hard jets, which (by definition)
have substantial transverse energy in particles nearby a $p_T<1\gev$ HIP.
\item Some background is removed by requiring that only 
one charged track is associated with a given impact
parameter (i.e., most $K_S^0\to\pi^+\pi^-$,
$\Lambda^0\to p^+\pi^-, \etc$, decays can be reconstructed and removed
when one of the tracks has a large $b$).  
\eit
Since heavy flavor
is always produced in pairs from the parton sea, it may be possible
to tag both hadrons and eliminate part of the background ($s\bar s\to
\Sigma^+ K^0_S+X$), but we have not included this in our analysis.
Nor have we used the fact that some of the decays with a single
charged track can be explicitly
reconstructed (e.g., $\Sigma^+ \to p^+\pi^0(\to\gamma\gamma)$).
After our cuts, of all the long--lived particles
noted earlier, only events containing
the baryons $\Sigma^+$, $\Sigma^-,\Xi$, and $\Omega$ survive.
Additional backgrounds arise from $\tau$ decays in the processes
$Z/\gamma^*(\to\tau^+\tau^-)+\gamma$
and $W(\to\tau\nu_\tau)+\gamma$, but these are insignificant
after the $\gamma+\etmiss$ cuts.

\begin{figure}[ht]
\leavevmode
\begin{center}
\epsfxsize=3.25in
\epsfysize=4.in
\hspace{0in}\epsffile{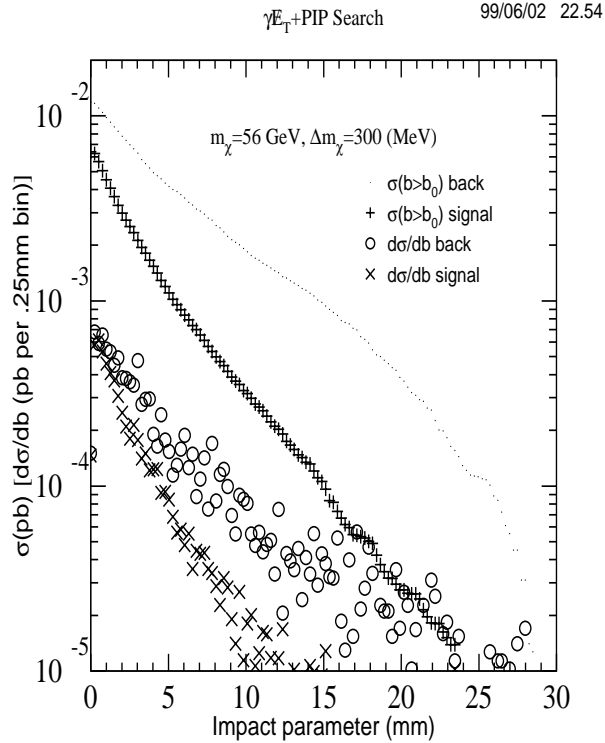}
\end{center}
\caption[]{Impact parameter distributions for the signal and background.
We plot the differential cross section, $d\sigma/db$ (in units of pb~/~0.25~mm),
and the integrated cross section, $\int_b^\infty (d\sigma/db^\prime)db^\prime$
(in pb units). The signal shown here is for $\mcpmone=56\gev$
and $\dmchi=300\mev$. The fluctuations in
the distributions are from the statistics of the Monte Carlo simulation.}
\label{bdist}
\end{figure}

After the cuts listed above, PYTHIA predicts
that about 14 fb of background remains in the single
HIP signal and a fraction of a fb in the double HIP signal, with a tail in the
impact parameter distribution extending out to the L0 radius.  
For any $\dmchi\gsim 200\mev$, the impact parameter distribution
for the signal is quite similar to that for the
background. This is illustrated in Fig.~\ref{bdist}
for the case of $\mcpmone=56\gev$ and $\dmchi=300\mev$.
As a result, no additional cuts on $b$ appear to be useful
and the HIP search is reduced to a simple counting experiment.
In order to check the PYTHIA computation of
the background from baryons with delayed
decays that dominate the impact parameter distribution, it will be very useful
to measure this same component of
the impact parameter distribution in $Z(\to e^+e^-,\mu^+\mu^-)+\gamma$.
This will allow considerable control of systematic errors in the background
predictions.

\begin{figure}[ht]
\leavevmode
\begin{center}
{\epsfxsize=3.25in
\epsfysize=4.in
\hspace{0in}\epsffile{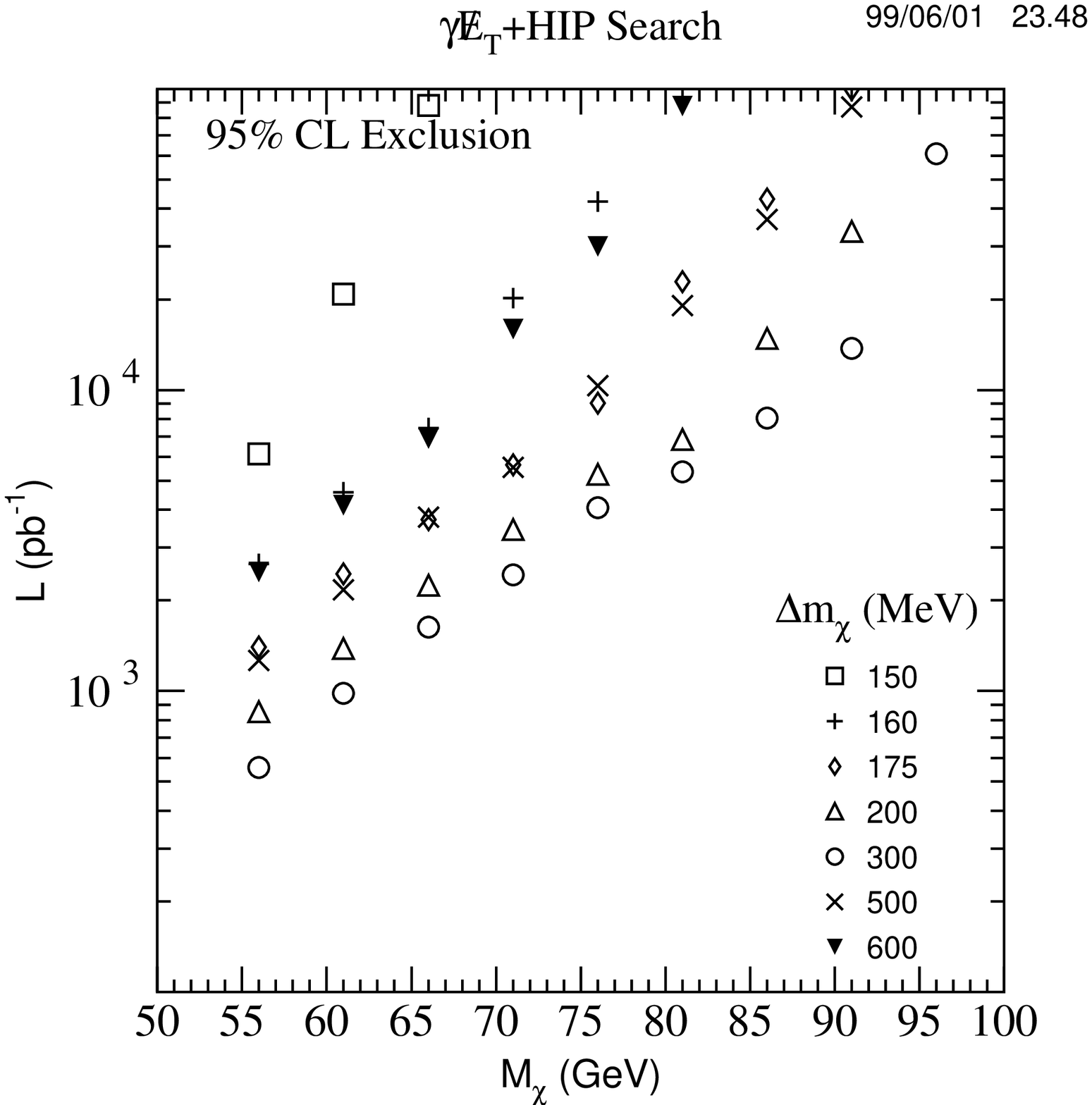}
\epsfxsize=3.25in
\epsfysize=4.in
\hspace{0in}\epsffile{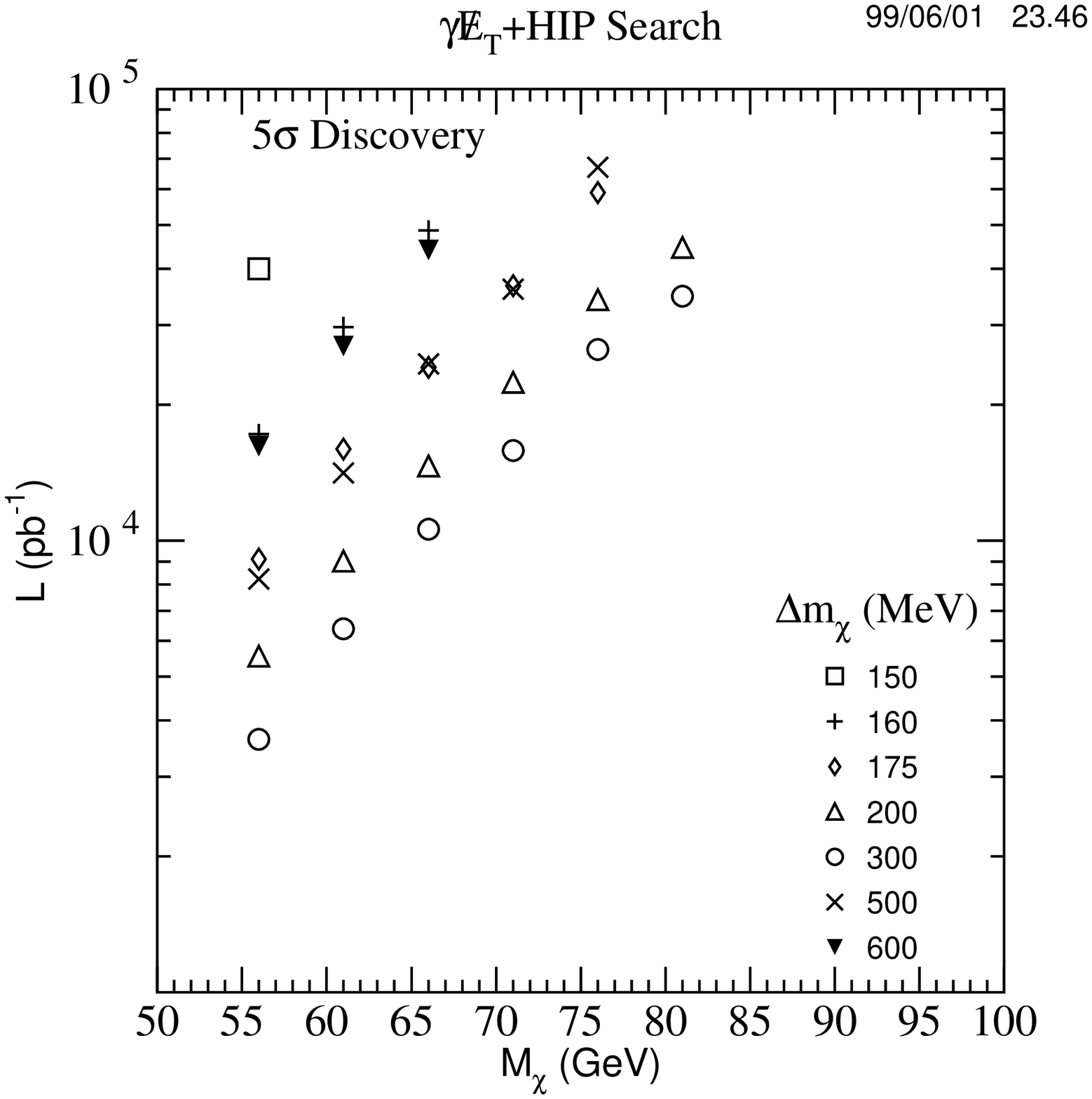}}
\end{center}
\caption[]{Reach for the $\gam+\etmiss$ searches
after requiring one or more HIP for
different mass splittings $\dmchi$.  The left
curve shows the 95\% C.L. exclusion; the right shows
the 5$\sigma$ discovery.  We require $S/B>.2$ and 
at least 3 (5) expected events for exclusion (discovery).}
\label{gammapip}
\end{figure} 

After requiring $S/B>0.2$,
the integrated luminosity required to either exclude 
a chargino of a given mass at the $1.96\sigma$
(95\% CL) level or discover it at the $5\sigma$ level is plotted in
Fig.~\ref{gammapip}(a) or (b), respectively. 
For $\mcpmone$ values above those plotted, $S/B$ falls below 
the $S/B>0.2$ criterion that we impose.
We note that the HIP signal for small $\dmchi$ is quite weak.  This
is because most decays are such that the chargino passes through
the SVX before decaying. In this case, one should look for the STUB
and STUB+KINK signatures discussed earlier, 
for which backgrounds are negligible and much better sensitivity is possible.
Clearly, the STUB and HIP signals are complementary
with viability for the latter rising with increasing $\dmchi$          
as mass reach for the former declines.
As $\dmchi$ increases, the HIP+$\gam+\etmiss$ signal increases in viability
until $\dmchi\gsim 300\mev$. By $\dmchi=600\mev$, the typical impact
parameter for the decay pion declines below $100~\mu$m, and cannot be
resolved by the SVX; the HIP signal can only probe $\mcpmone$ values
below the roughly $90\gev$ limit set by the DELPHI analysis for
$\dmchi\geq 600\gev$. 

Some further discussion of the difference between the STUB and HIP
signals is perhaps useful. First,
it is the large background that restricts
the mass reach of the HIP signal, whereas the biggest limitation
on the STUB signals is associated with the chargino lifetime.
Second, the STUB signals are background--free while the HIP signal is not.
The key to eliminating backgrounds
to STUB signatures is that the $\Sigma^\pm,\ldots$
hadrons that can give an SVX track will decay to particles
that pass all the way through the CT and give sizeable hadronic
calorimeter energy deposits; in addition, one or more of the decay products
are normally (i.e. except for distribution tails where
the charged decay products all have small $p_T$) visible as a full CT track.
If there is some remaining background, then one would have to also look 
to see if the STUB is heavily--ionizing. 
The $\Sigma^\pm,\ldots$ background tracks
would all be minimal--ionizing, so that a $\beta<0.6$ requirement
would certainly eliminate the remaining background.
The cross--over point between the signals depends
on whether the heavy--ionization requirement is necessary to
remove the background. For $\dmchi=180\mev$,
Fig.~\ref{limits:summary} shows that with $L=30\fbi$,
the SMET (SMET6) signals (which include the $\etmiss>35\gev$
trigger requirement) can be used to exclude at 95\% CL 
any $\mcpmone\lsim 240\gev$ ($\lsim 140\gev$). In contrast, for
$\dmchi=180\mev$ the HIP signal can only probe $\mcpmone\lsim 80\gev$.
For $\dmchi=200\mev$, the SMET signal still probes up to $\mcpmone\lsim
190\gev$, but the SMET6 signal falls just below the 95\% CL.
For $\dmchi=300\mev$, the SMET and HIP signals
both probe up to $\mcpmone\lsim 90\gev$.

Various exotic signals can be envisioned that might probe $\dmchi$
values above $600\mev$, but we only comment on them here.
For example, if $\dmchi>m_s+m_c$ the decay $\cpmone\to D_s^*\cnone$
may occur, leading to a $D_s$ meson that carries most of the $D_s^*$'s
momentum.  When combined with an $\etmiss$
trigger, the signature would be quite distinctive since the
$D_s$ will not be associated with a jet. However, the event rate for 
such an ``exclusive'' channel might be small.

\section{Summary of Results and Discussion}

In the previous sections, we considered several signatures for 
chargino production in models with
near mass degeneracy between the lightest chargino and neutralino.  
A brief summary of these signatures appears in Table~\ref{signals}.
We have seen that there is a natural boundary near a mass splitting 
of $\dmchi\sim 300\mev$. 
\bit
\item
For values of $\dmchi\lsim 300\mev$ (mass region A), 
one considers a set of signals based on observing a long--lived
chargino as a semi--stable, isolated track in the detector. 
The most unique signals are the long, heavily--ionizing--track
(LHIT) signal and the delayed time--of--flight (TOF) signal. These are present
for events in which the chargino does not decay 
before reaching the muon chambers. For events in which the chargino
decays before the muon chambers, but still produces a track of substantial
length, the relevant signals are 
the disappearing--isolated--track (DIT) signal 
and the short--isolated--SVX--track
(STUB) signal. The LHIT and TOF signals are dominant if $\dmchi$
is very small (implying a very long chargino lifetime), but the
latter signals quickly turn on as $\dmchi$ 
is increased, becoming the primary signals as $\dmchi$ is
increased to values above $\mpi$.

All these signals are
distinct enough to be possibly background--free.  
Because of the subtle nature of these signals, in estimating the
the range of $\mcpmone$ values to which they can be sensitive
for any given $\dmchi$, we have 
imposed cuts/requirements such that the backgrounds should be
negligible, even at the expense of some signal rate.
\item
For $\dmchi\gsim 300\mev$, the chargino has an average lifetime
such that the background--free signals 
have too low an event rate (at the Tevatron)
and we are forced to consider signals with substantial backgrounds from 
physics and mismeasurement sources. There are two primary
signals in this latter category, but the most sensitive one
can only be used for $300\mev\lsim \dmchi\lsim 600\mev$ (mass region B).
It relies on observation of a high--impact--parameter (HIP) pion
from the chargino decay in association with a photon tag/trigger
and large $\etmiss$.
For $600\mev\lsim 10-20\gev$ (mass region C), 
the chargino decay is essentially prompt, and we are forced to
use the very insensitive signal of a photon tag/trigger
plus large $\etmiss$ to search for chargino production.
\eit
We will now summarize the Tevatron mass reach in $\mcpmone$
that can be achieved in the $M_2<M_1\ll|\mu|$ scenario (1), assuming
that the gluino, squarks and sleptons are
all too heavy to have significant production rate (as is entirely possible).

\begin{figure}[ht]
\leavevmode
\begin{center}
\epsfxsize=6.5in
\hspace{0in}\epsffile{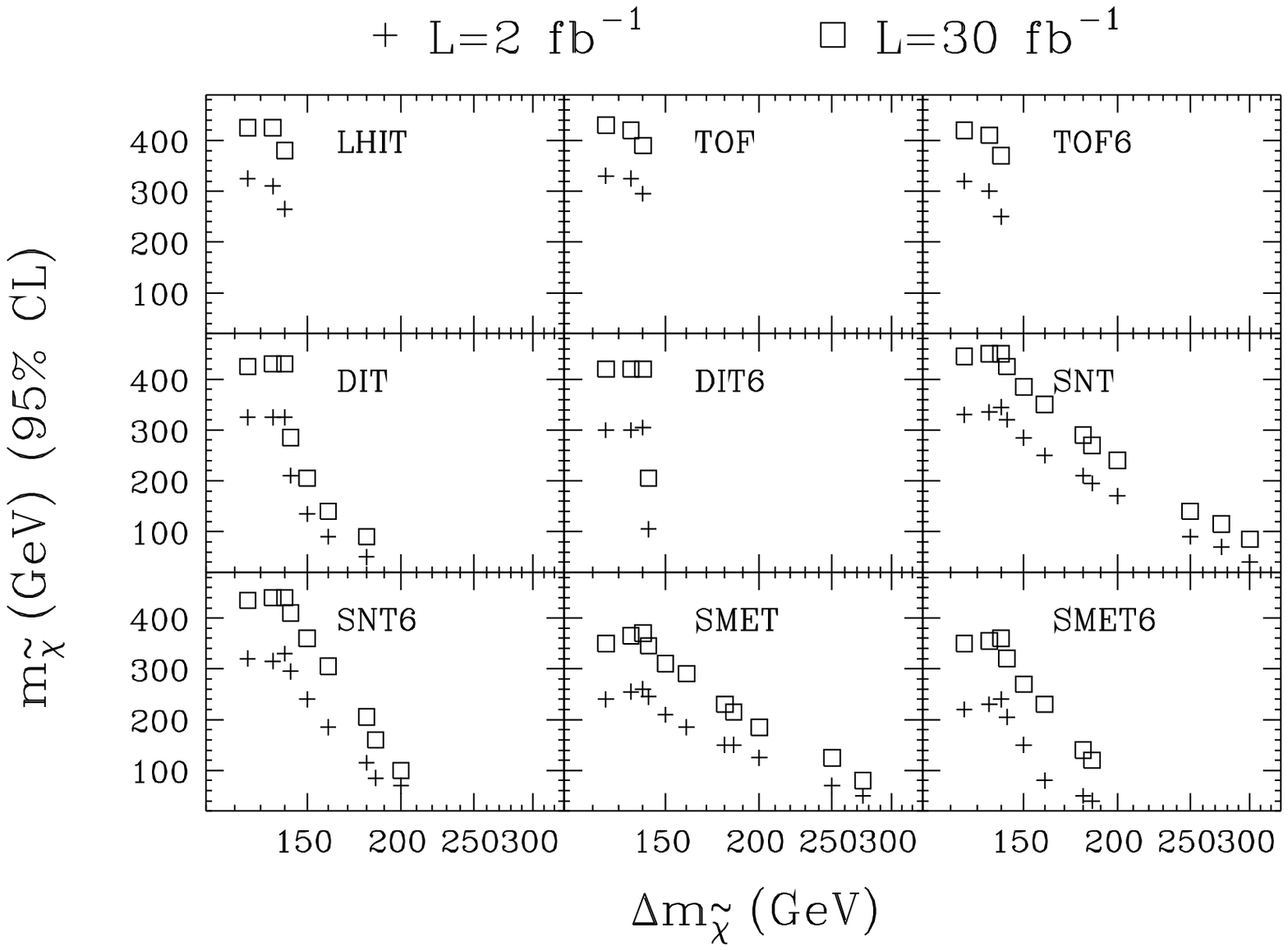}
\end{center}
\caption[]{95\% CL lower limits on $\mcpmone$ as a function of $\dmchi$
for ``background--free'' signatures at RunII with $L=2\fbi$
and $L=30\fbi$.}
\label{limits:summary} 
\end{figure}

\bigskip
\noindent {\bf Region (A)}
For $\dmchi$ values $\lsim 200-300\mev$, one considers
the background--free signals summarized above, which will have the most
substantial mass reach in $\mcpmone$.  
The $L=2\fbi$ and $L=30\fbi$ 95\% CL (3 events, no background) limits
on $\mcpmone$ deriving from these signals 
are summarized in Fig.~\ref{limits:summary}. We give a brief
verbal summary.
\bit
\item {\boldmath $\dmchi<\mpi$:}

For such $\dmchi$, the average $c\tau$ of the chargino
is of order a meter or more. The LHIT and TOF signals are prominent,
but the DIT and STUB signals appear if $\dmchi$ is not
extremely small. These arise as a result of the exponential
form of the $c\tau$ distribution in the chargino rest frame, which implies that
the chargino will decay over a range of radii within the detector.
One must also include the event--by--event variation of the
boosts imparted to the chargino(s) during production.
\bit
\item
The LHIT signature can probe masses in the range 
$260-325$ ($380-425)$ GeV for $L=2\fbi$ ($30\fbi$), the lower
reach applying for $\dmchi\sim\mpi$ and the highest reach applying
for any $\dmchi\lsim 125\mev$.  
The reach of the TOF signature is nearly identical to that of the LHIT
signature.\footnote{The primary difference between the LHIT and TOF signals
is that the LHIT signal requires $\beta\gamma<0.85$ for heavy ionization,
whereas the effective cut--off on $\beta\gamma$ imposed
by the TOF delay requirement allows much larger $\beta\gamma$ to also
contribute. That the maximum reach of the two signals is essentially
the same is somewhat accidental. It happens that the chargino
production cross sections are large enough that charginos with rather
large mass can be probed and such massive charginos are produced with low
$\beta$. Thus, a cut requiring low $\beta$ is highly efficient.
This is illustrated, for example, by comparing the TOF to the TOF6 results in
Fig.~\ref{limits:summary}. This same accident is generic to all
the signals discussed, so long as $\dmchi<\mpi$.}
\item
The DIT signature has a reach of 320 (425) GeV for 
$ 120\mev\leq\dmchi\leq\mpi$,\footnote{We did not study lower $\dmchi$
values since they are highly improbable after including
radiative correction contributions to $\dmchi$.} and, 
in particular, is more efficient than the LHIT and
TOF signals for $\dmchi\sim\mpi$.
The DIT signature reach drops by about 20 GeV with a $\beta<0.6$ cut
(DIT6) designed to require that the chargino track be heavily--ionizing.  
\item
The STUB signature with no additional 
trigger (SNT) can reach to $\simeq 340$ (450) GeV
for $120\mev\leq\dmchi\leq\mpi$, which mass reach drops
by $10-20$ GeV if $\beta<0.6$ is required.
However, neither D\O~nor CDF can use
STUB information at Level--1 in their current design.  
\item
With the addition of a standard $\etmiss$
trigger, the resulting STUB signature (SMET) will be viable with 
the present detectors, reaching to
$240-260$ ($350-375$) GeV for $120\mev\leq\dmchi\leq\mpi$, 
which numbers drop by about
10 GeV if $\beta<0.6$ is required (SMET6).
\eit
\item {\boldmath $\mpi\lsim\dmchi\lsim 200-300\mev$:}
\bit
\item
The LHIT and TOF signatures disappear,
since almost all produced charginos decay before reaching the MC or TOF.  
\item
The DIT signature
remains as long as the $\beta<0.6$ (heavily--ionizing)
requirement is not necessary to eliminate backgrounds.  
If we require $\beta<0.6$, there is a mismatch with the requirement
that the chargino pass through the CT -- once $\dmchi$
is above $145\mev$, the entire signal is
generated by large boosts in the production process which 
is in conflict with requiring small $\beta$.
\item
The SNT signature probes $\mcpmone\lsim 300\gev$ ($\lsim 400\gev$)
for $\dmchi\sim \mpi$ and $L=2\fbi$ ($L=30\fbi$). For $\dmchi$
as large as $300\mev$, it alone among the
background--free channels remains viable, probing $\mcpmone\lsim 70\gev$
($\lsim 95\gev$). Certainly, it would extend the $\sim 90\gev$ 
limit obtained by DELPHI at LEP2 that applies for $\dmchi<200\mev$
and the $\sim 45\gev$ limit from LEP data that is the only available
limit for $200\mev\leq\dmchi\leq 300\mev$.
But, as stated above, the SNT
signature will not be possible without a Level--1 SVX trigger.
\item The STUB+$\etmiss$, SMET and SMET6 signatures
are fully implementable at RunII and have a reach that
is only about 20 GeV lower than their SNT and SNT6 couterparts.
\eit
\eit

\bigskip
\noindent {\bf Region (B)}
For $300\mev\lsim \dmchi\lsim 600\mev$, the high--impact--parameter (HIP) 
signal (a $\gam+\etmiss$ tag for events yields
the smallest backgrounds) is very useful despite the large background from 
production of $\Sigma^\pm,\ldots$ hadrons. It would yield a 
95\% CL lower bound of $95\gev$ ($75\gev$) on $\mcpmone$ for 
$\dmchi=300\mev$ ($\dmchi=600\mev$) for $L=30\fbi$. This would represent
some improvement over the $\sim 60\gev$ lower bound
obtained in the current DELPHI analysis of their LEP2 data 
for this same range of $\dmchi$ if the sneutrino is heavy.
(If the $\snu$ is light, then there is no useful LEP2 limit
if $300\mev\leq\dmchi\leq 500\mev$, but LEP data requires $\mcpmone>45\gev$.)
With only $L=2\fbi$ of data, the HIP analysis would only exclude
$\mcpmone<68\gev$ ($<53\gev$) for $\dmchi=300\mev$ ($\dmchi=600\mev$).

\bigskip
\noindent {\bf Region (C)}
For $\dmchi\gsim 600\mev$, up to some fairly large value (we estimate
at least 10 to 20 GeV), the chargino decay products are effectively
invisible at a hadron collider and 
the most useful signal is $\gamma+\etmiss$. However, 
this signal at best probes $\mcpmone\lsim 60\gev$ (for any $L>2\fbi$), 
whereas the DELPHI analysis of their LEP2 data
already excludes $\mcpmone\leq 60\gev$ for $500\mev\leq\dmchi\leq 3\gev$
(if the sneutrino is heavy --- only $\leq 48\gev$
if the sneutrino is light) and $\mcpmone\leq 90\gev$ for $\dmchi>3\gev$.

\bigskip
All the above mass limits
assume that the gluino is quite heavy and rarely
produced at the Tevatron.  If it is not too much heavier
than the chargino, then all the above signals will have additional
event rate coming from $\gl\gl$
pair production followed by $\gl\to\cpmone q^\prime\anti q$ decays.
The effect of $\gl\gl$ pair production 
on the LHIT, TOF, DIT, SNT, SMET, and HIP
signatures depends strongly on the mass splitting between the gluino
and the chargino as well as on $\dmchi$, so we did not explicitly consider
possible enhancements to these signatures here. Instead, we focused
on the fact that gluino production will provide a critical increase
in the mass reach when neither the chargino track nor its decay products
are visible, and the only signatures are those dependent primarily
upon missing transverse energy. This is the case
if $\dmchi$ is above $600\mev$ but below the point at which the chargino
decay products can be seen as energetic jets or leptons.
For example, we explicitly considered the extreme of $\mgl\sim \mcpmone$
(which is quite natural in some string models --- see introduction).
In this case, we found that a monojet+$\etmiss$ signal will
probe (at 95\% CL) $\mgl\sim\mcpmone< 150\gev$,
while the $\gamma+\etmiss$ signal will probe $\mgl\sim\mcpmone<175\gev$.
(Both numbers assume that $S/B>0.2$ is required for a viable signal
in the presence of large background.)

Finally, we wish to note that the precise values of $\mcpmone$ and 
$\dmchi$ will be of significant theoretical interest. $\mcpmone$
will be determined on an event--by--event
basis if the chargino's momentum and velocity can both be measured. 
For the LHIT signal, $p$ will be measured by the curvature of the track
in the SVX and in the CT. The velocity will be measured 
by the ionization energy deposit in the SVX, CT and PS.
In the case of the TOF signal, there will be, in addition, an independent
time--of--flight determination of $\beta$.
For the DIT signal, the SVX+CT curvatures give a 
measurement of $p$ and the SVX+CT+PS
ionization energy deposits provide a determination of $\beta$.
For the SNT and SMET signals, $p$ and $\beta$ are measured 
by curvature and ionization (respectively) in the SVX.
(Note that, in all these cases,
accepting only events roughly consistent with a given value of
$\mcpmone$ will provide further discrimination against backgrounds.)
However, for the HIP and $\gam+\etmiss$ signals $\mcpmone$ can
only be estimated from the absolute event rate. As regards $\dmchi$,
it will be strongly constrained by knowing which signals
are present and their relative rates. In addition,
if the the soft charged pion can be detected,
its momentum distribution, in particular the end--point thereof,
would provide an almost direct determination of $\dmchi$.

\section{Conclusion} 

We have examined well--motivated scenarios of soft--SUSY--breaking with
high--scale boundary conditions that lead to
a lightest chargino and a lightest neutralino which are both 
wino--like and nearly degenerate in mass.  While it is not necessary, the
gluino can also be nearly degenerate with the $\cpmone$ and $\cnone$.
Typical values of the mass splitting $\dmchi\equiv\mcpmone-\mcnone$ in such models range from
$\lsim\mpi$ to several GeV.  If squarks, sleptons, and sneutrinos
are heavy (as is quite possible), then
the signals for supersymmetry at the Tevatron will be very limited
and very strongly--dependent upon the precise value of
$\dmchi$. In the very worst case, the gluino could also be heavy and 
the only substantial SUSY production cross sections would
be those for $\cpone\cmone$ and $\cpmone\cnone$. The mSUGRA--motivated
tri--lepton signal from $\cpmone\cntwo$ production is suppressed, because
$\cntwo$ has a small wino component and because
$\cpmone$ typically has a very small semi--leptonic branching ratio (for $\dmchi<\mpi$
the branching ratio can be large, but the lepton would
be very soft; for $\dmchi\ge\mpi$, the dominant decay is $\cpmone\to\pi^\pm \cnone$
until $\dmchi>1\gev$). The signatures that can be used to search for the 
$\cpone\cmone$ and $\cpmone\cnone$ production processes
are summarized in Table~\ref{signals}, given earlier in the paper.
We give a very brief summary of the results.

If $\dmchi$ is sufficiently small that the $c\tau$ for chargino decay 
is of order cm's or greater, then a promising signature 
is an isolated (chargino) track of significant length,
that possibly either penetrates to the  muon chambers or suddenly disappears
after appearing in the silicon vertex detector and/or central tracker.
The chargino track will not leave hadronic energy deposits and
will have a low--velocity, heavily--ionizing nature for many events.
In this case, one or more background--free signals will be
viable, allowing sensitivity to chargino masses competitive with, or in many
cases much superior to, those that can be probed for mSUGRA
boundary conditions. The importance of particular detector components
and triggers for such signatures, as outlined in
the main body of this paper, should be carefully reviewed by
the CDF and D\O~collaborations. The silicon tracker (SVX), central
tracking system (CT), pre--shower (PS), electromagnetic and hadronic
calorimeters (EC and HC), time--of--flight (TOF) measurement
and muon chambers (MC) all play crucial roles that change as a function
of $\dmchi$. (Our detector notation is summarized in Table~\ref{detector}.)

If $\dmchi\gsim 300\mev$, the only available signals have
substantial backgrounds.
For $300\mev\lsim \dmchi\lsim 600\mev$, the $c\tau$ of the chargino decay
is such that most events contain at least one
high--impact--parameter (HIP) pion track (the pion
coming from the decay of the chargino).  These HIP's can be observed
above the backgrounds provided that a radiated photon is present to
tag the $\etmiss$.
The HIP signature will probe up to $\mcpmone\sim 95-75\gev$
for $L=30\fbi$ ($\sim 68-53$ for $L=2\fbi$). The $L=30\fbi$ limits represent
some improvement over the DELPHI analysis based on LEP2 data, which currently
excludes $\mcpmone\lsim 60\gev$ in this same mass range.
For $\dmchi\gsim 600\mev$, the chargino decay will be effectively prompt,
and the main decay modes will be $\ell\nu\cnone$ and $q^\prime\anti q\cnone$.
If $\dmchi$ is at the same time 
too small for the jets or leptons coming from the chargino decays to 
be observable, then the best limits on $\mcpmone$
will come from the $\gam+\etmiss$ channel. The mass reach in $\mcpmone$
will not exceed about 60 GeV. This is about the same
as the DELPHI limit for $600\mev\leq\dmchi\leq 3\gev$
but is substantially below the DELPHI limit of $\sim 90\gev$ that
applies for $\dmchi>3\gev$.
In short, it is clear that a value of $\dmchi\geq 300\mev$ is 
both challenging and a real possibility.

\begin{table*}[p]
  \begin{center}
\sloppy
    \begin{tabular}[c]{|c|c|l|c|l|c|} 
\hline\hline\hline
~~$\dmchi$~~ & ~~$c\tau$~~ & \multicolumn{1}{c|}{Best RunII} & Trigger & \multicolumn{1}{c|}{Crucial measurements and} & ~~Reach~~ \\
(MeV) & (cm) & \multicolumn{1}{c|}{signature(s)} &   & \multicolumn{1}{c|}{associated detector components} & (GeV) \\
\hline\hline\hline
0     & $\infty$ & TOF   & MC  & TOF, $p_T$ (SVX+CT)  & 460 \\
      &          & LHIT  & MC  & $p_T$ (SVX+CT), $\dedx$ (SVX+CT+PS) & 450 \\
\hline\hline
125   &  1155    & TOF   & MC  & TOF, $p_T$ (SVX+CT)  & 430 \\
      &          & LHIT  & MC  & $p_T$ (SVX+CT),  $\dedx$ (SVX+CT+PS) & 425 \\
      &          & DIT   & CT  & $p_T$ (SVX+CT), HC veto &  425 \\
      &          & DIT6  & CT  & same + $\dedx$ (SVX+CT+PS), & 420 \\
\hline\hline
135   &   754    & LHIT  & MC  & $p_T$ (SVX+CT), $\dedx$ (SVX+CT+PS) & 425 \\
      &          & TOF   & MC  & TOF, $p_T$ (SVX+CT)  & 420 \\
      &          & DIT   & CT  & $p_T$ (SVX+CT), HC veto &  430 \\
      &          & DIT6  & CT  & same + $\dedx$ (SVX+CT+PS) & 420 \\
\hline\hline
140   &  317     & DIT   & CT  & $p_T$ (SVX+CT), HC veto &  430 \\
      &          & DIT6  & CT  & same + $\dedx$ (SVX+CT+PS) & 420 \\
\hline\hline
142.5 &   24     & SMET   & $\etmiss$ & $p_T$ (SVX), PS+EC+HC veto &  345 \\
      &          & SMET6  & $\etmiss$ & same + $\dedx$ (SVX) & 320 \\
\hline\hline
150   &   11     & SMET   & $\etmiss$ & $p_T$ (SVX), PS+EC+HC veto &  310 \\
      &          & SMET6  & $\etmiss$ & same + $\dedx$ (SVX) & 270 \\
\hline\hline
185   &   3.3    & SMET   & $\etmiss$ & $p_T$ (SVX), PS+EC+HC veto &  215 \\
      &          & SMET6  & $\etmiss$ & same + $\dedx$ (SVX) & 120 \\
\hline\hline
200 &   2.4      & SMET   & $\etmiss$ & $p_T$ (SVX), PS+EC+HC veto &  185 \\
\hline\hline
250 &   1.0     & SMET   & $\etmiss$ &  $p_T$ (SVX), PS+EC+HC veto  &  125 \\
\hline\hline
300 &   0.56    & HIP    & $\gamma,\etmiss$ & $b^\pi$ (SVX,L0), $p_T^\gam$, $\etmiss$, 
$p_T^\pi$ (CT),  EC+HC veto & 95 \\
\hline\hline
600 &   0.055   & HIP    & $\gamma,\etmiss$ & $b^\pi$ (SVX,L00), $p_T^\gam$, $\etmiss$, 
$p_T^\pi$ (CT), EC+HC veto & 75 \\
\hline\hline
$750-?$ &   $\sim 0$    & $\gam+\etmiss$ & $\gam,\etmiss$ & $p_T^\gam$, $\etmiss$ & $<60$ \\
\hline\hline
    \end{tabular}
    \caption{Summary of the best signals at RunII for $\cpone\cmone$ 
and $\cpmone\cnone$ production
and important detector components and measurements as a function of $\dmchi$.
Mass reaches quoted are 95\% CL for $L=30\fbi$. Detector component
notation is summarized in Table~\ref{detector}. Signal definitions
are summarized in Table~\ref{signals}. The PS, EC, or HC veto requires
no preshower, small EC, or small HC
energy deposits in a $\Delta R<0.4$ cone around the $\cpmone$ track candidate. 
$p_T$ ($p_T^\pi$) is the $p_T$
of the $\cpmone$ ($\pi^\pm$ from $\cpmone\to\pi^\pm\cnone$). $b^\pi$
is the $\pi^\pm$ impact parameter.}
    \label{tab:summary}
\fussy
  \end{center}
\end{table*}
An overall summary of the signals and their mass reach at the Tevatron
for detecting $\cpone\cmone$ and $\cpmone\cnone$ production
in the $M_2<M_1\ll|\mu|$ scenario (1) appears in Table~\ref{tab:summary}.

Mass reach in $\mcpmone$ is significantly improved if the gluino mass
is not so large that $\gl\gl$ production is suppressed.
In particular, we considered the other extreme
of $\mgl\sim \mcpmone$, as motivated in some of the models
for which the lightest chargino and neutralino are both wino--like. 
In this case, we found that
the $\gam+\etmiss$ signal will probe $\mgl\sim\mcpmone$ values as large as
$\sim 175\gev$ (for $L\geq 2\fbi$ at the Tevatron), while a monojet+$\etmiss$
signature can probe up to $\sim 150\gev$.

In some of the models in question, it is entirely
possible that $\dmchi$ is quite substantial ($>20\gev$).
The mass reach that can be achieved in this case requires further
study.  The signals considered in this paper are not very useful.
If the gluino is heavy, then one should
explore the potential of the tri--lepton signal coming from $\cpmone\cntwo$
production. However, this is a suppressed cross section when
both the lightest neutralino and lightest chargino are wino--like.
Standard mSUGRA studies do not apply without modification; the cross 
section must be rescaled and the lepton acceptance
recalculated as a function of $\dmchi$.
If the gluino is close in mass to the chargino, 
then the standard multi--jet+$\etmiss$
signal will be viable when $\dmchi$ is large enough for 
the jets in $\cpmone\to q^\prime\anti q\cnone$ decay to be visible. 
The like--sign di--lepton signal will also emerge for large $\dmchi$,
as the leptons in $\cpmone\to \ell\nu\cnone$ become energetic.
Since both signals will have substantial backgrounds,
a detailed study is required to determine
their exact mass reach as a function of $\dmchi$.
If the gluino has a moderate mass but 
$\mgl-\mcpmone$ is large enough, then the extra jets from $\gl\to
q^\prime\anti q\cpmone$ become visible and nearly all events contain
more than one jet. The multi--jet+$\etmiss$ signal becomes viable
as shown in Ref.~\cite{cdg2}.
(We note that the reach of the monojet+$\etmiss$ signature 
explored here deteriorates once the multi-jet+$\etmiss$ signal becomes
substantial. For instance, we find that the former signal is no longer
useful once $\mgl-\mcpmone>10\gev$.)

Of course, additional SUSY signals
will emerge if some of the squarks, sleptons and/or sneutrinos are light enough 
(but still heavier than the $\cpmone$) that 
their production rates are substantial.
In particular, leptonic signals
from the decays [\eg\ $\wtil \ell_L^{\pm}\to \ell^{\pm}\cnone$ or 
$\wtil\nu_{\ell} \to \ell^{\pm} \cmpone$ in scenario (1)] would be present.

Given the possibly limited reach of the Tevatron when the lightest
neutralino and chargino are nearly degenerate, it will be very important
to extend these studies to the LHC.  A particularly important issue
is the extent to which the large $c\tau$ tails of the $\cpmone$
decay distributions can yield a significant rate in 
the background--free channels studied here.
Hopefully, as a result of the very high
event rates and boosted kinematics expected at the LHC, 
the background--free
channels will remain viable for significantly larger $\dmchi$ 
and $\mcpmone$ values than those to which one has sensitivity at the Tevatron.
In this regard, a particularly important issue 
for maximizing the mass reach of these channels will be the
extent to which tracks in the silicon vertex detector and/or
in the central tracker can be used for triggering in
a high--luminosity enviroment. 

\bigskip

While finalizing the details of this study, another paper \cite{randall2}
appeared on the same topic.  Some of the signatures discussed here are
also considered in that paper. Our studies
are performed at the particle level and contain the most important
experimental details.

\bigskip

\centerline{\bf Acknowledgements}

This work was supported by the Department of Energy and by the Davis Institute
for High Energy Physics.

We benefited greatly from conversations with H. Frisch, G. Grim, R. Lander, 
G. Landsberg, D. Stuart, R. Van Kooten and J. Womersley.


\end{document}